\documentclass[aps,prb,twocolumn,a4paper,showpacs,groupedaddress,eqsecnum,fleqn,floatfix]{revtex4}
\usepackage{amssymb}
\usepackage{amsfonts}
\usepackage{amsmath}
\usepackage{bm}
\usepackage{natbib}
\usepackage{graphicx}
\usepackage{array}
\usepackage{srcltx}

\begin{document}

\title{The 2-site Hubbard and $t$-$J$ models}

\author{Adolfo Avella}
\email[E-mail: ]{avella@sa.infn.it}

\author{Ferdinando Mancini}
\email[E-mail: ]{mancini@sa.infn.it}

\author{Taiichiro Saikawa}

\affiliation{Dipartimento di Fisica ``E.R. Caianiello'' - Unit\`a INFM di Salerno\\
Universit\`a degli Studi di Salerno, I-84081 Baronissi (SA),
Italy}

\begin{abstract}
The fermionic and bosonic sectors of the 2-site Hubbard model have
been exactly solved by means of the equation of motion and Green's
function formalism. The exact solution of the $t$-$J$ model has
been also reported to investigate the low-energy dynamics. We have
successfully searched for the exact eigenoperators, and the
corresponding eigenenergies, having in mind the possibility to use
them as an operatorial basis on the lattice. Many local,
single-particle, thermodynamical and response properties have been
studied as functions of the external parameters and compared
between the two models and with some numerical and exact results.
It has been shown that the 2-site Hubbard model already contains
the most relevant energy scales of the Hubbard model: the local
Coulomb interaction $U$ and the spin-exchange one
$J=\frac{4t^2}U$. As a consequence of this, for some relevant
properties (kinetic energy, double occupancy, energy, specific
heat and entropy) and as regards the metal-insulator transition
issue, it has resulted possible to almost exactly mime the
behavior of larger systems, sometimes using a higher temperature
to get a comparable level spacing. The 2-site models have been
also used as toy models to test the efficiency of the Green's
function formalism for composite operators. The capability to
reproduce the exact solutions, obtained by the exact
diagonalization technique, gives a firm ground to the approximate
treatments based on this formalism.
\end{abstract}


\date{\today}

\maketitle

\section{Introduction}
\label{Sec:Intro}

Two are the aspects that gave so much popularity to the Hubbard
model: the richness of its dynamics that is thought to permit a
description of many puzzling issues like metal-insulator
transition, itinerant magnetism, electronic superconductivity, and
the simplicity of the Hamiltonian structure that let one speculate
about the possibility of finding the exact and complete solution
for any realization of the underlying lattice. Anyway, although
the model has been studied more than any other one in the last
fifty years, very few exact results are available and what we have
mainly regards either finite clusters or the infinite chain (i.e.,
the 1D case). For finite clusters of 2\cite{Harris:67,Shiba:72a}
or 4\cite{Shiba:72a,Heinig:72,Heinig:72a,Cabib:73,Schumann:01}
sites it is possible to find the complete set of eigenstates and
eigenvalues of the Hamiltonian and compute any quantity by means
of the thermal averages. However, it is not easy at all, although
possible in principle, to extract valuable and scalable (i.e.,
which can be used to find the solutions of bigger and bigger
clusters and, ultimately, of the infinite lattice cases)
information regarding the effective excitations present in the
system, the operators describing them and their dynamics. For the
infinite chain neither, we have all the information we wish; the
Bethe Ansatz is a very powerful tool, but is severely limited as
regards the range of applicability of the self-consistent
equations it supplies and the quantities for which it gives an
answer.

In this manuscript, in order to overcome the limitations discussed
above, we have exactly solved the Hubbard model, on a 2-site
cluster, completely within the equations of motion and the Green's
function formalism. By using this approach, we have had the
possibility to find the complete set of eigenoperators of the
Hamiltonian and the corresponding eigenenergies. This information
has been really fundamental as it permitted a deeper comprehension
of the features shown by the properties we have analyzed. It is
worth noting that the Hubbard model on a 2-site cluster is the
smallest system where both terms of the Hamiltonian (i.e., kinetic
and electrostatic) are effective and contributes to the dynamics.

By properly tuning the value of the temperature, we have found
that the 2-site system can almost perfectly mime, as regards
relevant properties such as the kinetic energy, the double
occupancy, the energy, the specific heat, the entropy and
fundamental issues such as the metal insulator transition, the
behavior of larger clusters and of the infinite chain. The tuning
of the temperature is necessary in order to get a comparable
effective level spacing (bigger the cluster, lower the spacing),
i.e. to excite the correct levels: the relevant energy scales are
present although the relative positions of the levels are affected
by the size of the system (only two k points!). The very positive
comparisons with exact results (Bethe Ansatz, exact
diagonalization) and numerical data (quantum Monte Carlo, Lanczos)
support the idea that a lot of physics can be described and
understood within this very small system, for which there is the
possibility to know the analytic expressions for all the
quantities under study. As regards the relevant scales of energy,
this 2-site system has demonstrated to contain all the necessary
ingredients to describe many features coming from the strong
electronic correlations and also appearing in the lattice case.
Two of the three relevant energy scales, which are thought to be
present in the Hubbard model, naturally emerge: the local Coulomb
interaction $U$ and the spin-exchange one $J=\frac{4t^2}U$ that
is, in principle, extraneous to the original purely electrostatic
Hamiltonian and is dynamically generated by the combined actions
of the two terms of the Hamiltonian. As useful guide to better
understand the low-energy dynamics we have also solved the $t$-$J$
model and presented the solution in parallel with the one found
for the Hubbard model.

This analysis, which have resulted to be really relevant by itself
as we have got a much better understanding of some energy scales
and internal parameter dynamics, has been worth to be performed
also as a prelude of the lattice analysis. In fact, the
eigenoperators we have found, both in the fermionic and the
bosonic sectors, can be used in the lattice case as a basis for
the Green's functions. In the strongly correlated systems, the
interactions can alter so radically the dynamics of the original
particles that these latter lose completely their own
identities\cite{Umezawa:93}. Actually, some new objects are
generated by the interactions and dictate the physical response.
They are not so easy to be identified: their number, exact
expression and relevance can only be suggested by the experience
and, when available, by exact and/or numerical results. For
instance, one can choose: the higher order fields emerging from
the equations of motion, the eigenoperators of some relevant
interacting terms, the eigenoperators of the problem reduced to a
small cluster, etc. In the last years, we have been focusing our
activity on the study of strongly correlated electronic models
like Kondo, $t$-$J$, $p$-$d$, Hubbard by means of the
\emph{Composite Operator
Method}\cite{COM1,COM2,Mancini:98b,Fiorentino:00a,Mancini:00} that
is based on two main ideas: one is the use of {\it composite
fields} as basis for our Green's functions, in accordance to what
has been discussed above, and the other one is the exploitation of
algebra constraints (e.g., the Pauli principle, the particle-hole
symmetry, the Ward-Takahashi identities, ...) to fix the correct
representation of the Green's functions and to recover the links
among the spin and charge configurations dictated by the
symmetries. It is worth noticing that the Composite Operator
Method is exact in itself. An additional approximation treatment
is needed when we deal with large or infinite degree-of-freedom
systems; in this case we have to treat in an approximate way the
otherwise intractable hierarchy of the equations of motion
generated by the projection procedure. If no approximation is
necessary (finite and reasonably small degree-of-freedom systems),
the COM cannot do else than give the exact solution. According to
this, the COM gives the exact solution also for the two systems
under analysis in this manuscript: the two-site Hubbard and t-J
models. Whenever, instead, we should resort to an approximate
treatment to close the hierarchy of the equations of motion
generated by the projection procedure, we expect some limitations
connected with the chosen approximation. For instance, we get only
the first moments correct if we truncate the equations of motion
hierarchy \cite{Mancini:98b}. On the other hand, it is really
worth noting that we properly take into account: the interaction
term of the Hamiltonian by using as basis operators its
eigenoperators \cite{Avella:02e}; the short-range correlations by
using as basic fields the eigenoperators of the problem reduced to
a small cluster \cite{Avella:03b}; the presence of a Kondo-like
singlet at low-energy by properly closing the equation of motion
of an \emph{ad hoc} chosen composite operator \cite{Villani:00}.

Obviously, we are aware that the exact diagonalization of this
very small system takes less than one afternoon to any graduate
student. Then, the reader could wonder why we decided to study
such a system so in detail. Well, the reasons are many and some
have been already pointed out above:
\begin{itemize}
 \item Any graduate student can surely compute eigenstates and
 eigenvalues of these Hamiltonians in one afternoon, but the analytic
 computation of Green's functions and correlation functions, in
 terms of the former eigen\emph{stuff}, is not that straightforward as
 one can think. At the end of the day, the time saved in computing
 eigenstates and eigenvalues instead of eigenoperators is
 almost fully recovered if you also take into account the time needed by the
 computation of the physical properties (Green's functions and
 correlations functions). Then, the possibility to have \emph{scalable}
 information putting, altogether, almost the same effort become quite tempting
 for anyone. At any rate and for the sake of completeness,
 we report in Appendix the expressions of Green's functions and
 correlations functions in terms of eigenstates and eigenvalues.
 \item The knowledge of the exact eigenoperators of a system is
 invaluable as they could be used as correct starting point for
 the application of the projection methods\cite{Mori,Rowe:68,Roth:69,Tserkovnikov:81,Fulde:95}
 to strongly correlated systems
 whose minimal model\cite{Mancini:00} is the exactly solved one.
 \item The Green's function formalism for composite operators is
 extremely complicated\cite{Mancini:00}. The 2-site Hubbard model
 can be used as toy model to fully explore this formalism and
 evidence the difficulties connected with the treatment of
 composite operators with non-canonical commutation relations.
 Within this system, the appearance of zero-frequency functions
 can be safely handled and resolved. The links among the different
 channels (fermion, charge, spin, pair) can be studied in detail.
 \item The 2-site Hubbard model, according to its status of
 minimal model\cite{Mancini:00}, contains the main scales of energy related with the
 interactions present in the Hamiltonian. The exact solution in
 terms of eigenoperators permits to individuate which are the
 composite fields responsible for the relevant transitions. These
 latter can be then used to efficiently study larger clusters.
 \item The absence of the three-site term, for obvious geometrical reasons, in the derivation of
 the 2-site $t$-$J$ model from the 2-site Hubbard one permits to
 push further the study of the relations between the two models.
 It is possible to exactly individuate the low-energy
 contributions and study them separately.
 \item Last, but not least, the capability of miming the numerical results
 for larger clusters (sometimes tuning the temperature and, consequently, the
 effective level spacing) opens the possibility to provide a
 testing-ground for the numerical techniques.
\end{itemize}

In the following, we define the models, give the self-consistent
solutions in the fermionic and bosonic sectors and study the
local, single-particle, thermodynamic and response properties of
the systems; the eigenoperators, eigenstates and eigenvalues of
the system are also given and analyzed in detail.

\section{The 2-site Hubbard and $\lowercase{t}$-$J$ models}
\label{Sec:Models}

The Hamiltonian of the Hubbard model\cite{Hubbard} for a $N$-site
chain reads as
\begin{equation} \label{HubHam}
H  =\sum_{ij}\left(t_{ij}-\mu  \delta _{ij}\right) c^{\dagger}(i)
c(j)+U\sum_{i}n_{\uparrow}(i) n_{\downarrow}(i)
\end{equation}
$\mu $ is the chemical potential, $c^{\dagger}(i)=\left(
c_{\uparrow}^{\dagger}(i),c_{\downarrow}^{\dagger}(i)\right) $ is
the electronic creation operator at the site $i$ in the spinorial
notation, $U$ is the on-site Coulomb interaction strength,
$n_{\sigma}(i)=c_{\sigma}^{\dagger}(i) c_{\sigma}(i)$ is the
charge density operator for spin $ \sigma $ at the site $i$ and
\begin{equation}
t_{ij}=-2t\alpha_{ij}=-t\sum_k \cos\left[k \left(i-j\right)\right]
\alpha(k)
\end{equation}
where $t$ is the hopping integral, $a$ is the lattice constant,
$\alpha_{ij}$ is the projection operator on the nearest-neighbor
sites and $\alpha(k)=\cos(k a)$. In the momentum representation,
the kinetic term $H_t$ of the Hamiltonian (\ref{HubHam}) reads as
\begin{equation}
H_t = -2t\sum_{k} \alpha(k) c^{\dagger}(k) c(k)
\end{equation}
where $k$ assumes the values $\frac{2\pi}{N a}l$ with
$l=0,\ldots,N-1$ for periodic boundary conditions and
$\frac{\pi}{(N+1)a}l$ with $l=1,\ldots,N$ for open boundary
conditions. We will study a 2-site cluster within periodic
boundary conditions: $N=2$ and $k=0, \frac{\pi}{a}$.

The Hamiltonian of the $t$-$J$
model\cite{Chao:77,Anderson:87,Zhang:88} for the same cluster and
boundary conditions reads as
\begin{multline}
H = \sum_{k} \left[-2t \alpha(k)-\mu \right] \xi^{\dagger}(k) \xi(k) \\
+\frac{1}{2} J \sum_{ij} \alpha_{ij}   \nu^{\mu}(i)   \nu_{\mu
}(j)
\end{multline} where $\xi (i)=\left[ 1-n(i)\right] c(i)$ is the
fermionic composite operator describing the transitions
$n=0\leftrightarrow n=1$, $\nu_{\mu}(i)=\xi ^{\dagger}(i) \sigma
_{\mu} \xi (i)$ is the total charge ($\mu =0$) and spin ($\mu =1,$
$2,$ $3$) density operator at the site $i$, $\sigma _{\mu
}=(1,\vec{\sigma})$, $\sigma ^{\mu}=(-1,\vec{\sigma})$ and $\vec{
\sigma}$ are the Pauli matrices.

We will extensively use the following definition for any field
operator $\Psi (i)$
\begin{equation}
\Psi ^{\alpha}(i) =\sum_{j}\alpha _{ij} \Psi (j)
\end{equation}
In particular, for the 2-site system we have $\Psi ^{\alpha
}\left( 0\right) =\Psi \left( a\right) $, $\Psi ^{\alpha}\left(
a\right) =\Psi \left( 0\right) $ and $\Psi ^{\alpha ^{2}}(i)
=\left( \Psi ^{\alpha}(i) \right) ^{\alpha}=\Psi (i) $.

\section{The fermionic sector}

\subsection{The equations of motion and the basis}

\subsubsection{The Hubbard model}

After the Hubbard Hamiltonian [Eq.~(\ref{HubHam})], the electronic
field $c(i)$ satisfies the following equation of motion
\begin{equation}
\mathrm{i}\frac{\partial}{\partial t}c(i)=-\mu  c(i)-2t c^{\alpha
}(i)+U \eta (i)
\end{equation}
with $\eta (i)=n(i) c(i)=-\frac13\sigma_k n_k(i) c(i)$. According
to this, we can decompose $c(i)$ as
\begin{equation}
c(i)=\xi (i)+\eta (i)
\end{equation}
where $\xi (i)$ and $\eta (i)$ are the Hubbard operators and
describe the transitions $n=0\leftrightarrow n=1$ and
$n=1\leftrightarrow n=2$, respectively. Moreover, they are the
local eigenoperators of the local term of the Hubbard Hamiltonian
and describe the original electronic field dressed by the on-site
charge and spin excitations. They satisfy the following equations
of motion
\begin{equation}
\begin{split}
& \mathrm{i}\frac{\partial}{\partial t}\xi (i) = -\mu \xi (i)-2t
c^{\alpha
}(i)-2t \pi (i) \\
& \mathrm{i}\frac{\partial}{\partial t}\eta (i) = (U-\mu )\eta
(i)+2t \pi (i)
\end{split}
\end{equation}
with
\begin{equation}
\pi (i)=\frac{1}{2}\sigma ^{\mu} n_{\mu}(i) c^{\alpha}(i)+\xi (i)
c^{\dagger \alpha}(i) \eta (i)
\end{equation}
where $n_{\mu}(i)=c^{\dagger}(i) \sigma _{\mu} c(i)$ is the total
charge ($\mu =0$) and spin ($\mu =1,$ $2,$ $3$) density operator
at the site $i$ in the Hubbard model. We use a different symbol to
distinguish it from the analogous operator, $\nu_{\mu}$, we have
defined in the $t$-$J$ model. The field $\pi (i)$ contains the
nearest neighbor charge, spin and pair excitations dressing the
electronic field $c(i)$.

\noindent The field $\pi (i)$ can be also decomposed as
\begin{equation}
\pi (i)=\xi _{s}(i)+\eta _{s}(i)
\end{equation}
where
\begin{equation}
\begin{split}
& \xi _{s}(i) = \frac{1}{2}\sigma ^{\mu} n_{\mu}(i) \xi ^{\alpha
}(i)+\xi
(i) {\eta ^{\alpha}}^{\dagger}(i) \eta (i) \\
& \eta _{s}(i) = \frac{1}{2}\sigma ^{\mu} n_{\mu}(i) \eta ^{\alpha
}(i)+\xi (i) {\xi ^{\alpha}}^{\dagger}(i) \eta (i)
\end{split}
\end{equation}
These latter, which are non-local eigenoperators of the local term
of the complete Hubbard Hamiltonian, satisfy the following
equations of motion
\begin{equation}
\begin{split}
& \mathrm{i}\frac{\partial}{\partial t}\xi _{s}(i) = -\mu  \xi
_{s}(i)+4t \eta
(i)+2t \xi _{s}^{\alpha}(i)+4t \eta_{s}^{\alpha}(i) \\
& \mathrm{i}\frac{\partial}{\partial t}\eta _{s}(i)  = (U-\mu
)\eta _{s}(i)+2t \eta (i)+2t \xi _{s}^{\alpha}(i)
\end{split}
\end{equation}

By choosing these four operators as components of the basic field
\begin{equation}
\psi (i)=\left(
\begin{array}{c}
\xi (i) \\
\eta (i) \\
\xi _{s}(i) \\
\eta _{s}(i)
\end{array}
\right)
\end{equation}
we obtain a closed set of equations of motion. In the momentum
space, we have
\begin{equation}
\mathrm{i}\frac{\partial}{\partial t}\psi (k)=\varepsilon (k) \psi
(k) \label{Hubpsieq}
\end{equation}
where the energy matrix $\varepsilon (k)$ is
\begin{equation}
\varepsilon (k) = \left(
\begin{array}{cccc}
-\mu -2t \alpha(k) & -2t \alpha(k) & -2t & -2t \\
0 & U-\mu & 2t & 2t \\
0 & 4t & -\mu +2t \alpha(k) & 4t \alpha(k) \\
0 & 2t & 2t \alpha(k) & U-\mu
\end{array}
\right)
\end{equation}

It is worth noting that $\psi (i)$ is an eigenoperator of the
Coulomb term of the Hubbard Hamiltonian for any lattice structure
according to the local nature of the interaction\cite{Avella:02e}.

\subsubsection{The $t$-$J$ model}

For the $t$-$J$ model we can obtain a closed set of equations of
motion by choosing as basic field
\begin{equation}
\psi (i)=\left(
\begin{array}{c}
\xi (i) \\
\zeta (i)
\end{array}
\right)
\end{equation}
where
\begin{equation}
\zeta (i)=\frac12 \sigma^\mu \nu_\mu(i) \xi^{\alpha}(i)
\end{equation}

\noindent We have
\begin{equation}
\mathrm{i}\frac{\partial}{\partial t}\psi (k)=\varepsilon (k) \psi
(k)
\end{equation}
with
\begin{equation}
\varepsilon (k)=\left(
\begin{array}{cc}
-\mu -2t \alpha (k) & -2t+2J \alpha
(k) \\
0 & -\mu +2t \alpha (k) -4J
\end{array} \right)
\end{equation}

It should be noted that the $t$-$J$ model exactly reproduces the
Hubbard model in the regime of very strong coupling (i.e., $U\gg
t$). The three-site term, appearing in the derivation of the
$t$-$J$ model from the Hubbard one, is completely absent in the
2-site system for obvious geometric reasons and the given
Hamiltonians are exactly equivalent in the strong coupling limit.
In particular, we note the following limits: $\lim_{U\gg
t}\xi_{s}(i)=\zeta (i)$ and  $\lim_{U\gg t}\eta(i)=\lim_{U\gg
t}\eta_{s}(i)=0$.

\subsection{The Green's function}

\subsubsection{The Hubbard model}

Let us now compute the thermal retarded Green's function $G\left(
k,\omega \right) ={\cal F}\left\langle {\cal R}\left[ \psi (i)
 \psi ^{\dagger}(j) \right] \right\rangle $ that, after the
equation of motion of $\psi (k)$ (\ref{Hubpsieq}), satisfies the
following equation
\begin{equation} \label{eqGF}
\left[\omega-\varepsilon (k)\right]G\left( k,\omega \right) =I(k)
\end{equation}
where $I(k) ={\cal F}\left\langle \left\{ \psi (i) ,\psi ^{\dagger
}(j) \right\} \right\rangle_{E.T.} $ is the normalization matrix.
The subscript $E.T.$ means that the anticommutator $\{\ldots\}$ is
evaluated at equal times. ${\cal F}$ and ${\cal R[ \ldots ]}$
stand for the Fourier transformation and the usual retarded
operator, respectively. $\langle \ldots \rangle$ indicates the
thermal average in the grand canonical ensemble. The solution of
Eq.~\ref{eqGF}, by taking into account the retarded boundary
conditions, is the following
\begin{equation}
G\left( k,\omega \right) =\sum_{n=1}^{4}\frac{\sigma ^{(n)}(k)}{
\omega -E_{n}(k)+{\mathrm{i} \delta}} \label{GFequ}
\end{equation}

The spectral weights $\sigma ^{(n)}(k)$ can be computed by means
of the following expression
\begin{equation}\label{sigmafor}
\sigma_{ab}^{(n)}(k)=\Lambda_{an}\sum_{c}\Lambda_{nc}^{-1}
I_{cb}(k)
\end{equation}
where $\Lambda$ is a matrix whose columns are the eigenvectors of
the energy matrix $\varepsilon (k) $.

The energy spectra $E_{n}(k)$ are the eigenvalues of the energy
matrix $\varepsilon (k) $
\begin{equation}
\begin{split}
& E_{1}(k) = -\mu -2t \alpha (k) \\
& E_{2}(k) = -\mu -2t \alpha (k)+U \\
& E_{3}(k) = -\mu +2t \alpha (k)-4J_{U} \\
& E_{4}(k) = -\mu +2t \alpha (k)+4J_{U}+U
\end{split}
\end{equation}
where $J_{U}=\frac{1}{8}\left( \sqrt{U^{2}+64t^{2}}-U\right) $. It
is worth noting that $\lim_{U\gg t}J_{U}=\frac{4t^{2}}{U}$ which
is the value of $J$ we get from the derivation of the $t$-$J$
model from the Hubbard one in the strong coupling regime. One can
easily check that the energy spectra $E_{n}(k)$ represent the
energy of the single-particle transitions among the eigenstates
given in App.~\ref{App:Eigen}.

The normalization matrix $I(k)$ has the structure
\begin{equation}
I(k)=\left(
\begin{array}{cccc}
I_{11} & 0 & I_{13}(k) & 0 \\
0 & I_{22} & 0 & I_{24}(k) \\
I_{13}(k) & 0 & I_{33}(k) & 0 \\
0 & I_{24}(k) & 0 & I_{44}(k)
\end{array}
\right)
\end{equation}
The explicit form of its entries are
\begin{equation}
\begin{split}
&I_{11} = 1-\frac{n}{2} \\
&I_{22} = \frac{n}{2} \\
&I_{13}(k) = \Delta +\alpha (k)(p-I_{22}) \\
&I_{24}(k) = -\Delta -\alpha (k) p \\
&I_{33}(k) = -2(p-I_{22})-2\alpha (k) \Delta \\
&I_{44}(k) = I_{22}
\end{split}
\end{equation}
with
\begin{subequations}
\begin{align}
& n =\left\langle n(i)\right\rangle \\
& \Delta =\left\langle \xi ^{\alpha}(i) \xi ^{\dagger
}(i)\right\rangle
-\left\langle \eta ^{\alpha}(i) \eta ^{\dagger}(i)\right\rangle  \\
& p =\frac{1}{4} \chi^{\alpha} - d \label{pequ}\\
& \chi^{\alpha} = \sum_{\mu=0}^3 \left\langle n_{\mu}^{\alpha}(i) n_{\mu}(i)\right\rangle \\
& d = \left\langle \xi _{\uparrow}(i)\eta _{\downarrow}(i)[\eta
_{\downarrow}^{\dagger}(i)\xi _{\uparrow}^{\dagger}(i)]^{\alpha
}\right\rangle
\end{align}
\end{subequations}
$\Delta $ gives a measure of the difference in mobility between
the two Hubbard subbands. $p$ contains charge, spin and pair
correlation functions. In order to compute the electronic Green's
function $G_{cc}=\left\langle {\cal R}\left[ c c^{\dagger}\right]
\right\rangle =G_{11}+2G_{12}+G_{22}$ we only need the following
quantities
\begin{equation}
\begin{split}
&\sigma _{11}^{(1)}(k) =I_{11}+\frac{1}{2}\alpha (k) I_{13}(k) \\
&\sigma _{11}^{(2)}(k) =0 \\
&\sigma _{11}^{(3)}(k) =-\frac{1}{2}\alpha (k) I_{13}(k)\left(
1-\frac{
4J_{U}}{U+8J_{U}}\right) \\
&\sigma _{11}^{(4)}(k) =-\frac{1}{2}\alpha (k)
I_{13}(k)\frac{4J_{U}}{ U+8J_{U}}
\end{split}
\end{equation}
\begin{equation}
\begin{split}
&\sigma _{12}^{(1)}(k) =\sigma _{12}^{(2)}(k)=0 \\
&\sigma _{12}^{(3)}(k) =-\sigma
_{12}^{(4)}(k)=-I_{13}(k)\frac{2t}{U+8J_{U}}
\end{split}
\end{equation}
\begin{equation}
\begin{split}
&\sigma _{22}^{(1)}(k) =0 \\
&\sigma _{22}^{(2)}(k) =I_{22}+\frac{1}{2}\alpha (k) I_{13}(k) \\
&\sigma _{22}^{(3)}(k) =-\frac{1}{2}\alpha (k)
I_{13}(k)\frac{4J_{U}}{
U+8J_{U}} \\
&\sigma _{22}^{(4)}(k) =-\frac{1}{2}\alpha (k) I_{13}(k)\left(
1-\frac{ 4J_{U}}{U+8J_{U}}\right)
\end{split}
\end{equation}

Equation (\ref{GFequ}) does not define uniquely the Green's
function, but only its functional dependence\cite{Mancini:00}. The
knowledge of the spectral weights $\sigma ^{(n)}(k)$ and the
spectra $E_{n}(k)$ requires the determination of the parameters
$\Delta $ and $p$, together with that of the chemical potential
$\mu $. The connection between the parameter $\Delta$ and the
elements of the Green's function and the equation fixing the
filling allow us to determine these two parameters as functions of
the parameter $p$
\begin{equation}
\left\{
\begin{array}{l}
n=2(1-C_{11}-C_{22}) \\
\Delta =C_{11}^{\alpha}-C_{22}^{\alpha}
\end{array}
\right.  \label{SelfHub1}
\end{equation}
where the correlation functions $C_{ab}=\left\langle \psi _{a}(i)
 \psi _{b}^{\dagger}(i) \right\rangle $ and $ C_{ab}^{\alpha
}=\left\langle \psi _{a}^{\alpha}(i)  \psi _{b}^{\dagger}(i)
\right\rangle $ can be computed, after the spectral theorem, by
means of the following equations
\begin{equation}
\begin{split}
&C_{ab} =\frac{1}{4}\sum_{k}\sum_{n=1}^{4}[1+T_{n}(k)]\sigma
_{ab}^{(n)}(k) \\
&C_{ab}^{\alpha} =\frac{1}{4}\sum_{k}\sum_{n=1}^{4}\alpha
(k)[1+T_{n}(k)]\sigma _{ab}^{(n)}(k)
\end{split}
\end{equation}
with
\begin{equation}
T_{n}(k)=\tanh \left( \frac{E_{n}(k)}{2T}\right)
\end{equation}

The presence of parameters related to bosonic correlation
functions within the fermionic dynamics (e.g., the parameter $p$)
is characteristic of strongly correlated systems. In these
systems, the elementary fermionic excitations are described by
composite operators whose non-canonical anticommutation relations
contain bosonic operators. Then, according to the general
projection procedure, which is approximate for an \emph{infinite}
(i.e., with infinite degrees of freedom) system and coincident
with the exact diagonalization for a \emph{finite} system, the
energy and normalization matrices generally contain correlation
functions of these bosonic operators.

The parameter $p$ could be computed through its definition (see
Eq.~\ref{pequ}). In this case we need to open the bosonic sector
(i.e., the charge, spin and pair channels), to determine the
eigenoperators and the normalization and energy matrices and to
require the complete self-consistency between the fermionic and
the bosonic sectors. Moreover, we run into the problem of fixing
the value of the zero-frequency constants\cite{ZFC}. This
procedure will be widely discussed in the next section where the
bosonic sector and the relative channels will be opened and
completely solved. However, we can use another procedure to fix
the parameter $p$ without resorting to the bosonic sector. We have
to think over the reason why the parameters $\mu$, $\Delta$ and
$p$ appear into our equations: we have not fixed yet the
representation of the Hilbert space where the Green's functions
are realized. The proper representation is the one where all the
relations among operators coming from the algebra (e.g.,
$n_\sigma^2=n_\sigma$) and the symmetries (e.g., particle-hole,
spin rotation, ...) are verified as relations among matrix
elements. In the case of usual electronic operators the
representation is fixed by simply determining the value of the
chemical potential. In the case of composite fields, which do not
satisfy canonical anticommutation relations, fixing the
representation is more involved and the presence of internal
parameters (e.g., $\mu$, $\Delta$ and $p$) is essential to the
process of determining the proper representation. The requirement
that the algebra is satisfied also at macroscopic level generates
a set of self-consistent equations, \emph{the local algebra
constraints}, that will fix the value of the internal
parameters\cite{Mancini:00}.

In particular, for the 2-site system the equation
\begin{equation}\label{SelfHub2}
\langle\xi(i) \eta^{\dagger}(i)\rangle=0
\end{equation}
coming from the algebra constraint $\xi(i) \eta^{\dagger}(i)=0$,
together with Eqs.~\ref{SelfHub1} constitute a complete set of
self-consistent equations which exactly solves the fermionic
dynamics allowing to compute the internal parameters (i.e., $\mu$,
$\Delta$ and $p$) for any value of the model ($t$, $U$) and
thermodynamical ($n$, $T$) parameters. This procedure is clearly
extremely simpler than that requiring the opening of the bosonic
sector. It is worth mentioning that the system of self-consistent
equations can be analytically solved as regards the parameters
$\Delta $ and $p$ as functions of the chemical potential $\mu$. It
is possible to show that the equation for this latter parameter,
as function of the internal (i.e., model and thermodynamical)
parameters, exactly agrees with that coming from the thermal
averages (cfr. Eq.~\ref{eqmun}) as it should be since the model is
exactly solved. This further confirms the validity, the
effectiveness and the power of the method used to fix the
representation.

In the next section, we will see that also in the case of the
bosonic sector, the determination of the proper representation
will be obtained by means of the local algebra constraints which
will fix the value of the zero-frequency constants. It will be
also shown that the two procedures for computing the parameter $p$
are equivalent, as it should be, and give exactly the same results
for the fermionic dynamics although with remarkably different
effort. It is worth noticing that the use of the local algebra
constraints, which is unavoidable to fix the representation both
in the fermionic and in the bosonic sectors\cite{Mancini:00},
permits to close the fermionic sector on itself without resorting
to the bosonic one also in the lattice case\cite{COM2} where a
fully self-consistent solution of both sectors, although
approximate, is very difficult to obtain.

\subsubsection{The $t$-$J$ model}

In the $t$-$J$ model, the energy spectra (i.e., the eigenvalues of
$\varepsilon (k)$) read as
\begin{equation}
\begin{split}
&E_{1}(k) = -\mu -2t \alpha (k) \\
&E_{2}(k) = -\mu +2t \alpha (k) -4J
\end{split}
\end{equation}
In particular, $E_{2}(k)$ corresponds to the single-particle
transitions between the single occupied states and the 2-site
singlet. The latter is the ground state at half filling.
$\zeta(k)$ drives the corresponding fermionic excitation between
the single occupied states and the 2-site singlet and the
appearance of the most relevant scale of energy at low
temperatures $J$.

The normalization matrix $I(k)$ has the following entries
\begin{equation}
\begin{split}
&I_{11} = 1-\frac{n}{2} \\
&I_{12}(k) = C_{11}^{\alpha}+\alpha
(k)\left(\frac14 \chi ^{\alpha} -\frac n2\right) \\
&I_{22}(k) = -2I_{12}(k)
\end{split}
\end{equation}
In the $t$-$J$ model the spin and charge correlator
$\chi^{\alpha}$ is defined as
$\chi^{\alpha}=\sum_{\mu=0}^3\left\langle \nu_{\mu}^{\alpha}(i)
\nu_{\mu}(i)\right\rangle$ in agreement with the strong coupling
nature of the model. The spectral weights $\sigma ^{(n)}(k)$ have
the following expressions
\begin{equation}
\begin{split}
&\sigma _{11}^{(1)}(k) = I_{11}+\frac{1}{2}I_{12}(k) \\
&\sigma _{11}^{(2)}(k) = -\frac{1}{2}I_{12}(k)
\end{split}
\end{equation}
\begin{equation}
\begin{split}
&\sigma _{12}^{(1)}(k) = 0 \\
&\sigma _{12}^{(2)}(k) = I_{12}(k)
\end{split}
\end{equation}
\begin{equation}
\begin{split}
&\sigma _{22}^{(1)}(k) = 0 \\
&\sigma _{22}^{(2)}(k) = I_{22}(k)
\end{split}
\end{equation}

Also in this case, as for the Hubbard model, we could compute
$\chi^{\alpha}$ opening the bosonic sector for the $t$-$J$ model.
Again, following the reasoning given in the previous section, a
simpler and completely equivalent procedure relies on the
exploitation of the algebra constraint $\zeta(i)
\zeta^\dagger(i)=-2\zeta(i) \xi^\dagger(i)$. According to this,
the parameters $\mu$ and $\chi^{\alpha}$ can be computed by means
of the following set of self-consistent equations
\begin{equation}
\left\{
\begin{array}{l}
n=1-C_{11} \\
C_{22}=-2C_{12}
\end{array}
\right.
\end{equation}

\section{The bosonic sector}

\subsection{The equations of motion and the basis}

\subsubsection{The Hubbard model}

\paragraph{The spin and charge sectors.}

After the Hubbard Hamiltonian [Eq.~(\ref{HubHam})], the charge
($\mu =0$) and spin ($\mu =1$, $2$, $3$) density operator $n_{\mu
}(i)=c^{\dagger}(i) \sigma _{\mu} c(i)$ satisfies a closed set of
equations of motion, which describes the spin and charge dynamics
in the system under study, once we choose as basic field
\begin{equation}
\phi _{\mu}(i)=\left(
\begin{array}{c}
\phi^{(1)} _{\mu}(i) \\
\phi^{(2)} _{\mu}(i) \\
\phi^{(3)} _{\mu}(i) \\
\phi^{(4)} _{\mu}(i) \\
\phi^{(5)} _{\mu}(i) \\
\phi^{(6)} _{\mu}(i)
\end{array}
\right)=\left(
\begin{array}{c}
n_{\mu}(i) \\
g_{\mu}(i) \\
w_{\mu}(i)+w_{\mu}^{\dagger}(i) \\
w_{\mu}(i)-w_{\mu}^{\dagger}(i) \\
h_{\mu}(i)-h_{\mu}^{\dagger}(i) \\
h_{\mu}(i)+h_{\mu}^{\dagger}(i)
\end{array}
\right)
\end{equation}
where
\begin{align}
g_{\mu}(i)& =c^{\dagger}(i) \sigma _{\mu} c^{\alpha}(i)-c^{\alpha
\dagger}(i) \sigma _{\mu} c(i) \\
w_{\mu}(i)& =d_{\mu}(i)-d_{\mu}^{\alpha}(i) \\
h_{\mu}(i)& =f_{\mu}(i)-f_{\mu}^{\alpha}(i) \\
d_{\mu}(i)& =\xi ^{\dagger}(i) \sigma _{\mu} \eta ^{\alpha}(i) \\
f_0(i)& =-\eta ^{\dagger}(i) \eta (i)-d_0^{\dagger}(i)
d_0^{\alpha}(i)  \notag
\\
& +\eta ^{\dagger}(i) \eta (i) \xi ^{\alpha \dagger}(i) \xi
^{\alpha
}(i) \\
f_{k}(i)& =\xi ^{\dagger}(i) \xi (i) n_{k}^{\alpha
}(i)-\frac{1}{2} \mathrm{i} \varepsilon^{kpq} n_{q}(i)
n_{q}^{\alpha}(i)
\end{align}
The components of $\phi _{\mu}(i)$, suggested by the hierarchy of
the equations of motion, are either hermitian or anti-hermitian as
densities or currents should be at any order in time
differentiation. $\phi _{\mu}(i)$ satisfies the following equation
of motion in momentum space
\begin{equation}\label{emcs}
\mathrm{i}\frac{\partial}{\partial t}\phi _{\mu}(k)=\omega_b (k)
\phi _{\mu}(k)
\end{equation}
where
\begin{equation}
\omega_b (k)=\left(
\begin{array}{cccccc}
0 & -2t & 0 & 0 & 0 & 0 \\
-4t[1-\alpha(k)] & 0 & U & 0 & 0 & 0 \\
0 & 0 & 0 & U & 2t & 0 \\
0 & 0 & U & 0 & 0 & 2t \\
0 & 0 & 8t & 0 & 0 & 0 \\
0 & 0 & 0 & 8t & 0 & 0
\end{array}
\right)
\end{equation}

\paragraph{The pair channel.}

Within the analysis of the dynamics of the Hubbard model, another
relevant bosonic operator is the \textit{pair} operator
$p(i)=c_{\uparrow}(i) c_{\downarrow}(i)$. The set of composite
fields
\begin{equation}
P(i)=\left(
\begin{array}{c}
P^{(1)}(i) \\
P^{(2)}(i) \\
P^{(3)}(i) \\
P^{(4)}(i) \\
P^{(5)}(i) \\
P^{(6)}(i)
\end{array}
\right)
\end{equation}
where
\begin{align}
P^{(1)}(i)&=p(i) \\
P^{(2)}(i)&=c_{\uparrow}(i) c_{\downarrow}^{\alpha}(i) \\
P^{(3)}(i)&=c_{\uparrow}(i) \eta _{\downarrow}^{\alpha}(i)+\eta
_{\uparrow
}(i) c_{\downarrow}^{\alpha}(i) \\
P^{(4)}(i)&=p(i) n^{\alpha}(i) \\
P^{(5)}(i)&=2\eta _{\uparrow}(i) \eta _{\downarrow}^{\alpha}(i) \\
P^{(6)}(i)&=p(i) \eta^{\alpha \dagger}(i) \eta^{\alpha}(i)
\end{align}
satisfies the following closed set of equations of motion
\begin{equation}\label{emp}
\mathrm{i}\frac{\partial}{\partial t}P(k)=\omega_p(k) P(k)
\end{equation}
where
\begin{widetext}
\begin{equation}
\omega_p(k)=\left(
\begin{array}{cccccc}
U-2\mu  & -2t\left[ 1+\alpha (k)\right]  & 0 & 0 & 0 & 0 \\
-2t\left[ 1+\alpha (k)\right]  & -2\mu  & U & 0 & 0 & 0 \\
0 & 0 & U-2\mu  & -2t\left[ 1+\alpha (k)\right]  & U & 0 \\
0 & 0 & -2t\left[ 1+\alpha (k)\right]  & U-2\mu  & 0 & 0 \\
0 & 0 & 0 & 0 & 2(U-\mu ) & -2t\left[ 1+\alpha (k)\right]  \\
0 & 0 & 0 & 0 & -2t\left[ 1+\alpha (k)\right]  & U-2\mu
\end{array}
\right)
\end{equation}
\end{widetext}

\subsubsection{The $t$-$J$ model}

The basis  for the bosonic sector in the $t$-$J$ model is given by
\begin{equation}
\varphi(i)=\left(
\begin{array}{c}
\varphi^{(1)}_{0}(i) \\
\varphi^{(2)}_{0}(i)
\end{array}
\right)
\end{equation}
for the charge channel, and by
\begin{equation}
\varphi_{k}(i)=\left(
\begin{array}{c}
\varphi^{(1)}_{k}(i) \\
\varphi^{(2)}_{k}(i) \\
\varphi^{(3)}_{k}(i) \\
\varphi^{(4)}_{k}(i)
\end{array}
\right)
\end{equation}
for the spin channel. We have defined
\begin{align}
\varphi^{(1)}_{\mu}(i) &= \nu_{\mu}(i) \\
\varphi^{(2)}_{\mu}(i) &= \xi^{\dagger}(i) \sigma_{\mu}
\xi^{\alpha}(i)-\xi
^{\dagger\alpha}(i) \sigma_{\mu} \xi(i) \\
\varphi^{(3)}_{k}(i) &= \nu(i) \nu_{k}^{\alpha}(i) \\
\varphi^{(4)}_{k}(i) &= \mathrm{i} \varepsilon^{kpq} \nu_{p}(i)
\nu_{q}^{\alpha}(i)
\end{align}

In the momentum space we have the following closed sets of
equations of motion
\begin{align}\label{emcsJ}
\mathrm{i}\frac{\partial}{\partial t}\varphi(q) &
=\omega_{c}(q) \varphi(q) \\
\mathrm{i}\frac{\partial}{\partial t}\varphi_{k}(q) &
=\omega_{s}(q) \varphi _{k}(q)
\end{align}
with
\begin{equation}
\omega_{c}(q)=\left(
\begin{array}{cc}
0 & -2t \\
-4t[1-\alpha(q)] & 0
\end{array}
\right)
\end{equation}
and
\begin{equation}
\omega_{s}(q)=\left(
\begin{array}{cccc}
0 & -2t & 0 & -2J \\
-4t[1-\alpha(q)] & 0 & -4t[1-\alpha(q)] & 0 \\
0 & 0 & 0 & 2J \\
0 & 0 & 4J[1-\alpha(q)] & 0
\end{array}
\right)
\end{equation}

\subsection{The Green's function}

In Ref.~\onlinecite{Mancini:00}, we have shown that the retarded
and causal Green's functions contain substantially different
information according to the unavoidable presence of the zero
frequency constants (ZFC). In particular, we have reported on the
relations between different types of Green's functions and on the
correct order in which they should be computed. According to this,
in the bosonic sector, we have to start from the causal Green's
function and not from the retarded one, as we correctly did in the
fermionic sector.

\subsubsection{The Hubbard model}

\paragraph{The spin and charge sectors.}\label{suscss}

After Eq.~\ref{emcs}, the thermal causal Green's function $G_{\mu
}\left( k,\omega \right) =\mathcal{F}\left\langle
\mathcal{T}\left[ \phi _{\mu}(i) \phi _{\mu}^{\dagger}(j)\right]
\right\rangle $ satisfies the following equation
\begin{equation}\label{grecs}
\left[ \omega -\omega_b \left( k\right) \right] G_{\mu}\left(
k,\omega \right) =I_{\mu}(k)
\end{equation}
where the relevant entries of the normalization matrix
$I_{\mu}(k)= \mathcal{F} \left\langle \left[ \phi _{\mu}(i),\phi
_{\mu}^{\dagger}(j)\right] \right\rangle_{E.T.}$ are
\begin{align}
I_{11\mu}(k) & =0 \\
I_{12\mu}(k) & =4[1-\alpha(k)]C_{cc}^{\alpha} \\
I_{13\mu}(k) & =0 \\
I_{14\mu}(k) & =8[1-\alpha(k)]C_{12}^{\alpha} \\
I_{15\mu}(k) & =-[1-\alpha(k)]q^{(\mu)} \\
I_{16\mu}(k) & =0
\end{align}
with
\begin{align}
& C_{cc}^{\alpha} =\left\langle c^{\alpha}(i) c^{\dagger}(i)\right\rangle \\
& q^{(0)} =16d \\
& q^{(k)} =\frac{8}{3}\chi_{s}^{\alpha}\\
& \chi_{s}^{\alpha}=\sum_{k=1}^3 \left\langle n_{k}^{\alpha}(i)
n_{k}(i)\right\rangle
\end{align}
$\mathcal{T}[\ldots]$ stands for the usual time ordering operator.

The solution of Eq.~\ref{grecs} is the following
one\cite{Mancini:00}
\begin{multline}\label{greenz}
G_{\mu}\left( k,\omega \right) =
-2\mathrm{i} \pi \Gamma_\mu(k) \delta(\omega)\\
+\sideset{}{'}\sum_{i=1}^6 \frac{\sigma_\mu
^{(i)}(k)}{1-\mathrm{e}^{-\beta \omega}}\left[\frac{1}{\omega
-E_{ib}(k)+ \mathrm{i} \delta} - \frac{\mathrm{e}^{-\beta
\omega}}{\omega -E_{ib}(k)-\mathrm{i} \delta}\right]
\end{multline}
where
\begin{multline}
\Gamma_\mu(k)= \frac12 \lim_{\omega \rightarrow 0}
\omega\\
\times \mathcal{F}\left[\theta(t_i-t_j)\left\langle \phi _{\mu
}(i) \phi _{\mu}^{\dagger
}(j)\right\rangle-\theta(t_j-t_i)\left\langle \phi _{\mu
}^{\dagger}(j) \phi _{\mu}(i)\right\rangle\right]
\end{multline}
is the zero frequency function and is undetermined at this level
unless to compute the quite \emph{anomalous} Green's function
appearing in its definition, which involves anticommutators of
bosonic operators \cite{Mancini:00}. The general definition of a
zero frequency function in terms of eigenvectors and eigenvalues
is given in Eq.~\ref{Gdef}. The primed sum is restricted to values
of $i$ for which $E_{ib}(k) \neq 0$.

The $E_{ib}(k)$ are the eigenvalues of the energy matrix $\omega_b
\left( k\right)$
\begin{align}
E_{1b}(k)& =-2t\sqrt{2[1-\alpha (k)]} \\
E_{2b}(k)& =2t\sqrt{2[1-\alpha (k)]} \\
E_{3b}(k)& =-U-4J_{U} \\
E_{4b}(k)& =-4J_{U} \\
E_{5b}(k)& =4J_{U} \\
E_{6b}(k)& =U+4J_{U}
\end{align}
According to this, the primed sum in Eq.~\ref{greenz} does not
contain the $i=1$ and $i=2$ elements for $k=0$ ($E_{1,2b}(0)=0$)
and the zero frequency function $\Gamma_\mu(k)$ reduces to the
constant $\Gamma_\mu=\Gamma_\mu(0)$. The spectral weights $\sigma
_{ab\mu}^{(i)}(k)$ with $a$ and $b=1$, $2$, computed through
Eq.~\ref{sigmafor} have the following expressions
\begin{widetext}
\begin{align}
\sigma _{11\mu}^{(1)}(k)& =-\sigma _{11\mu}^{(2)}(k)=
\frac{\sqrt{1-\alpha (k)}\left[U\left\{t q^{(\mu )}\left[ 1+\alpha
(k)\right] +4U\left[ 1-\alpha (k)\right] C_{12}^{\alpha
}\right\}+2C^{\alpha}\left\{8\left[ 1+\alpha (k) \right]
^{2}t^{2}-U^{2}\left[ 1-\alpha
(k)\right]\right\}\right]}{\sqrt{2}\left\{8\left[ 1+\alpha
(k)\right] ^{2}t^{2}-U^{2}\left[ 1-\alpha (k)\right]\right\}} \\
\sigma _{11\mu}^{(3)}(k)& =-\sigma _{11\mu}^{(6)}(k)=
\frac{4tU\left[ 1-\alpha (k)\right]\left[4(U+4J_{U})C_{12}^{\alpha
}-t q^{(\mu )}\right]}{\left(U+8J_{U}\right) \left\{ 16
t^{2}\left[ 1+\alpha (k)\right]+2U(U+4J_{U})\right\}} \\
\sigma _{11\mu}^{(4)}(k)& =-\sigma _{11\mu}^{(5)}(k)=
-\frac{4tU\left[ 1-\alpha (k)\right]\left[16J_{U}C_{12}^{\alpha
}+t q^{(\mu )}\right]}{\left(U+8J_{U}\right) \left\{ 16
t^{2}\left[ 1+\alpha (k)\right]-8U J_{U}\right\}}
\end{align}
\end{widetext}
and
\begin{align}\label{Ward}
\sigma _{12\mu}^{(i)}(k)& =-\frac{E_{ib}(k)}{2t}\sigma _{11\mu}^{(i)}(k) \\
\sigma _{22\mu}^{(i)}(k)& =-\frac{E_{ib}(k)}{2t}\sigma _{12\mu
}^{(i)}(k)=\left( \frac{E_{ib}(k)}{2t}\right) ^{2}\sigma _{11\mu
}^{(i)}(k)
\end{align}

We note the sum rules
\begin{equation}
\sum_{i=1}^{6}\sigma_{ab\mu}^{(i)}(k) = I_{ab\mu}(k)
\end{equation}

In order to finally compute the Green's function $G_{11\mu}\left(
k,\omega \right)$ we should fix the internal parameters
$\chi_{s}^{\alpha}$, $d$ and the zero frequency constant
$\Gamma_{11\mu}$.

The parameter $\chi_{s}^{\alpha}$ is directly connected to the
Green's function. Let us consider the correlation function
$D_{\mu}\left( k,\omega\right) = \mathcal{F}\left\langle
\phi_{\mu}(i) \phi_{\mu}^{\dagger}(j)\right\rangle $ which is
linked to the causal Green's function through the spectral theorem
\begin{multline}
D_{\mu}\left( k,\omega\right) =-\left(
1+\tanh\frac{\omega}{2T}\right) \Im\left[G_{\mu}\left( k,\omega\right)\right]\\
=2\pi \Gamma_\mu \delta(\omega)+ 2 \pi \sideset{}{'}\sum_{i=1}^6
\delta\left[\omega -E_{ib}(k)\right]\frac{\sigma_\mu^{(i)}(k)}
{1-\mathrm{e}^{-\beta E_{ib}(k)}}
\end{multline}
Then,
\begin{equation}\label{cs1}
\chi_{s}^{\alpha}=3\left\langle \phi^{(1)}_{3}(i)
\phi^{(1)\alpha}_3(i)\right\rangle
\end{equation}

The computation of the parameter $d$ requires instead the opening
of the pair sector. Otherwise, it could be computed through the
following relation with the parameter $p$
\begin{equation}\label{cs2}
d=\frac{1}{4}\chi^{\alpha}-p
\end{equation}
already given in the fermionic sector (see Eq.~\ref{pequ}), once
we note that $\chi^{\alpha}=\chi_{s}^{\alpha}+\left\langle
\phi^{(1)}_0(i) \phi^{(1)\alpha}_0(i)\right\rangle$.

The zero frequency constant $\Gamma_{11\mu}$ cannot be directly
connected to any correlation function at this level and its
determination requires the use of local algebra constraints. In
particular, we can use the following relation
\begin{equation}\label{cs3}
\left\langle \phi^{(1)}_\mu(i)
\phi^{(1)}_\mu(i)\right\rangle=\left\langle n_{\mu}(i)
n_{\mu}(i)\right\rangle =n+2D (2\delta_{\mu 0}-1)
\end{equation}
where the double occupancy $D=\langle n_\uparrow(i)
n_\downarrow(i)\rangle$ is given by $D=I_{22}-C_{22}$.

Equations~\ref{cs1}, \ref{cs2} and \ref{cs3} constitute a complete
set of self-consistent equations which allow to compute the
parameters $\chi_{s}^{\alpha}$, $d$ and $\Gamma_{11\mu}$ and then
to determine the Green's function $G_{11\mu}\left( k,\omega
\right)$.

It is worth noting that, once $C_{cc}^{\alpha}$,
$C_{12}^{\alpha}$, $p$ and $D$ are computed in the fermionic
sector, the bosonic correlation functions can be easily obtained.
We have the following expressions for the relevant correlators and
zero-frequency constants
\begin{align}
& \chi_{s}^{\alpha} = 3\frac{n-2D-2a}{1+2b} \\
& \chi_{c}^{\alpha} = \left\langle
\phi^{(1)}_0(i) \phi^{(1)\alpha}_0(i)\right\rangle=n+2D-2a\notag \\
& -\frac{b}{1+b}\left[ \frac{3(n-2D-2a)}{1+2b}
+n+2D-2a-4p\right] \\
& d =\frac{1}{4(1+b)}\left[
\frac{3(n-2D-2a)}{1+2b}+n+2D-2a-4p\right] \\
& \Gamma_{110} =\frac{1}{1+b}\left[2n+4D-2a\right.\notag \\
& \left.+b(n+2D-3\frac{n-2D-2a}{1+2b} +4p\right] \\
& \Gamma_{113} =\frac{2(n-2D)(1+b)-2a}{1+2b}
\end{align}
where
\begin{align}
a & =-(C_{cc}^{\alpha}-2C_{12}^{\alpha})\coth\frac{2t}{T}\notag \\
& +\frac{ 8t C_{12}^{\alpha}}{U+8J_{U}}\left[
\coth\frac{E_{3b}(\pi)}{2T}+\coth\frac{
E_{4b}(\pi )}{2T}\right] \\
b &
=-\frac{4J_{U}}{U+8J_{U}}\coth\frac{E_{3b}(\pi)}{2T}+\frac{U+4J_{U}}
{U+8J_{U}}\coth\frac{E_{4b}(\pi)}{2T}
\end{align}

Taking into account the following local algebra constraints
\begin{align}
D_{120}\left( i,i\right) & =\left\langle n(i)
g^{\dagger}(i)\right\rangle
=2C_{cc}^{\alpha} \\
D_{220}\left( i,i\right) & =\left\langle g(i)
g^{\dagger}(i)\right\rangle
=2n-\chi^{\alpha}-4d   \notag \\
&=2(n-4d-2p)=2(n-\chi^{\alpha}+2p) \\
D_{123}\left( i,i\right) & =\left\langle n_3(i)
g_3^{\dagger}(i)\right\rangle = 2C_{cc}^{\alpha} \\
D_{223}\left( i,i\right) & =\left\langle g_3(i)
g_3^{\dagger}(i)\right\rangle
=2n+\frac13\chi^\alpha_s-\chi^\alpha_c+4d\notag \\
&=2(n-\frac23\chi^\alpha_s-2p)
\end{align}
and performing similar calculations we can also obtain (we omit
the expressions of $\Gamma_{22\mu}$ for the sake of brevity)
\begin{align}
&\left\langle n^{\alpha}(i)  g^{\dagger}(i)
\right\rangle =-2C_{cc}^{\alpha} \\
&\left\langle g^{\alpha}(i)  g^{\dagger}(i) \right\rangle =-2(n-4d-2p) \\
&\Gamma_{12\mu} = 0
\end{align}

\paragraph{The pair channel.}

After Eq.~\ref{emp}, the thermal causal Green's function
$G_p\left( k,\omega \right) =\mathcal{F}\left\langle
\mathcal{T}\left[ P(i) P^{\dagger}(j)\right] \right\rangle $
satisfies the following equation
\begin{equation}\label{grep}
\left[ \omega -\omega_p \left( k\right) \right] G_p\left( k,\omega
\right) =I_p(k)
\end{equation}
where the relevant entries of the normalization matrix $I_p(k)=
\mathcal{F} \left\langle \left[ P(i),P^{\dagger}(j)\right]
\right\rangle_{E.T.}$ are
\begin{align}
I_{11p}(k) & = 1-n \\
I_{12p}(k) & = [1+\alpha(k)]C_{cc}^{\alpha} \\
I_{13p}(k) & = I_{12p}(k) \\
I_{14p}(k) & = n-\chi_{c}^{\alpha}+2d\alpha(k) \\
I_{15p}(k) & = 2[1+\alpha(k)]C_{12}^{\alpha} \\
I_{16p}(k) & = 2D-2\gamma+2d\alpha(k)
\end{align}
where $\gamma=\left\langle
n^{\alpha}(i)n_{\uparrow}(i)n_{\downarrow}(i)\right\rangle $.

The solution of Eq.~\ref{grep} is the following
one\cite{Mancini:00}
\begin{multline}\label{greenpz}
G_p\left( k,\omega \right) =
-2\mathrm{i} \pi \Gamma_p(k) \delta(\omega)\\
+\sideset{}{'}\sum_{i=1}^6 \frac{\sigma_p
^{(i)}(k)}{1-\mathrm{e}^{-\beta \omega}}\left[\frac{1}{\omega
-E_{ip}(k)+ \mathrm{i} \delta} - \frac{\mathrm{e}^{-\beta
\omega}}{\omega -E_{ip}(k)-\mathrm{i} \delta}\right]
\end{multline}
where
\begin{multline}
\Gamma_p(k)= \frac12 \lim_{\omega \rightarrow 0}
\omega\\
\times \mathcal{F}\left[\theta(t-t')\left\langle P(i) P^{\dagger
}(j)\right\rangle-\theta(t'-t)\left\langle P^{\dagger}(j)
P(i)\right\rangle\right]
\end{multline}
The primed sum is again restricted to values of $i$ for which
$E_{ip}(k) \neq 0$.

The $E_{ip}(k)$ are the eigenvalues of the energy matrix $\omega_p
\left( k\right)$
\begin{align}
E_{1p}(k) & =-2\mu+U-2t[1+\alpha(k)] \label{enpin}\\
E_{2p}(k) & =-2\mu+U+2t[1+\alpha(k)] \\
E_{3p}(k) & =-2\mu+\frac{1}{2}[U-Q(k)] \\
E_{4p}(k) & =-2\mu+\frac{1}{2}[3U-Q(k)] \\
E_{5p}(k) & =-2\mu+\frac{1}{2}[U+Q(k)] \\
E_{6p}(k) & =-2\mu+\frac{1}{2}[3U+Q(k)] \label{enpfin}
\end{align}
with $Q(k)=\sqrt{U^{2}+16t^{2}[1+\alpha(k)]^{2}}$. The $E_{ip}(k)$
are zero only in isolated points of the parameter space ($n$, $T$,
$U$) (see Eqs.~\ref{enpin} - \ref{enpfin}). According to this, the
zero frequency function $\Gamma_p(k)$ is identically zero except
in these points, where it could be finite. In this treatment, for
the sake of simplicity, we neglect these isolated points.

The spectral weights $\sigma _{11p}^{(i)}(k)$, computed through
Eq.~\ref{sigmafor}, have the following expressions
\begin{align}
\sigma_{11p}^{(1)}(k) &
=\frac{1}{2}[I_{13p}(k)+I_{14p}(k)-I_{15p}(k)-I_{16p}(k)] \\
\sigma_{11p}^{(2)}(k) &
=\frac{1}{2}[-I_{13p}(k)+I_{14p}(k)+I_{15p}(k)-I_{16p}(k)] \\
\sigma_{11p}^{(3)}(k) & =\frac{t[1+\alpha(k)]}{Q(k)}I_{15p}(k) \notag \\
& +\frac{Q(k)-U}{4Q(k)}[2I_{11p}(k)-2I_{14p}(k)+I_{16p}(k)] \\
\sigma_{11p}^{(4)}(k) & =\frac{t [1+\alpha(k)]}{Q(k)}I_{15p}(k)
+ \frac{Q(k)+U}{4Q(k)}I_{16p}(k) \\
\sigma_{11p}^{(5)}(k) & =-\frac{t[1+\alpha(k)]}{Q(k)}I_{15p}(k) \notag \\
&+\frac{Q(k)+U}{4Q(k)}[2I_{11p}(k)-2I_{14p}(k)+I_{16p}(k)] \\
\sigma_{11p}^{(6)}(k) & =-\frac{t [1+\alpha(k)]}{Q(k)}I_{15p}(k)
+\frac{Q(k)-U}{4Q(k)}I_{16p}(k)
\end{align}

In order to finally compute the Green's function $G_{11p}\left(
k,\omega \right)$ we should fix the internal parameter $\gamma$.
The determination of the parameter $\gamma$ requires the
computation of another Green's function
$G_D(k,\omega)=\mathcal{F}\left\langle \mathcal{T}[n(i)
D(j)]\right\rangle $, where $D(i)=n_{\uparrow}(i)n_{\downarrow
}(i)$. Otherwise, we could resort to the following local algebra
constraint
\begin{equation}
\left\langle p(i) p^{\dagger}(i)\right\rangle =1-n+D
\end{equation}

It is worth noting that within the pair channel we can also
compute the parameter $d$ directly from its definition:
$d=\left\langle p^{\alpha}(i) p^{\dagger}(i)\right\rangle$. This
finally opens the possibility to compute fully self-consistently
the fermionic and bosonic sectors at once. It can be shown that
the two procedures (the fully self-consistent one and the one
presented at length in the previous sections) give exactly the
same results, as it should be according to the exact nature of the
proposed treatment.

\subsubsection{The $t$-$J$ model}

After Eqs.~\ref{emcsJ}, the thermal causal Green's functions
$G_c\left( q,\omega \right) =\mathcal{F}\left\langle
\mathcal{T}\left[ \varphi(i) \varphi^{\dagger}(j)\right]
\right\rangle $ and $G_{sk}\left( q,\omega \right)
=\mathcal{F}\left\langle \mathcal{T}\left[ \varphi _k(i) \varphi
_k^{\dagger}(j)\right] \right\rangle $ satisfy the following
equations
\begin{align}\label{grecsJ}
&\left[ \omega -\omega_c \left( q\right) \right] G_c\left(
q,\omega \right) = I_c(q) \\
&\left[ \omega -\omega_s \left( q\right) \right] G_{sk}\left(
q,\omega \right) = I_s(q)
\end{align}
where the relevant entries of the normalization matrices $I_c(q)=
\mathcal{F} \left\langle \left[ \varphi(i),\varphi^{\dagger
}(j)\right] \right\rangle_{E.T.}$ and $I_{s}(q)= \mathcal{F}
\left\langle \left[ \varphi _k(i),\varphi _k^{\dagger}(j)\right]
\right\rangle _{E.T.}$ are
\begin{align}
I_{11c}(q) & =0 \\
I_{12c}(q) & =4\left[ 1-\alpha(q)\right] C_{11}^{\alpha} \\
I_{22c}(q) & =0
\end{align}
and
\begin{align}
I_{11s}(q) & =0 \\
I_{12s}(q) & =4\left[ 1-\alpha(q)\right] C_{11}^{\alpha} \\
I_{13s}(q) & =0 \\
I_{14s}(q) & =\frac{4}{3}\left[ 1-\alpha(q)\right]
\chi_{s}^{\alpha}
\end{align}
with
\begin{equation}
\chi_{s}^{\alpha}=\sum_{k=1}^3 \left\langle \nu_{k}^{\alpha}(i)
\nu_{k}(i)\right\rangle
\end{equation}
The spin rotational invariance makes $G_{sk}$ independent from the
index $k$. According to this and for the sake of simplicity, we
have omitted in the text the index $k$ in the expressions of the
related quantities: $I_s$, $E_s$, $\sigma_s$ and $\Gamma_s$.

The general solutions of Eqs.~\ref{grecsJ} are the following
ones\cite{Mancini:00}
\begin{align}\label{greenJz}
&G_c\left(q,\omega\right)=-2\mathrm{i} \pi \Gamma_c(q) \delta(\omega) \notag\\
&+\sideset{}{'}\sum_{i=1}^2\frac{\sigma_c^{(i)}(q)}{1-\mathrm{e}^{-\beta\omega}}\left[\frac{1}{\omega
-E_{ic}(q)+\mathrm{i} \delta}-\frac{\mathrm{e}^{-\beta \omega}}{\omega-E_{ic}(q)-\mathrm{i} \delta}\right]\\
&G_{sk}\left(q,\omega\right)=-2\mathrm{i} \pi \Gamma_{s}(q) \delta(\omega) \notag\\
&+\sideset{}{'}\sum_{i=1}^4
\frac{\sigma_s^{(i)}(q)}{1-\mathrm{e}^{-\beta\omega}}\left[\frac{1}{\omega
-E_{is}(q)+\mathrm{i} \delta}-\frac{\mathrm{e}^{-\beta
\omega}}{\omega-E_{is}(q)-\mathrm{i} \delta}\right]
\end{align}
where
\begin{align}
&\Gamma_c(q)= \frac12 \lim_{\omega \rightarrow 0}
\omega \notag\\
&\times \mathcal{F}\left[\theta(t-t')\left\langle
\varphi(i) \varphi^\dagger(j)\right\rangle-\theta(t'-t)\left\langle \varphi^\dagger(j) \varphi(i)\right\rangle\right] \\
&\Gamma_{s}(q)= \frac12 \lim_{\omega \rightarrow 0}
\omega \notag\\
&\times \mathcal{F}\left[\theta(t-t')\left\langle \varphi_k(i)
\varphi^\dagger_k(j)\right\rangle-\theta(t'-t)\left\langle
\varphi^\dagger_k(j) \varphi_k(i)\right\rangle\right]
\end{align}
The primed sum is again restricted to values of $i$ for which
$E_{ic,s}(k) \neq 0$.

The $E_{ic}(q)$ and $E_{is}(q)$ are the eigenvalues of the energy
matrices $\omega_c (q)$ and $\omega_s (q)$, respectively
\begin{align}
E_{1c}(q) &= 2t\sqrt{2[1-\alpha(q)]} \\
E_{2c}(q) &= -2t\sqrt{2[1-\alpha(q)]}
\end{align}
and
\begin{align}
E_{1s}(q) &= 2t\sqrt{2[1-\alpha(q)]} \\
E_{2s}(q) &= -2t\sqrt{2[1-\alpha(q)]} \\
E_{3s}(q) &= 2J\sqrt{2[1-\alpha(q)]} \\
E_{4s}(q) &= -2J\sqrt{2[1-\alpha(q)]}
\end{align}
According to this, the primed sums in Eqs.~\ref{greenJz} do not
contain the elements for $q=0$ and the zero frequency functions
$\Gamma_{c,s}(q)$ reduce to the constants $\Gamma_c=\Gamma_c(0)$
and $\Gamma_{s}=\Gamma_{s}(0)$, respectively. The spectral weights
$\sigma _{c,s}^{(i)}(k)$, computed through Eq.~\ref{sigmafor} have
the following expressions
\begin{align}
\sigma_{11c}^{(1)}(q) &
=-\sigma_{11c}^{(2)}(q)=-C_{11}^{\alpha}\sqrt {2}
\sqrt{1-\alpha(q)} \\
\sigma_{12c}^{(1)}(q) &
=\sigma_{12c}^{(2)}(q)=2C_{11}^{\alpha}\left[
1-\alpha(q)\right] \\
\sigma_{22c}^{(1)}(q) &
=-\sigma_{22c}^{(2)}(q)=-2\sqrt{2}C_{11}^{\alpha}
\left[ 1-\alpha(q)\right] ^{\frac{3}{2}} \\
\sigma_{11s}^{(1)}(q) &
=-\sigma_{11s}^{(2)}(q)=-C_{11}^{\alpha}\sqrt {2}
\sqrt{1-\alpha(q)} \\
\sigma_{11s}^{(3)}(q) &
=-\sigma_{11s}^{(4)}(q)=-\frac{1}{3}\chi_{s}^{\alpha}
\sqrt{2}\sqrt{1-\alpha(q)}
\end{align}

In order to finally compute the Green's function
$G_c\left(q,\omega\right)$ we should fix the zero frequency
constant $\Gamma_c$. This latter cannot be directly connected to
any correlation function at this level and its determination
requires the use of local algebra constraints. Let us consider the
correlation function $D_c\left(q,\omega\right)=
\mathcal{F}\left\langle \varphi(i) \varphi^\dagger(j)\right\rangle
$ which is linked to the causal Green's function through the
spectral theorem
\begin{multline}
D_c\left(q,\omega\right)= -\left(1+\tanh\frac{\omega}{2T}\right)
\Im\left[G_c\left(q,\omega\right)\right]\\
=2\pi \Gamma_c \delta(\omega)+ 2 \pi \sideset{}{'}\sum_{i=1}^2
\delta\left[\omega-E_{ic}(q)\right]\frac{\sigma_c^{(i)}(q)}
{1-\mathrm{e}^{-\beta E_{ic}(q)}}
\end{multline}
Then, we can use the following relations in order to compute
$\Gamma_c$
\begin{align}\label{ccJ}
D_{11c}\left(i,i\right) & =\left\langle \nu(i)  \nu(i)
\right\rangle =n \\
D_{12c}\left( i,i\right) & =\left\langle \nu(i)  \Pi
^{\dagger}(i) \right\rangle =2C_{11}^{\alpha} \\
D_{22c}\left( i,i\right) & =\left\langle \Pi(i) \Pi^{\dagger}(i)
\right\rangle =2\left( n-\chi_{c}^{\alpha}\right)
\end{align}
$\chi_{c}^{\alpha}$ is directly connected to the Green's function
\begin{equation}
\chi_{c}^{\alpha}=\left\langle \nu(i) \nu^{\alpha}(i)\right\rangle
\end{equation}

In order to finally compute the Green's function $G_{11s}\left(
k,\omega \right)$ we should fix the internal parameter
$\chi_{s}^{\alpha}$ and the zero frequency constant
$\Gamma_{11s}$.

The parameter $\chi_{s}^{\alpha}$ is directly connected to the
Green's function. Let us consider the correlation function
$D_{sk}\left(q,\omega\right) = \mathcal{F}\left\langle
\varphi_k(i) \varphi^\dagger_k(j)\right\rangle $ which is linked
to the causal Green's function through the spectral theorem
\begin{multline}
D_{sk}\left(q,\omega\right) =-\left(
1+\tanh\frac{\omega}{2T}\right) \Im\left[G_{sk}\left(q,\omega\right)\right]\\
=2\pi \Gamma_{s} \delta(\omega)+ 2 \pi \sideset{}{'}\sum_{i=1}^4
\delta\left[\omega -E_{is}(q)\right]\frac{\sigma_{s}^{(i)}(q)}
{1-\mathrm{e}^{-\beta E_{is}(q)}}
\end{multline}
Then,
\begin{equation}\label{csJ1}
\chi_{s}^{\alpha}=3\left\langle \nu_3(i)
\nu^{\alpha}_3(i)\right\rangle
\end{equation}

The zero frequency constant $\Gamma_{s}$ cannot be directly
connected to any correlation function at this level and its
determination requires the use of local algebra constraints. In
particular, we can use the following relation
\begin{equation}\label{csJ2}
D_{11sk}\left(i,i\right) = \left\langle \nu_3(i)   \nu_3(i)
\right\rangle = n
\end{equation}

Equations~\ref{csJ1} and \ref{csJ2} constitute a complete set of
self-consistent equations which allow to compute the Green's
function $G_{11sk}\left(q,\omega \right)$.

We have the following expressions for the relevant zero frequency
constants and correlators
\begin{align}
& \Gamma_{11c} = n+\coth\frac{2t}{T}C_{11}^{\alpha} \\
& \Gamma_{12c} = 0 \\
& \Gamma_{22c} = 0 \\
& \Gamma_{11s} = 3n+ 3\coth\frac{2t}{T}C_{11}^{\alpha}+
\coth\frac{2J}{T}\chi _{s}^{\alpha} \\
& \chi_{c}^{\alpha} = n+2\coth\frac{2t}{T}C_{11}^{\alpha} \\
& \left\langle \nu^{\alpha}(i) \Pi^{\dagger}(i)\right\rangle = -2C_{11}^{\alpha} \\
& \left\langle \Pi^{\alpha}(i) \Pi^{\dagger}(i)\right\rangle = 4\coth\frac{2t}{T}C_{11}^{\alpha} \\
& \chi_{s}^{\alpha} = \frac{3n+6\coth\frac{2t}{T}
C_{11}^{\alpha}}{1-2\coth\frac{2J}{T}} =
\frac{3\chi_{c}^{\alpha}}{1-2\coth\frac{2J}{T}}
\end{align}

\section{The Results}

In the previous two sections, we have given a detailed summary of
the analytical calculations which lead to computable expressions
for internal parameters, zero frequency constants, correlation and
response functions for the fermionic and bosonic (charge, spin and
pair) sectors of the Hubbard and $t$-$J$ models. In this section,
we analyze the related results for the local, single-particle,
thermodynamical and response properties. Hereafter, $t$ will be
used as energy scale.

\begin{figure}[tb!!]
\begin{center}
\includegraphics[width=8cm,keepaspectratio=true]{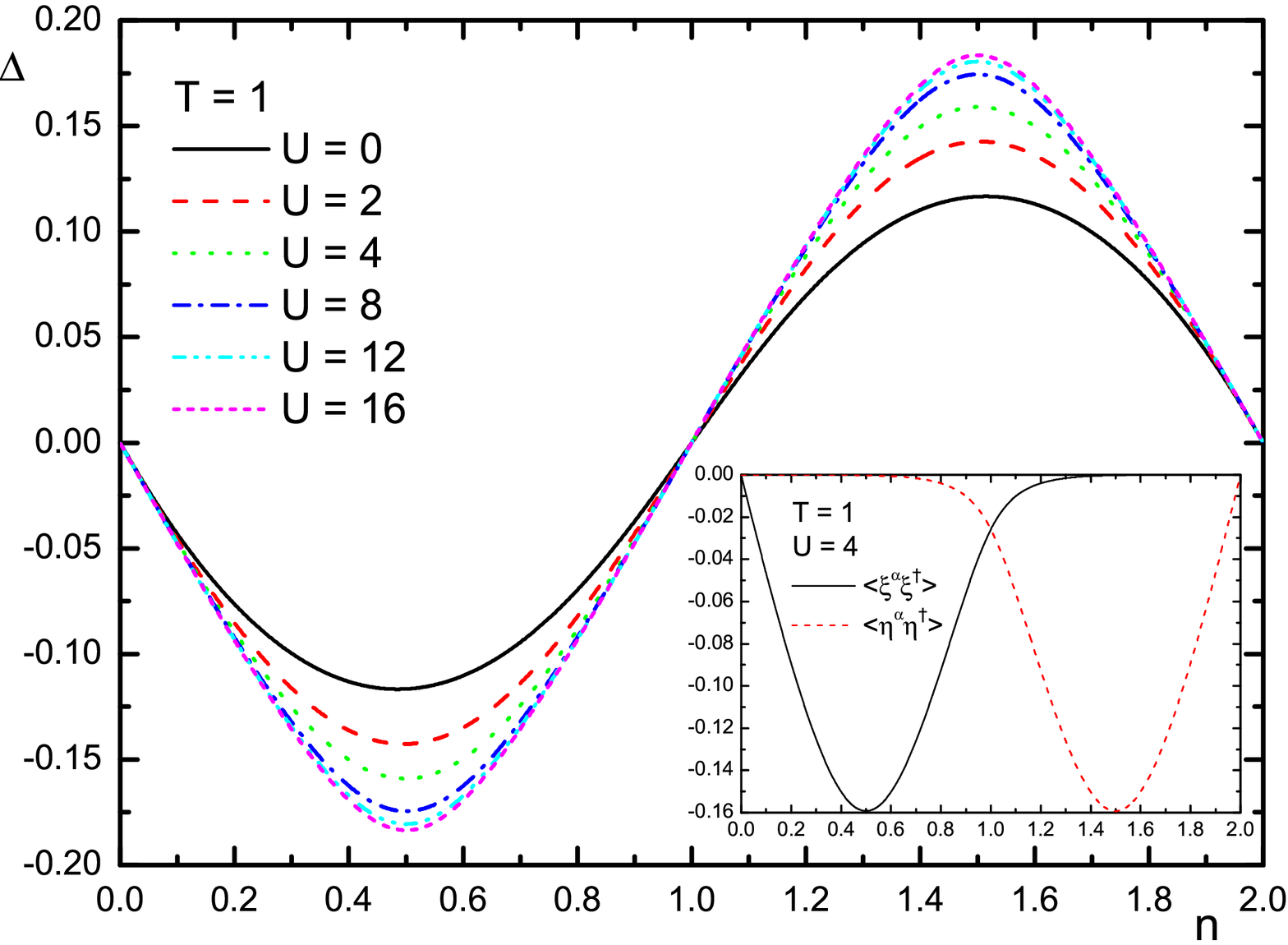}
\includegraphics[width=8cm,keepaspectratio=true]{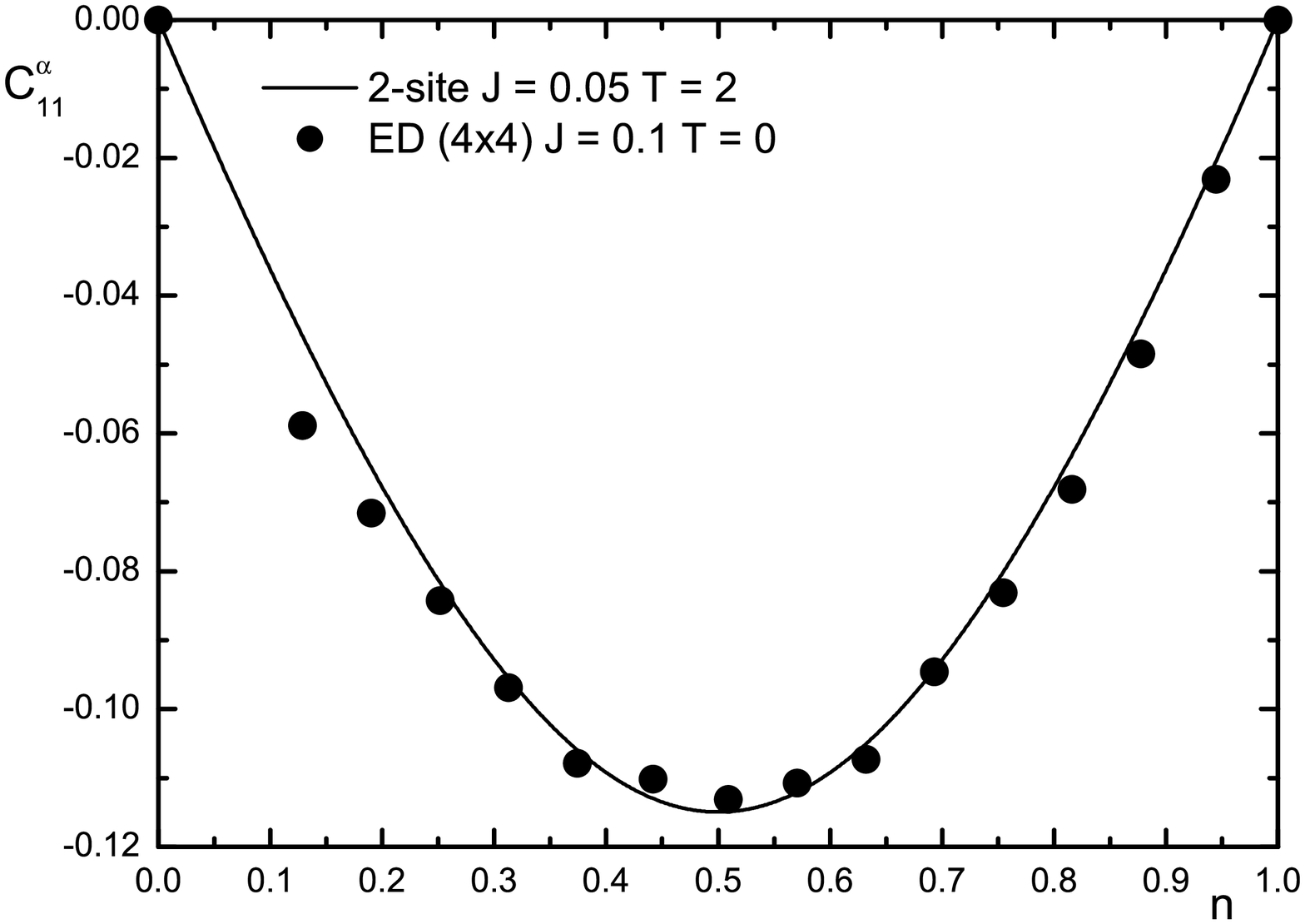}
\end{center}
\caption{(top) The $\Delta$ parameter as function of the filling
$n$ for $T=1$ and $U=0$, $2$, $4$, $8$, $12$ and $16$ (inset:
$\xi$ and $\eta$ components for $T=1$ and $U=4$); (bottom)
$C^{\alpha}_{11}$ as function of the filling $n$ for $T=2$ and
$J=0.05$, the ED data ($4 \times 4$) are for $T=0$ and from
Ref.~\protect\onlinecite{Dagotto:92a}.} \label{Fig5}
\end{figure}

\subsection{The $\Delta$ and $p$ parameters} \label{Deltap}

The $\Delta$ and $p$ parameters are computed by solving the system
of self-consistent Equations~\ref{SelfHub1} and \ref{SelfHub2}.
The behavior of $\Delta$, as function of the filling $n$, is shown
in Fig.~\ref{Fig5} (top panel). $\Delta$ is defined as the
difference between the hopping amplitudes computed between empty
and singly occupied sites (i.e., $\langle\xi(i)
\xi^\dagger(i)\rangle$) and between singly and doubly occupied
sites (i.e., $\langle\eta(i) \eta^\dagger(i)\rangle$). In this
sense, $\Delta$ gives a measure of the electron mobility. The
following features have been observed: the hopping amplitude
coming from $\xi$ ($\eta$) prevails below (above) half-filling and
is always negative [see inset Fig.~\ref{Fig5} (top panel)]; the
absolute value of $\Delta$ diminishes on increasing the
temperature $T$ and on decreasing the Coulomb repulsion $U$. Below
half filling, the electrons move preferably among empty and single
occupied sites; the same happens to the holes above half filling.
This explains the prevalence of one hopping amplitude per each
region of filling. In addition, by decreasing the temperature or
by increasing the Coulomb repulsion we can effectively reduce the
number of doubly occupied sites present in the system and,
consequently, favor the mobility of the electrons and increase the
absolute value of $\Delta$. These are also the reasons behind the
filling dependence: its absolute value first increases with the
number of available almost free moving particles ($n\lesssim
0.5$), then decreases when the number of particles is such to
allow some double occupancies and to force the reduction of the
number of empty sites ($0.5\lesssim n\lesssim 1$) and finally
vanishes when, in average, the sites are all singly occupied
($n=1$); the behavior above half filling is equivalent and is
related to the holes.

The behavior of $C_{11}^{\alpha}$ within the $t$-$J$ model, as
function of the filling $n$, is shown in Fig~\ref{Fig5} (bottom
panel). Below half filling and for rather high values of the
Coulomb repulsion $U$, $\Delta$ and $C_{11}^{\alpha}$ exactly
coincide as the $\eta$ component of $\Delta$ completely vanishes.
According to this, all the features discussed above for the
parameter $\Delta$ are also observed for $C_{11}^{\alpha}$ ($J$
acts like $1/U$). Anyway, we have to report a really small
dependence on the exchange interaction $J$ that is fully effective
only at $n=1$ where the mobility is already zero as no empty site
at all is left. It is worth noting that the exact result for the
2-site system at $T=2$ almost exactly reproduces the Exact
Diagonalization (ED)\cite{Dagotto:92a} data for a $4\times 4$
cluster at $T=0$. This shows that, by increasing the temperature
in a 2-site system it is possible to mime a cluster of bigger size
at a lower temperature, at least as regards some of its
properties. This can be understood by thinking to the level
spacing in the two cases (i.e., bigger the cluster lower the
spacing) and to the value of temperature needed to excite those
levels. Clearly, not all the properties are in a so strong
relation with the relative level positions, but depend on the
absolute positions of them. Anyway, this also shows that the
relative energy scales/levels are already present in the 2-site
system. We have used a value of $J$ twice smaller than the one
used within the numerical analysis according to the difference,
between the 2-site system and any larger system, regarding the
value of the exchange energy $J$ appearing in the derivation of
the $t$-$J$ model from the Hubbard one\cite{Castellani:79}.
Actually, the derivation is pathological just for the 2-site
system and gives a $J$ larger of a factor $2$.

\begin{figure}[tb!!]
\begin{center}
\includegraphics[width=8cm,keepaspectratio=true]{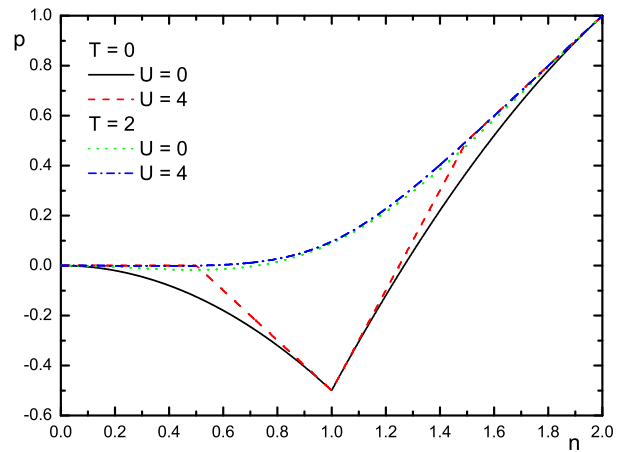}
\end{center}
\caption{The $p$ parameter as function of the filling $n$ for
$T=0$ and $T=2$ at $U=0$ and $U=4$.} \label{Fig6}
\end{figure}

As regards the $p$ parameter, in Fig~\ref{Fig6} we report its
dependence on filling $n$ for two values of the on-site Coulomb
repulsion $U=0$ and $U=4$ and for two temperatures $T=0$ and
$T=2$. At low temperatures and for $n\lesssim 1.26$, the value of
the $p$ parameter is mainly negative according to the {\it
antiferromagnetic} nature of the spin correlations in the ground
state; at higher temperatures the {\it antiferromagnetic} spin
correlations get weaker and the sign of the parameter changes. The
filling $n$, the temperature $T$ and the Coulomb interaction $U$
rule the balance between the spin, the charge and the pair
correlations.

\begin{figure}[tb!!]
\begin{center}
\includegraphics[width=8cm,keepaspectratio=true]{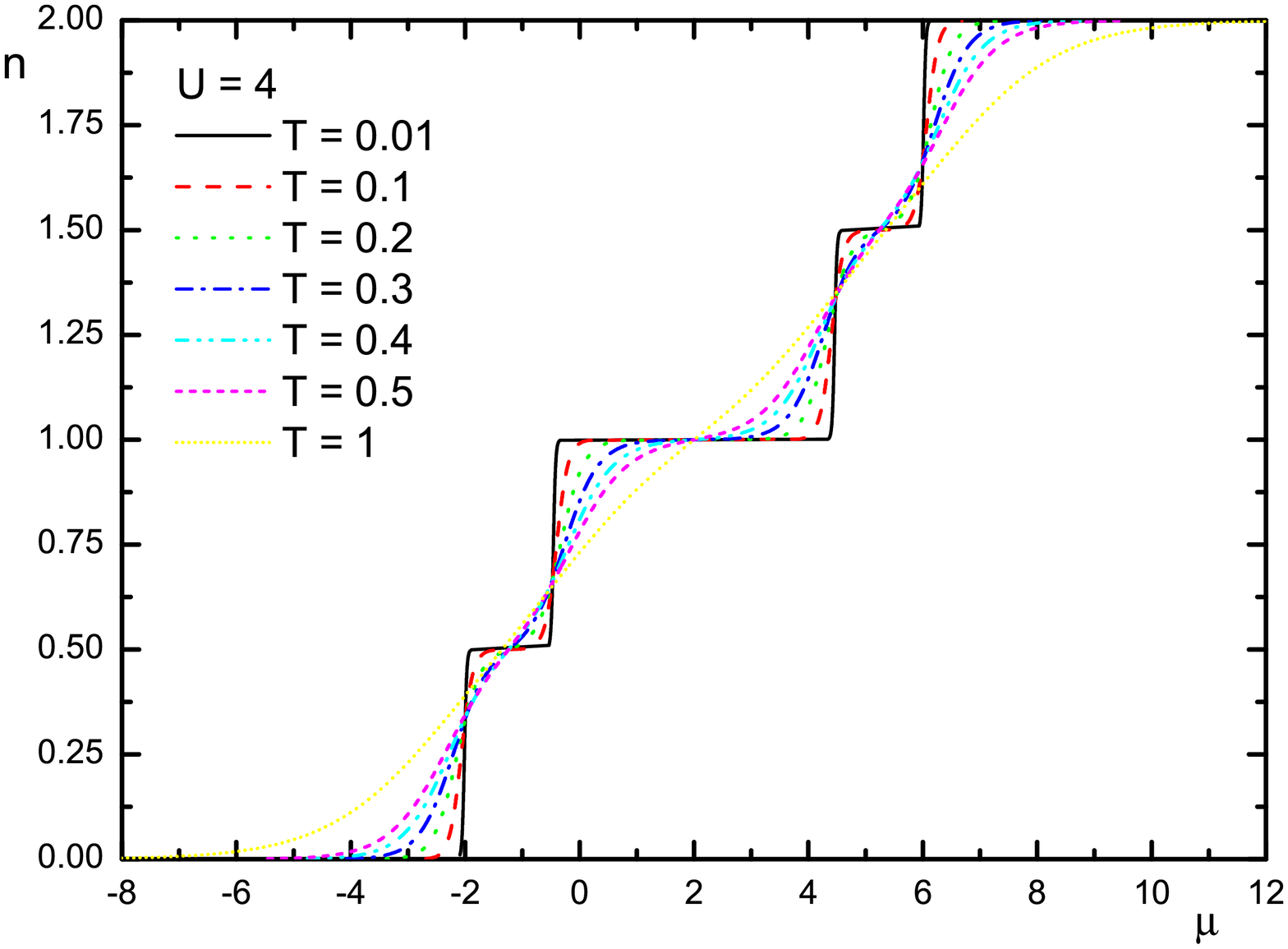}
\includegraphics[width=8cm,keepaspectratio=true]{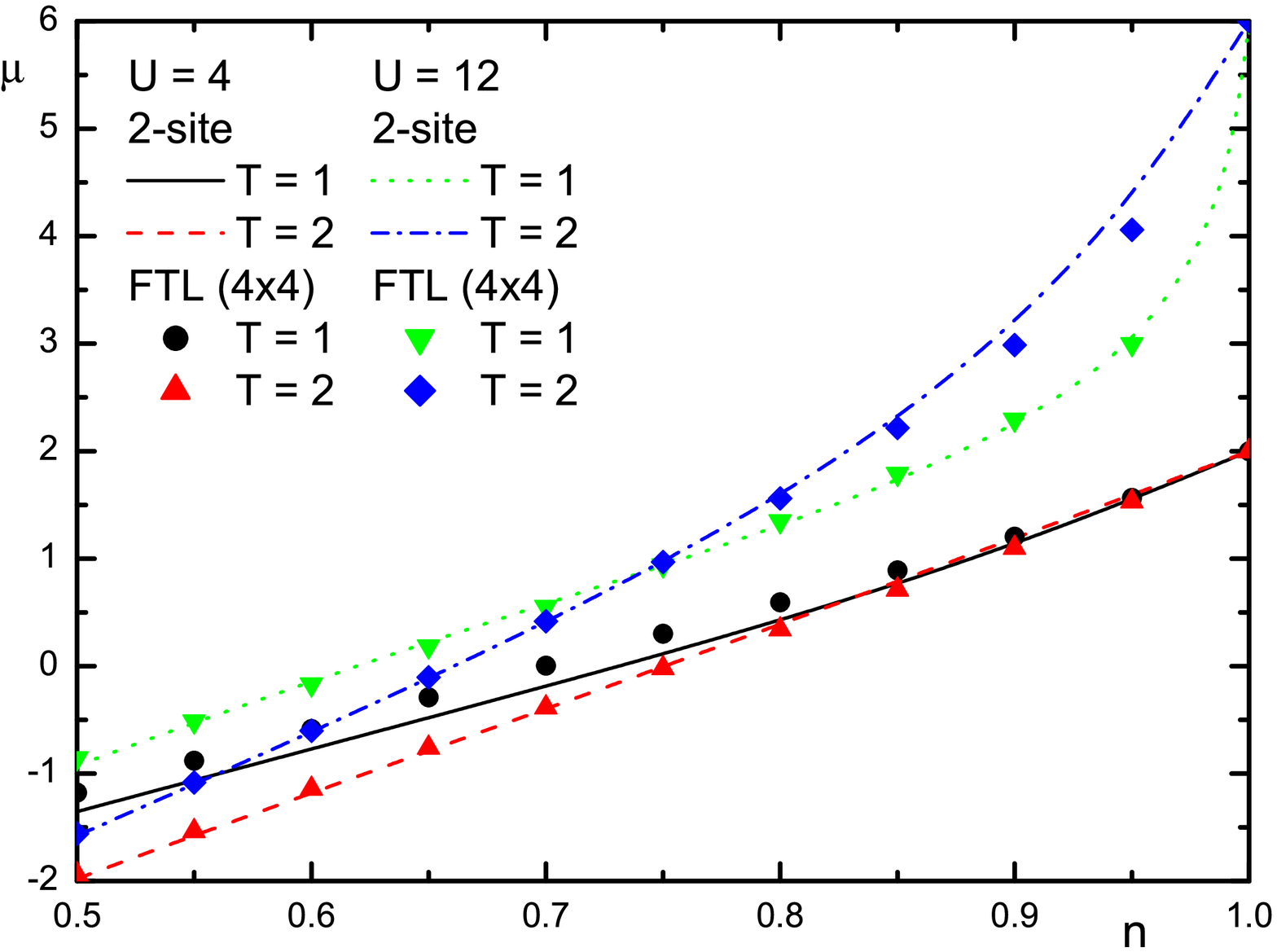}
\end{center}
\caption{The chemical potential $\mu$ as function of the filling
$n$ (top) for $U=4$ and $T=0.01$, $0.1$, $0.2$, $0.3$, $0.4$,
$0.5$ and $1$ and (bottom) for $U=4$, $12$ and $T=1$, $2$. The FTL
data ($4 \times 4$) are taken from
Ref.~\protect\onlinecite{Prelovsek}.} \label{Fig1}
\end{figure}

\begin{figure}[tb!!]
\begin{center}
\includegraphics[width=8cm,keepaspectratio=true]{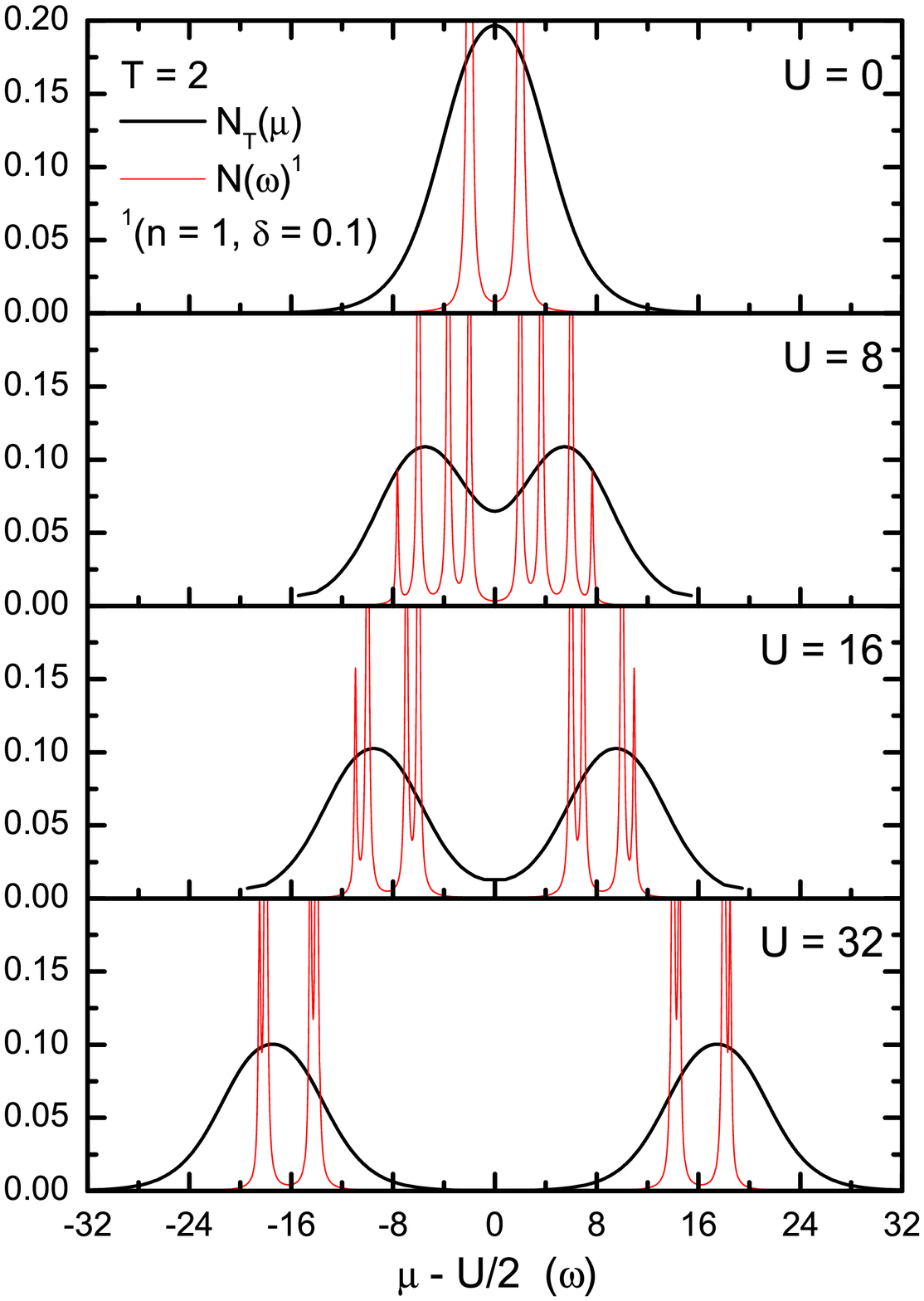}
\end{center}
\caption{The thermodynamic $N_T(\mu)$ and the traditional
$N(\omega)$ ($n=1$ and $\delta=0.1$) densities of states as
functions of the scaled chemical potential $\mu-U/2$ and the
frequency $\omega$, respectively, for $T=2$ and $U=0$, $8$, $16$
and $32$.} \label{Fig3}
\end{figure}

\begin{figure}[tb!!]
\begin{center}
\includegraphics[width=8cm,keepaspectratio=true]{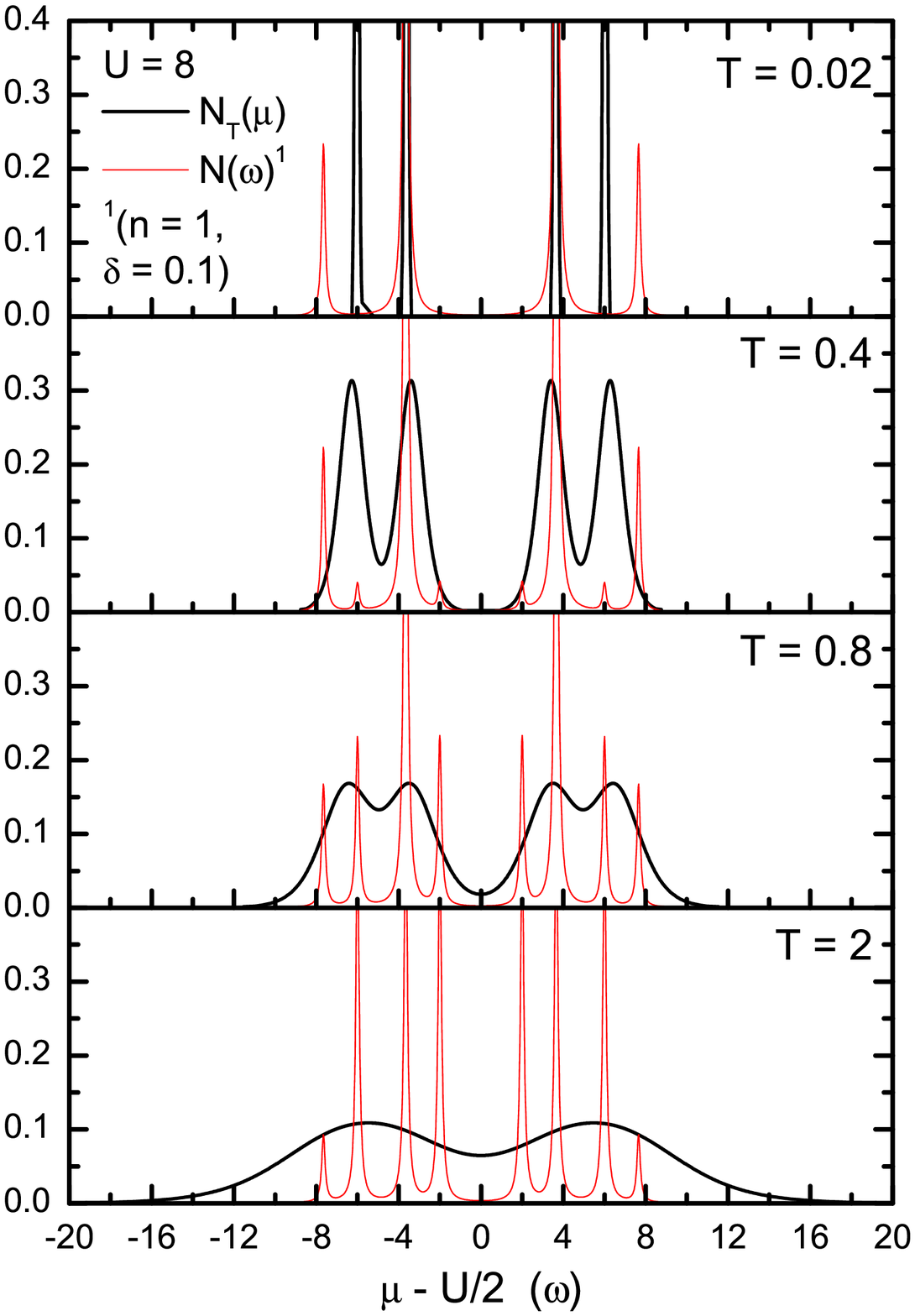}
\end{center}
\caption{The thermodynamic $N_T(\mu)$ and the traditional
$N(\omega)$ ($n=1$ and $\delta=0.1$) densities of states as
functions of the scaled chemical potential $\mu-U/2$ and the
frequency $\omega$, respectively, for $U=8$ and $T=0.02$, $0.4$,
$0.8$ and $2$.} \label{Fig4}
\end{figure}

\subsection{The chemical potential and the density of states}

The chemical potential $\mu$ is one of the self-consistent
parameters and has been computed by solving the system of
self-consistent Equations~\ref{SelfHub1} and \ref{SelfHub2}. As a
function of $U$ the chemical potential increases for all studied
values of temperature $T$ and filling $n$. In particular, we have
the maximum increment for $U\lesssim 20$; for higher values of $U$
the chemical potential saturates. A similar behavior has been
reported by numerical data on larger
clusters\cite{Dagotto:91,Dagotto:92a,Furukawa:92,Furukawa:93}. In
Fig.~\ref{Fig1}, we show the filling dependence of the chemical
potential for some values of the temperature $T$ at $U=4$ (top
panel) and for $U=4$, $12$ and $T=1$, $2$ (bottom panel). The
Finite Temperature Lanczos (FTL) data ($4 \times 4$) are taken
from Ref.~\protect\onlinecite{Prelovsek}. At low temperatures, the
chemical potential has a step-like behavior which recalls the
discreteness of the energy levels. At higher temperatures, the
step-like feature is smeared out due to the thermal hybridization
of the energy levels. It is again really noteworthy the agreement
between the 2-site results and the numerical data. No temperature
adjustment has shown to be necessary in this case as the
temperature is already quite high [cfr.~Fig.~\ref{Fig2} (bottom)].
Generally, the agreement is better for high temperatures and high
values of the Coulomb repulsion. Anyway, there is a competition
between the temperature smearing effects on the level spacing and
the possibility to access the states with energy $J_U$ at lower
temperatures for higher values of $U$ ($J_U \propto 1/U$).
Actually, the number of accessible states is comparable for medium
temperatures and too large for very high temperatures. This should
explain the deviations for $U=4$ and $T=1$, and $U=12$ and $T=2$
at low doping.

By looking at these plots we could be induced to consider the
possibility of a metal-insulator transition driven by the Coulomb
repulsion $U$ and controlled by the temperature $T$. Obviously, no
such transition is possible in a finite system that is always in a
metal paramagnetic state for any finite or zero value of the
Coulomb repulsion $U$; we will show, in Sec.~\ref{Dw}, that the
Drude weight is always finite. Such tricky behavior of the
chemical potential can be understood by thinking at the nature of
the grand canonical ensemble we are using for our thermal
averages. As we are very far from the thermodynamic limit, there
is no equivalence at all among the micro-, the grand- and the
canonical ensembles. In particular, whenever we speak about
temperature and chemical potential in a finite system we have just
to think in terms of mixtures of quantum mechanical states with
different energies and numbers of particles. To get a deeper
comprehension of this issue we can define a thermodynamic density
of states $N_T(\mu)$ through the following equation
\begin{equation} \label{TDOS}
n=\int_{-\infty}^{\mu}\!d\omega N_T(\omega)
\end{equation}
which is a generalization to finite temperatures of the identical
relation existing, at zero temperature, between the filling $n$
and the usual density of states $N(\omega)$. In general, the two
densities of states coincide only at zero temperature and for
systems, like the non-interacting ones, whose density of states is
independent on the chemical potential $\mu$. From the
definition~(\ref{TDOS}), we can simply compute the thermodynamic
density of states $N_T(\mu)$ by differentiating the filling $n$
with respect to the chemical potential $\mu$
[$N_T(\mu)=\emph{d}n/\emph{d}\mu$]. It is worth mentioning that
the thermodynamic density of states $N_T(\mu)$, apart from a
factor $\frac{1}{n^2}$, is simply the compressibility of the
system. A vanishing value of this quantity denotes the
impossibility for the system to accept more particles and,
obviously, the presence of a gap. The thermodynamic density of
states $N_T(\mu)$ enormously facilitates the comprehension of the
chemical potential features that are singled out by the
differentiation procedure. The usual density of states $N(\omega)$
is computed as follows
\begin{equation}
N(\omega)=\frac12 \sum_{n=1}^4 \sum_k
\delta\!\left[\omega-E_n(k)\right]   \sigma_{cc}^{(n)}(k)
\end{equation}
Both densities of states are reported in Figs.~\ref{Fig3} and
\ref{Fig4} as function of the chemical potential ($\mu$) and the
frequency ($\omega$), respectively. We have computed $N(\omega)$
by substituting, for obvious graphical reasons, the Dirac deltas
($\delta(\omega)$) with Lorentzian functions
($\frac1\pi\frac{\delta}{\omega^2+\delta^2}$) with $\delta=0.1$.
In reporting the thermodynamic density of states $N_T(\mu)$ we
have scaled the chemical potential $\mu$ by its value at half
filling ($U/2$) as we want to make the comparison for this
particular value of $n$. While $N(\omega)$ always presents a gap
at $\omega=0$ (the overlapping tails are due to the finite
$\delta$ and the non-zero kinetic energy ensures a metallic
behavior), $N_T(\mu)$ shows a gap only above a certain value of
the Coulomb repulsion for a fixed $T$ (see Fig.~\ref{Fig3}) and
below a certain temperature for a fixed $U$ (see Fig.~\ref{Fig4}).
We want to emphasize that, at high temperature, $N_T(\mu)$ is
capable to mime the behavior expected for the lattice system. In
particular, $N_T(\mu)$ presents two well defined structures (i.e.,
the two Hubbard subbands) that continuously separate on increasing
$U$ (see Fig.~\ref{Fig3}). This behavior is the one expected,
after many analytical and numerical results, for the bulk at
dimension greater than one and it is a realization of the
metal-insulator transition according to the Mott-Hubbard
mechanism. It is worth noting that, for this small system, the
critical Coulomb repulsion, at which the gap appears, is a
function of the temperature $T$.

The density of states $N(\omega)$ is also reported in order to
give an idea of the positions of the poles and of the relative
intensity of the spectral weights, although we had to cut the
highest peaks in drawing the pictures. We can observe two relevant
features. In the non-interacting case ($U=0$) only two poles/peaks
are present; those coming from the band $E_1(k)$ and separated by
the \emph{bandwidth} $W=4t$. On increasing the Coulomb potential
$U$ the three scales of energy present in the system clearly
manifest themselves: the exchange interaction $J_U$, the
\emph{bandwidth} $W$ and the Coulomb repulsion $U$. The eight
poles group in two main structures separated by $U$. Within any
structure the four poles are separated according to the
combination of $\pm 2t$ (\emph{bandwidth} separation) and the
presence or absence of the exchange interaction $J_U$. The
\emph{bandwidth} $W$ is obviously $U$ independent and generates a
rigid two peak structure for any band $E_n(k)$. On the contrary,
the exchange energy $J_U$ decreases on increasing $U$ with a
consequent reduction of the resolution of the peaks within the
Hubbard subbands. The other relevant feature is the redistribution
of the total spectral weight on increasing the temperature. At low
temperatures the poles/peaks that do not contain the exchange
interaction $J_U$ have negligible spectral weights; the 2-site
singlet is the ground state at half-filling. On increasing
temperature the total spectral weight redistributes and the other
poles get more and more spectral weight due to the thermal
hybridization of the energy levels.

\begin{figure}[tb!!]
\begin{center}
\includegraphics[width=8cm,keepaspectratio=true]{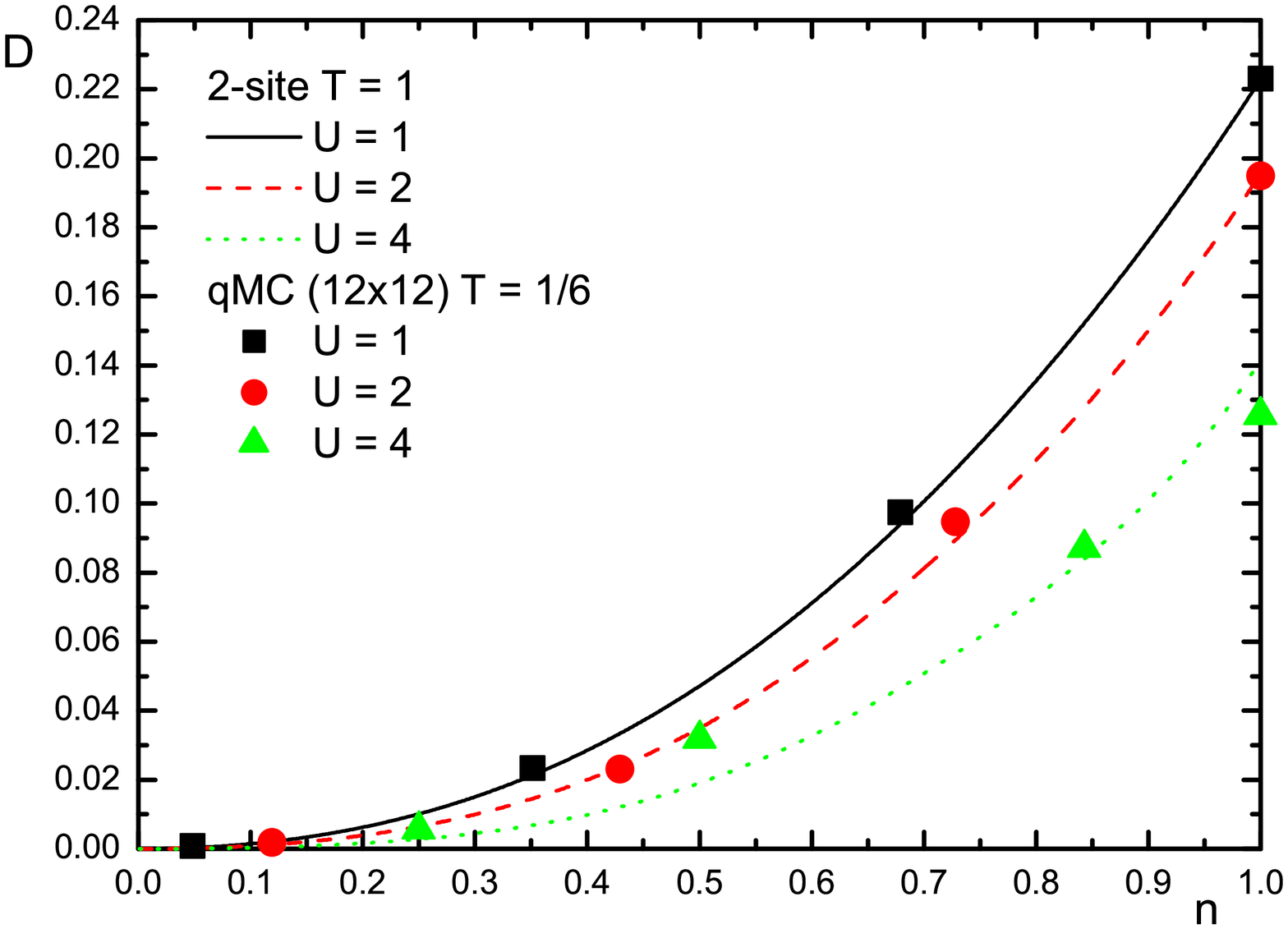}
\includegraphics[width=8cm,keepaspectratio=true]{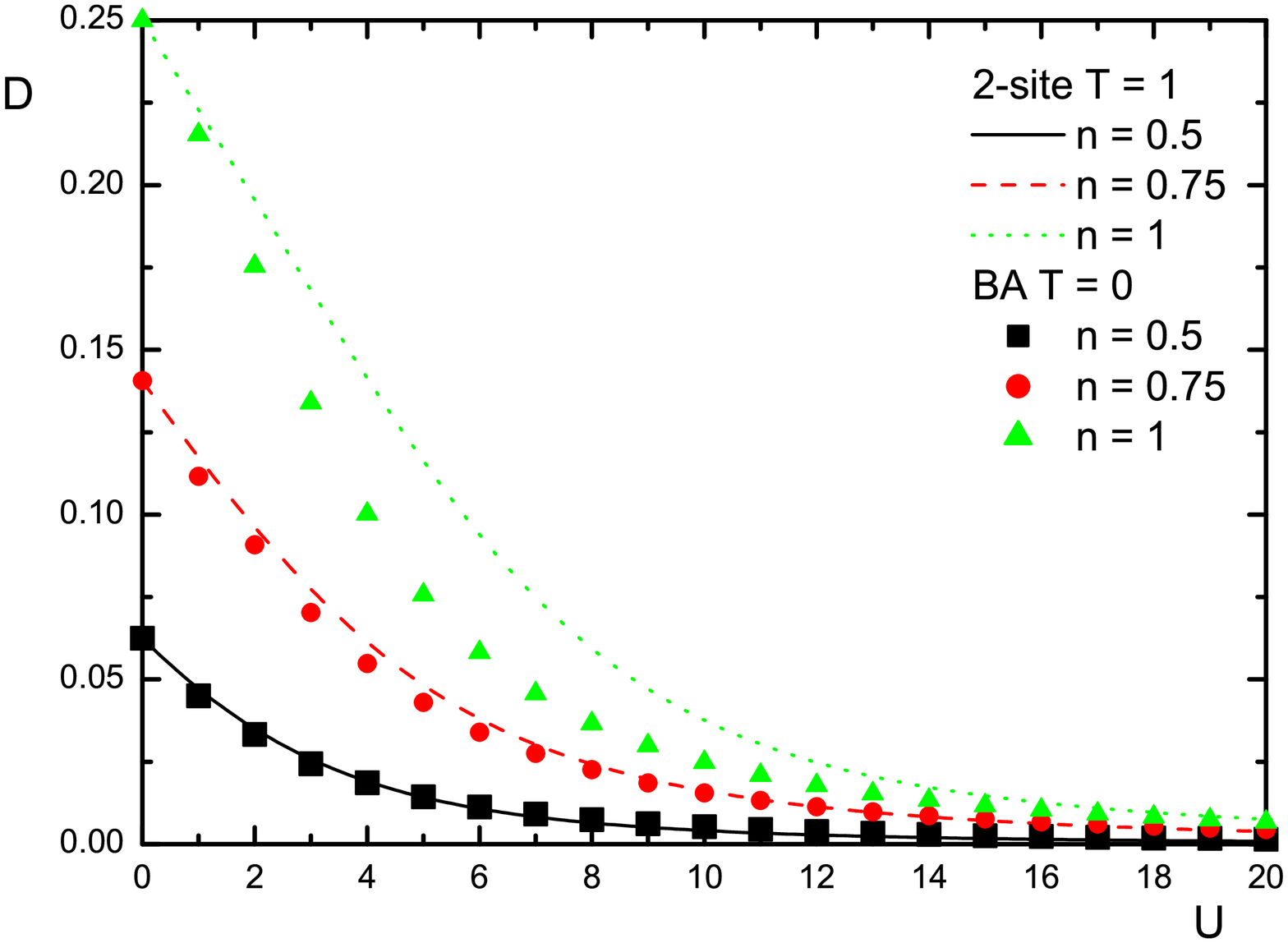}
\end{center}
\caption{The double occupancy $D$: (top) as function of the
filling $n$ for $U=1$, $2$ and $4$ at $T=1$, the qMC data ($12
\times 12$) are for $T=1/6$ and from
Ref.~\protect\onlinecite{Moreo:90}; (bottom) as function of the
Coulomb repulsion $U$ for $n=0.5$, $0.75$ and $1$ at $T=1$, the BA
data are for $T=0$.} \label{Fig7}
\end{figure}

\subsection{The double occupancy $D$}

In the non-interacting case ($U=0$) we have $D=n^{2}/4$. At zero
temperature, the double occupancy $D$ vanishes for $n\leq0.5$;
below this value of filling we have, in average, less than one
electron in the system and only for a finite temperature we can
get some contributions by states with a finite double occupancy.
At half filling we have the following exact formula for the double
occupancy at zero temperature
\begin{equation}
D=\frac{2J_U}{U+8J_U}
\end{equation}

The filling dependence of the double occupancy $D$ (top panel) is
reported in Fig.~\ref{Fig7} for several values of the Coulomb
repulsion $U$ at temperature $T=2$. There are also reported some
quantum Monte Carlo (qMC) data\cite{Moreo:90} for a bigger cluster
($12 \times 12$). As in the case of $C_{11}^{\alpha}$ within the
$t$-$J$ model, we note that the exact results for the 2-site
system at temperature $T=1$ very well reproduce the quantum Monte
Carlo (qMC) data for the bigger cluster at $T=1/6$. The same
explanation obviously holds. A comparison with the exact results
from Bethe Ansatz (BA) are also reported in Fig.~\ref{Fig7}
(bottom panel). Also in this case the 2-site system manage to
reproduce the \emph{bigger} cluster data (actually, the Bethe
Ansatz (BA) system is the 1D bulk) by increasing the temperature.
The discrepancy at half filling can be understood as a consequence
of the difference in the definition of the exchange energy $J$
discussed in the Section regarding the internal energy. Obviously,
the discrepancy is larger where the exchange interaction is mainly
effective (i.e., at half filling and for intermediate-strong
values of the Coulomb interaction).

\subsection{Thermodynamics}

\begin{figure}[tb!!]
\begin{center}
\includegraphics[width=8cm,keepaspectratio=true]{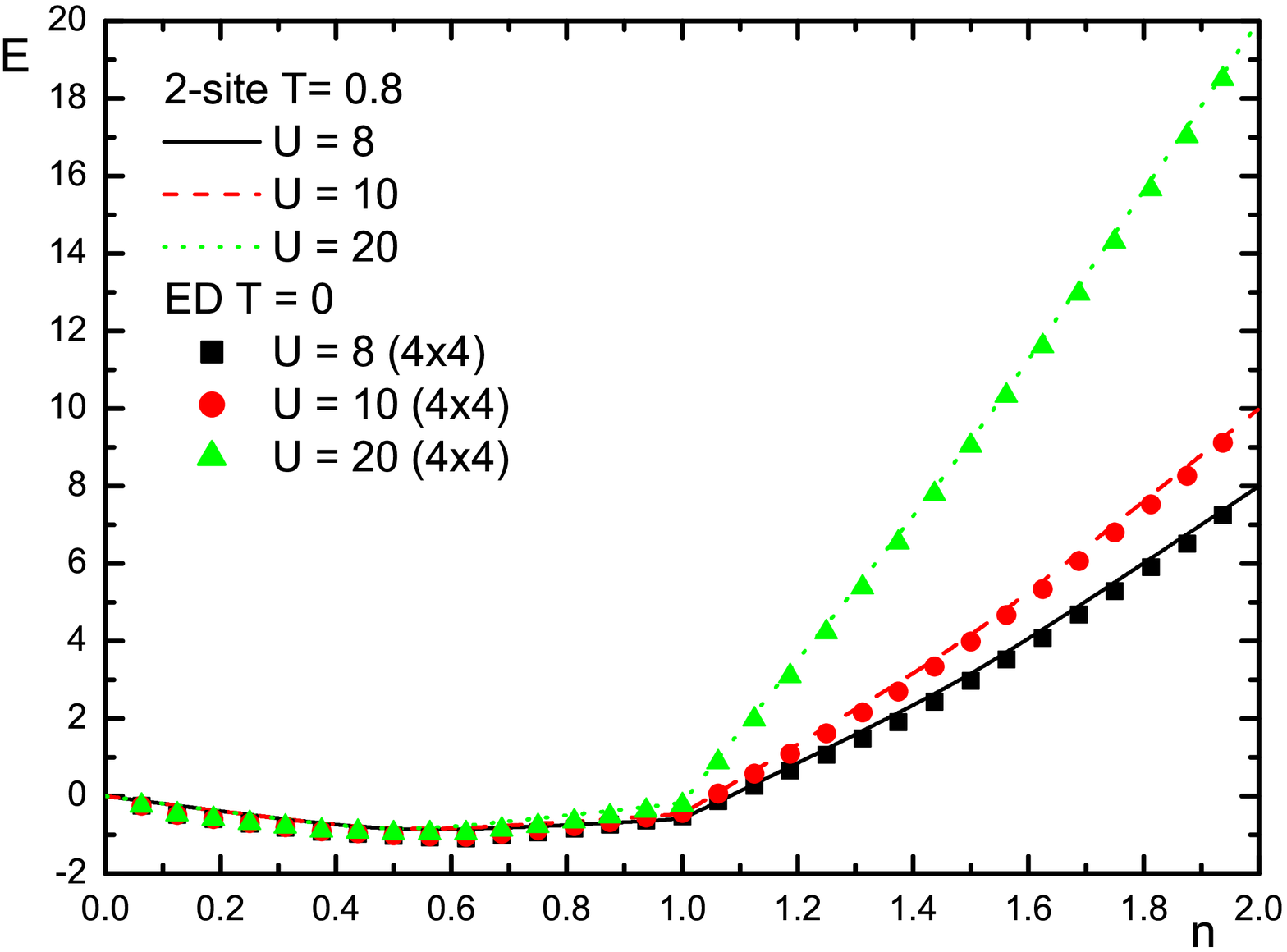}
\includegraphics[width=8cm,keepaspectratio=true]{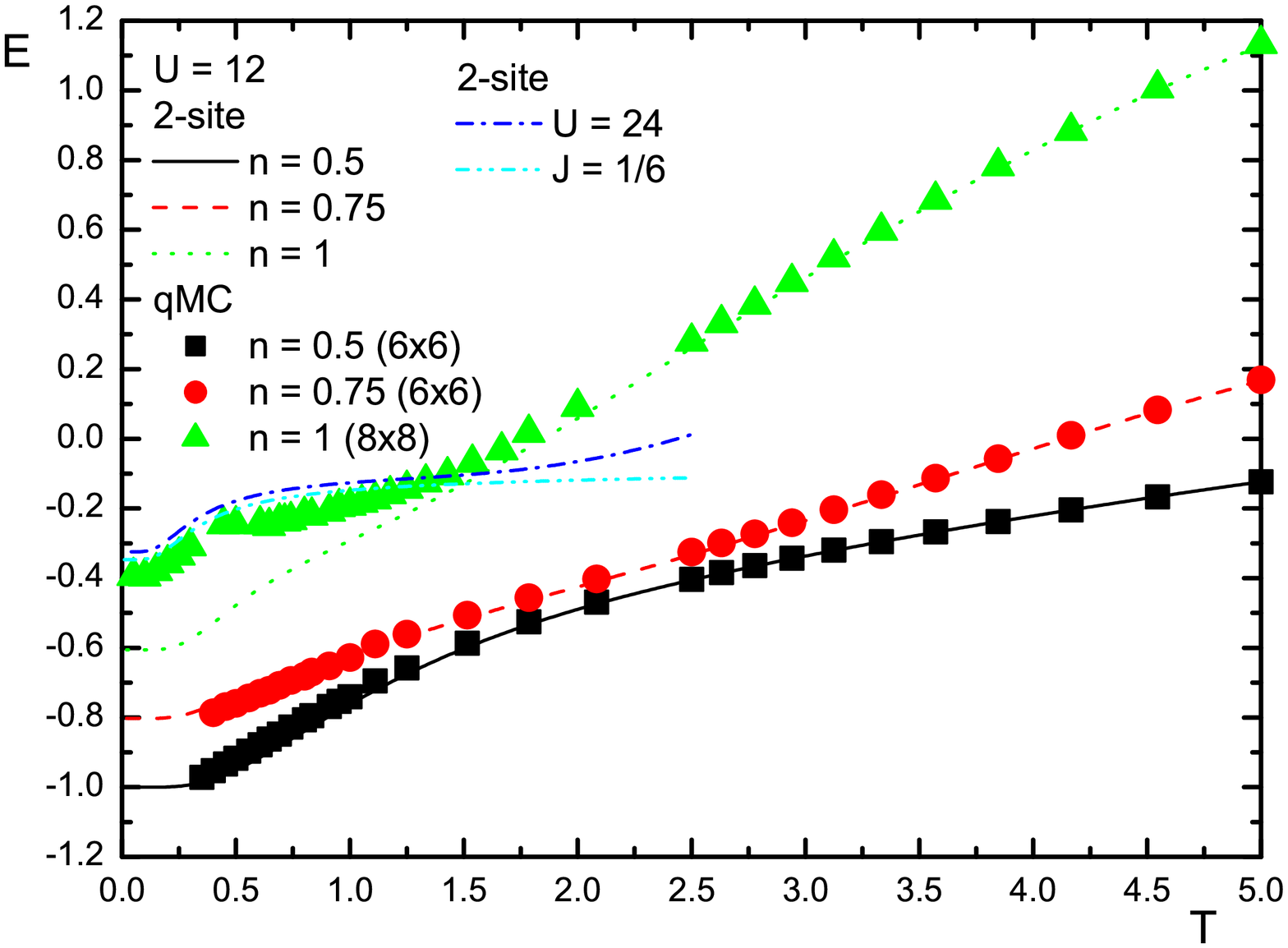}
\includegraphics[width=8cm,keepaspectratio=true]{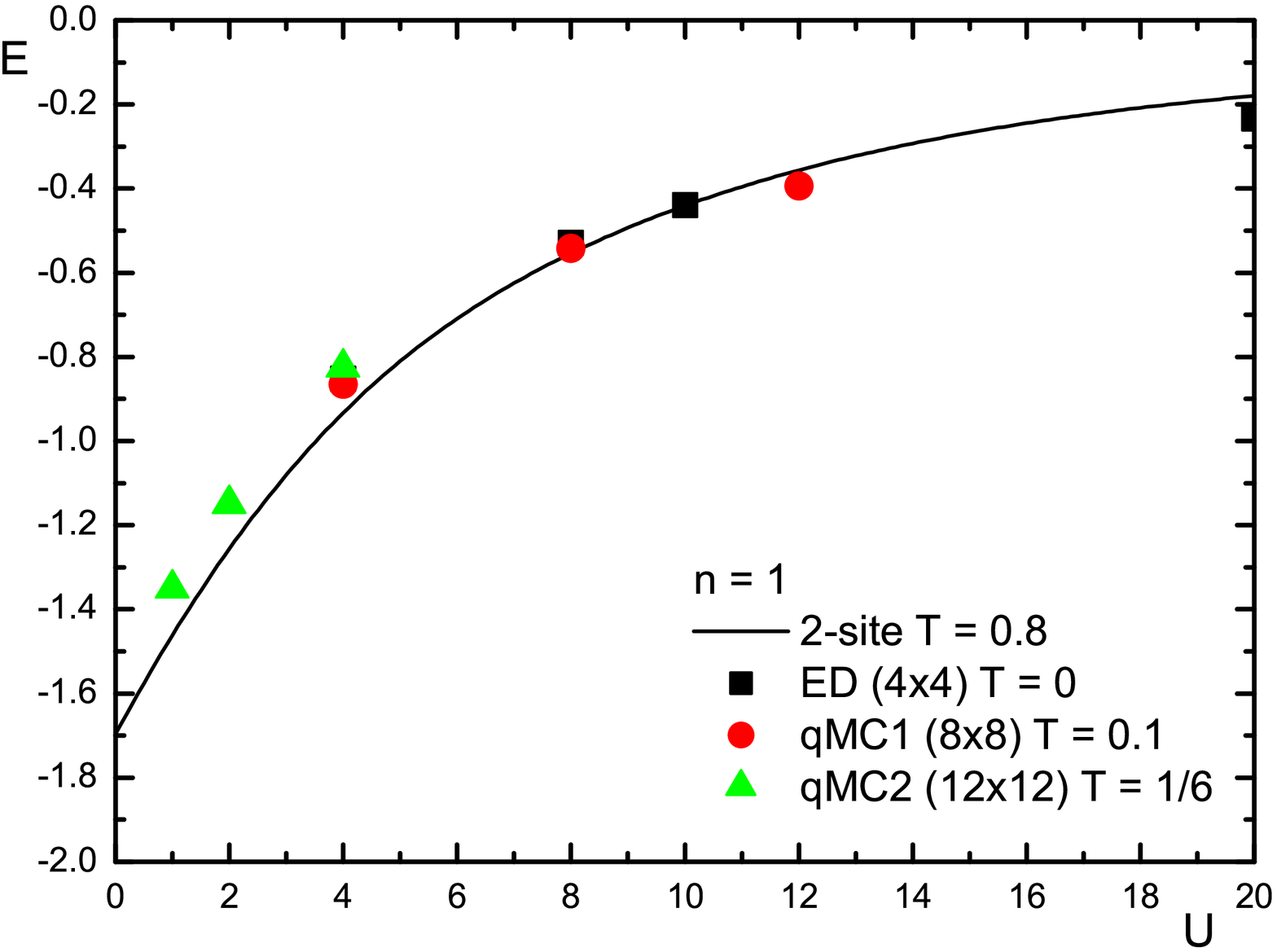}
\end{center}
\caption{The energy  $E$: (top) as function of the filling $n$ for
$T=0.8$ and $U=8$, $10$ and $20$, the ED data ($4\times4$) are for
$T=0$ and from Ref.~\protect\onlinecite{Dagotto:92a}; (middle) as
function of the temperature $T$ for $U=12$ and $n=0.5$, $0.75$ and
$1$ [$U=24$ and $J=1/6$ ($t$-$J$ mode) data are also reported],
the qMC data ($6\times6$ and $8\times8$) are from
Ref.~\protect\onlinecite{Duffy:97b}; (bottom) as function of the
Coulomb repulsion $U$ for $n=1$ and $T=0.8$, the ED data
($4\times4$) are for $T=0$ and from
Ref.~\protect\onlinecite{Dagotto:92a}, the qMC1 data ($8\times8$)
are for $T=0.1$ and from Ref.~\protect\onlinecite{Duffy:97b}, the
qMC2 data ($12\times12$) are for $T=1/6$ and from
Ref.~\protect\onlinecite{Moreo:90}.} \label{Fig8}
\end{figure}

\subsubsection{The energy $E$, the specific heat $C$ and the entropy $S$}

The energy per site $E$ is computed as thermal average of the
Hamiltonian divided by the number of sites. Its filling dependence
is reported in Fig.~\ref{Fig8} (top panel) together with some
Exact Diagonalization (ED) data for bigger
clusters\cite{Dagotto:92a}. As expected, the energy $E$ increases
as the Coulomb repulsion $U$ increases. For $n<1$ and high values
of the Coulomb repulsion the kinetic term prevails on the Coulomb
one. The opposite behavior is observed for $n>1$. The behavior as
a function of the temperature $T$ and Coulomb repulsion $U$ are
reported in Fig.~\ref{Fig8} (middle and bottom panels,
respectively) together with some data coming from numerical
analysis for bigger clusters\cite{Moreo:90,Dagotto:92a,Duffy:97b}.
By using a higher temperature for the 2-site data is again
possible to get an extremely good agreement with the numerical
results [see Fig.~\ref{Fig8} (top and bottom panels)]. In the case
of the temperature behavior instead, the agreement is obtained
without tuning any parameter [see Fig.~\ref{Fig8} (middle panel)].

The discrepancy at half filling and low temperatures [see
Fig.~\ref{Fig8} (middle panel)] is a consequence of the different
value of the exchange energy $J$ appearing in the derivation of
the 2-site $t$-$J$ model from the Hubbard one (see detailed
discussion about the behavior of $C^\alpha_{11}$ reported in
Fig.~\ref{Fig1} (bottom panel), Sec.~\ref{Deltap}). According to
this, at half-filling and low temperatures we have also reported
the 2-site results for $U=24$ and $J=1/6$ ($t$-$J$ model) and
obtained the expected agreement. The reason why no effect is
evident for lower fillings and high temperatures is that the
exchange interaction is really effective only at half-filling (in
average one spin per site is necessary) and low temperatures (the
fluctuations should be small).

\begin{figure}[tb!!]
\begin{center}
\includegraphics[width=8cm,keepaspectratio=true]{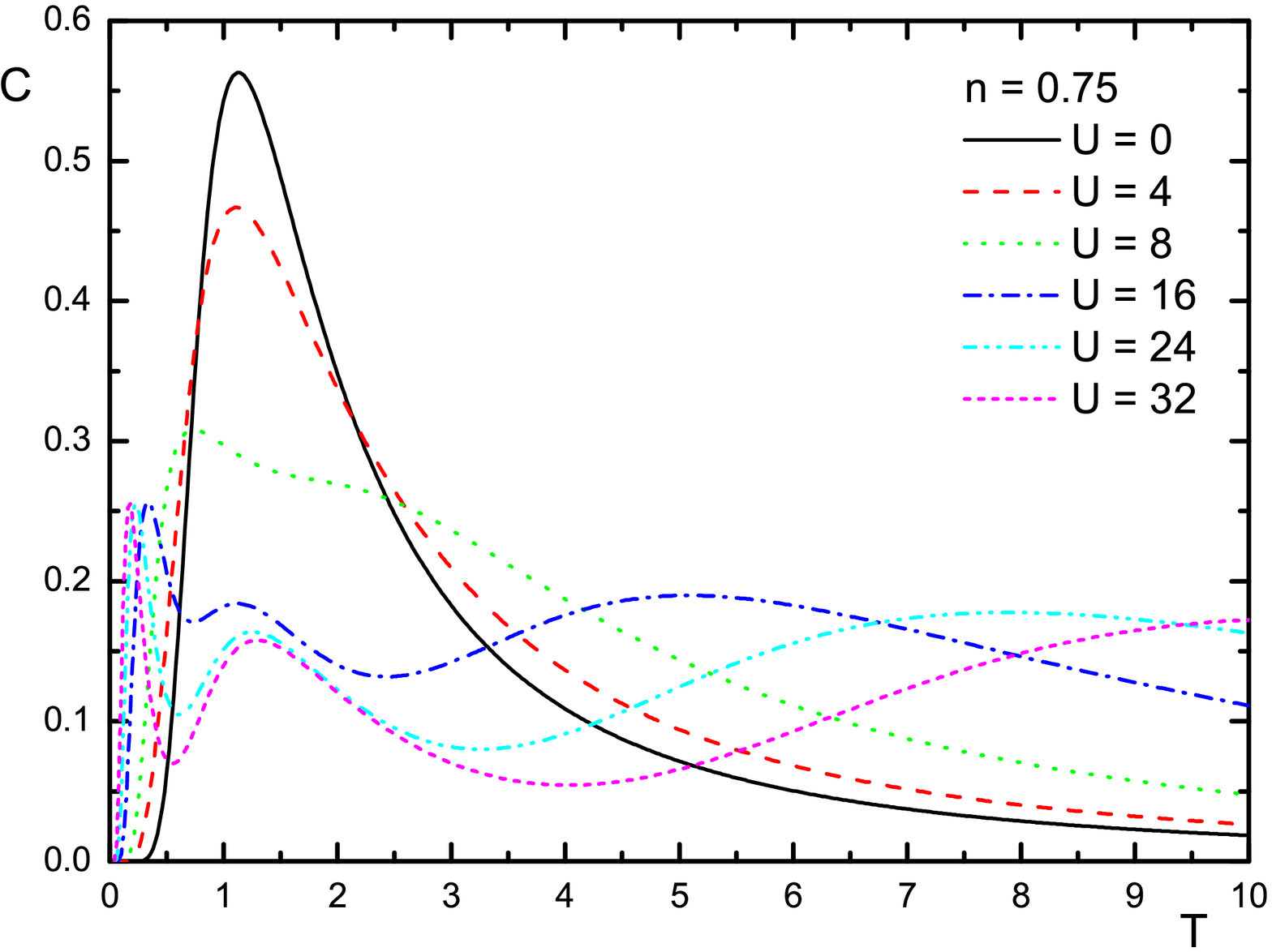}
\includegraphics[width=8cm,keepaspectratio=true]{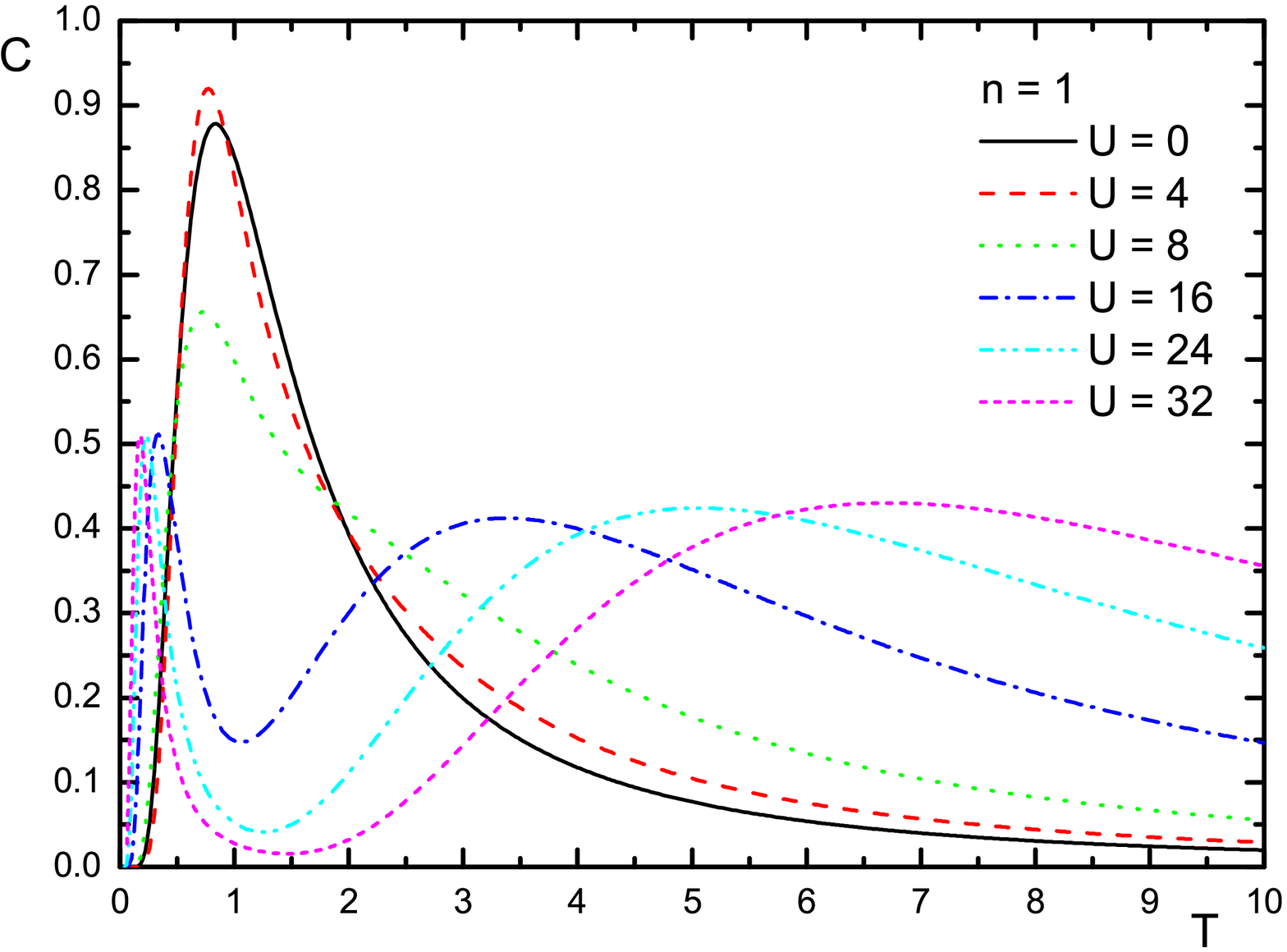}
\end{center}
\caption{The specific heat $C$: as function of the temperature $T$
for $n=0.75$ (top) [$n=1$ (bottom)] and $U=0$, $4$, $8$, $16$,
$24$ and $32$.} \label{Fig9}
\end{figure}

In Fig.~\ref{Fig9} we show the temperature dependence of the
specific heat $C=dE/dT$ for several values of $U$ at $n=0.75$ and
$n=1$. For $n\neq1$ we have three peaks at temperatures of the
order half ($0.3\div0.6$) the scales of energies $J_U$, $t$ and
$U$. This is in agreement with the behavior of a \emph{pure} two-
or three- gap system in the canonical ensemble. The effect of
increasing the Coulomb repulsion $U$ is a better resolution of the
three peaks. In fact, the first one moves towards lower
temperatures, the second one is stable and the third moves to
higher temperatures; this is in perfect agreement with their
origins: $1/U$, $t$ and $U$, respectively. At half filling nothing
changes except for the absence of the middle peak: the single
occupied states do not contribute. It is worth mentioning that the
{\it kinetic} peak appears as the energy levels corresponding to
different values of the momentum are here discrete as in any
finite system. No such a peak is present in infinite systems where
the kinetic energy just spreads over a band the energy levels
coming from the interactions.

\begin{figure}[tb!!]
\begin{center}
\includegraphics[width=8cm,keepaspectratio=true]{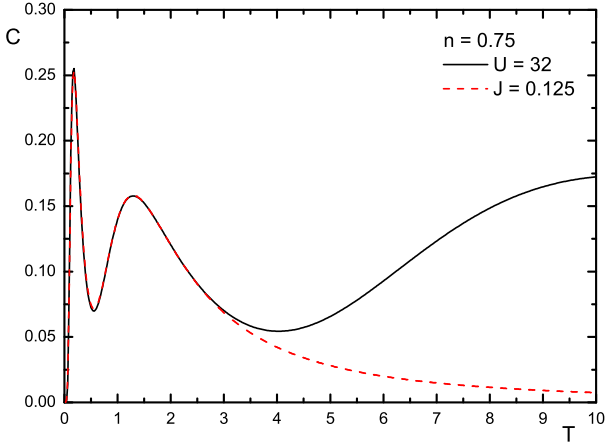}
\end{center}
\caption{The specific heat $C$ as function of the temperature $T$
for $n=0.75$ and $U=32$ and $J=0.125$.} \label{Fig10}
\end{figure}

\begin{figure}[tb!!]
\begin{center}
\includegraphics[width=8cm,keepaspectratio=true]{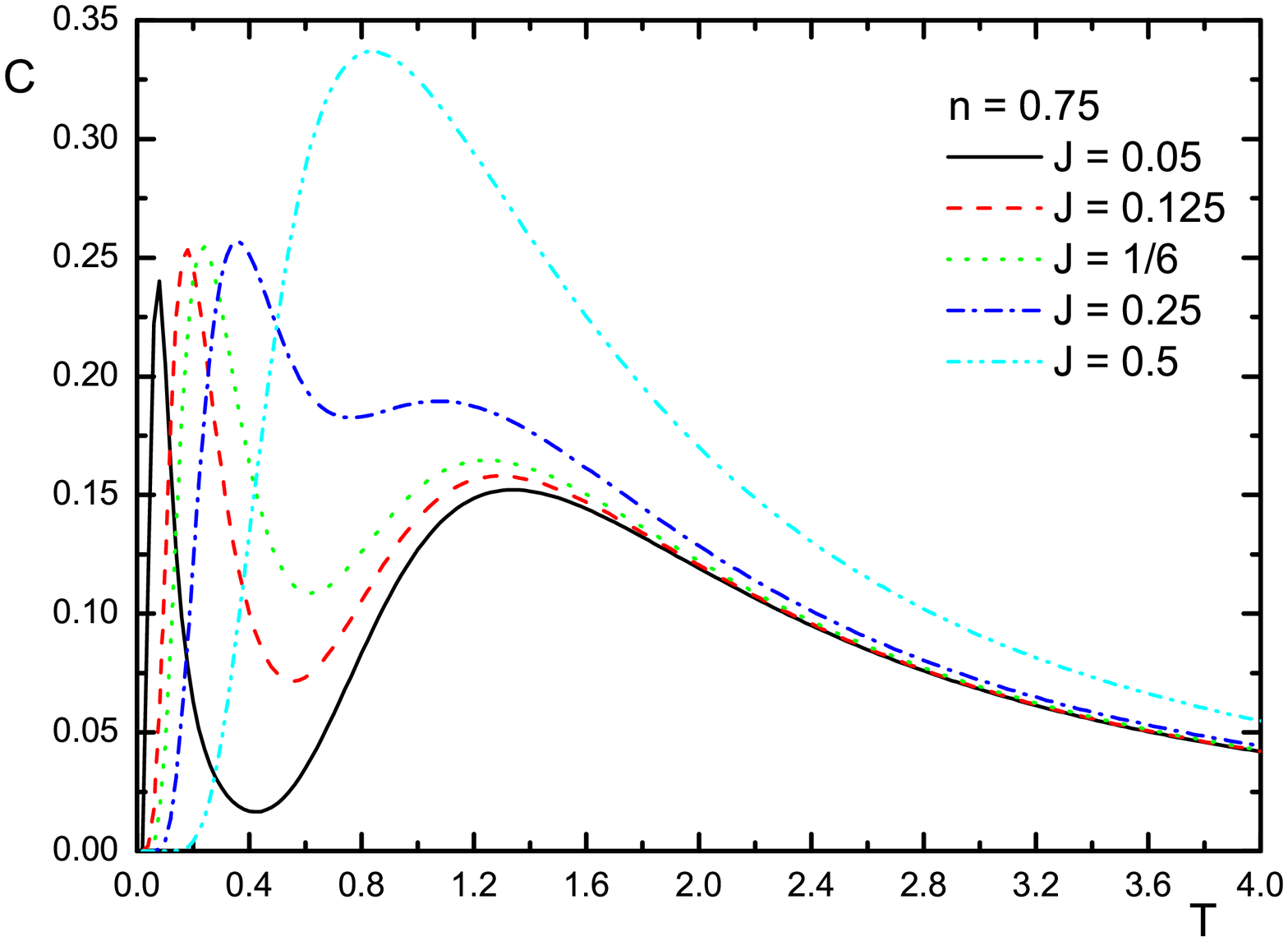}
\includegraphics[width=8cm,keepaspectratio=true]{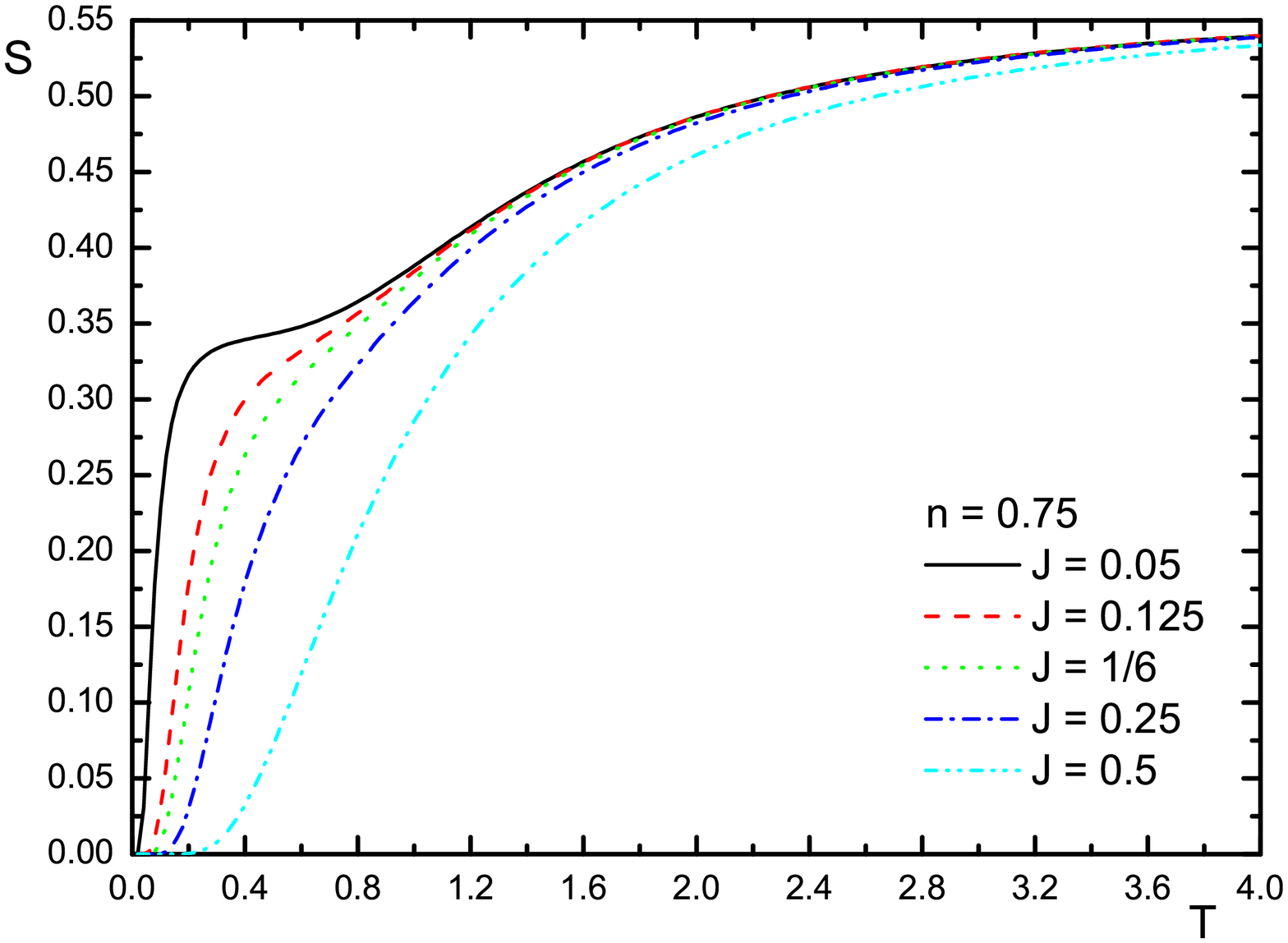}
\end{center}
\caption{(top) The specific heat $C$ as function of the
temperature $T$ for $n=0.75$ and $J=0.0.5$ , $0.125$, $1/6$,
$0.25$ and $0.5$; (bottom) The entropy $S$ as function of the
temperature $T$ for $n=0.75$ and $J=0.0.5$ , $0.125$, $1/6$,
$0.25$ and $0.5$.} \label{Fig11}
\end{figure}

\begin{figure}[tb!!]
\begin{center}
\includegraphics[width=8cm,keepaspectratio=true]{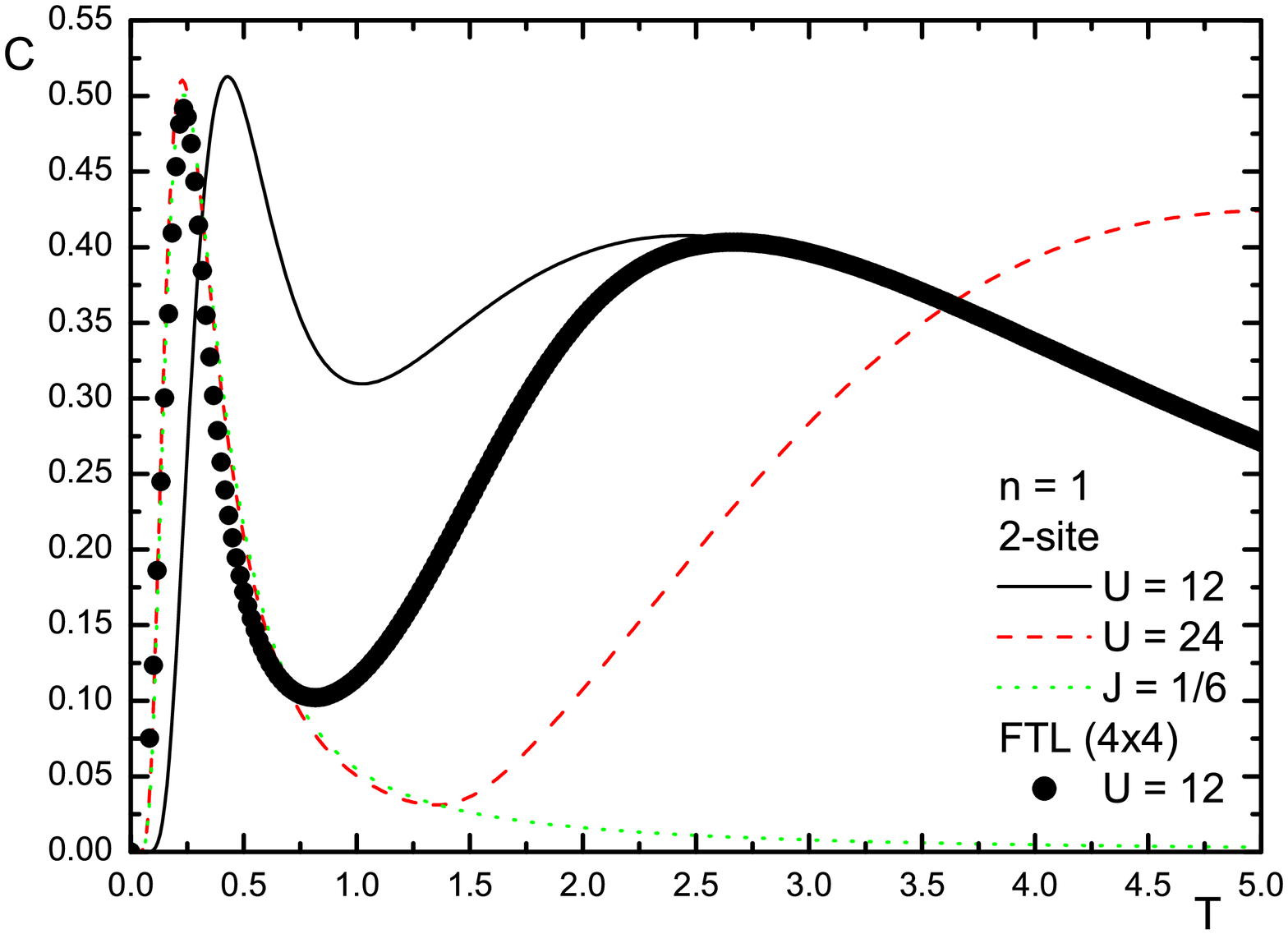}
\includegraphics[width=8cm,keepaspectratio=true]{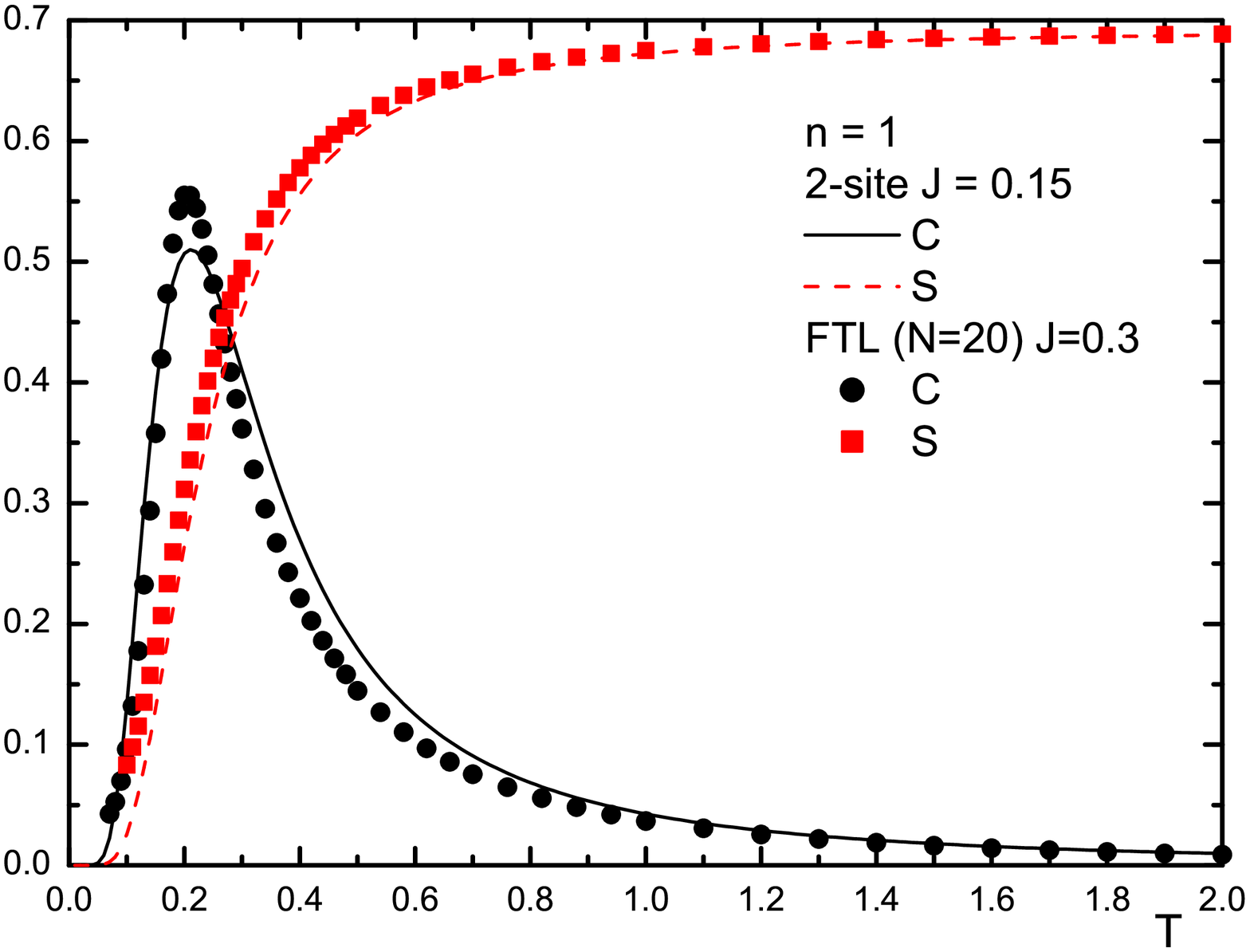}
\end{center}
\caption{(top) The specific heat $C$ as a function of the
temperature $T$ for $U=12$ at $n=1$. The $U=24$ and $J=1/6$
($t$-$J$ mode) data are also reported. The FTL data ($4 \times 4$)
are taken from Ref.~\protect\onlinecite{Bonca:02}. (bottom) The
specific heat $C$ and the entropy $S$ as functions of the
temperature $T$ for $J=0.15$ at $n=1$. The FTL data ($N=20$) are
from Ref.~\protect\onlinecite{Jaklic:00}.} \label{Fig13}
\end{figure}

The specific heat $C$ is a really valuable property to quantify
how good is the $t$-$J$ model to describe the low energy dynamics
of the Hubbard model. We have studied $C$ as a function of both
$4t^{2}/U$ and $J$ in the Hubbard and $t$-$J$ model, respectively.
We need a Coulomb repulsion $U\geq24t$ ($4t^{2}/U \leq \frac16 t$)
to have the possibility of a faithful mapping at low temperatures;
only for such values of $U$ we manage to sufficiently resolve the
three scales of energy in the Hubbard model. For instance, at
$U=32 \Leftrightarrow J=0.125$ the mapping can be absolutely
trusted as shown in Fig.~\ref{Fig10} where we report the specific
heat for $n=0.75$. As regards the {\it exchange} and {\it kinetic}
peaks the agreement is perfect; obviously the $t$-$J$ model cannot
reproduce the {\it Coulomb} peak, which is extraneous to its
dynamics. The absence of the {\it kinetic} peak at half filling is
also perfectly reproduced (see Fig.~\ref{Fig13}).

In Fig.~\ref{Fig11} (top panel) we report the behavior, in the
$t$-$J$ model, of the specific heat as function of the temperature
$T$ for different values of the exchange constant $J$. It is worth
mentioning that in the $t$-$J$ model the two peaks in the specific
heat (i.e., the {\it exchange} and {\it kinetic} peaks) can be
easily studied separately as they just come from the corresponding
terms of the Hamiltonian. This simple analysis cannot be performed
in the Hubbard model where the {\it exchange} peak gets
contribution from both terms of the Hamiltonian.

In Fig.~\ref{Fig11} (bottom panel) we report the entropy $S$ of
$t$-$J$ model computed by means of the usual thermodynamic
relations from the specific heat $C$
\begin{equation}
S(T)=\int_0^T\frac{d\tilde{T}}{\tilde{T}}C(\tilde{T})
\end{equation}
We can easily put in correspondence the peaks of the specific heat
[see Fig.~\ref{Fig11} (middle panel)] and the change in the slope
of the entropy. Again, a lowering of the exchange energy $J$ helps
resolving the scales of energy and makes much more visible the
difference in the slopes.

In Fig.~\ref{Fig13}, it is reported the specific heat as a
function of the temperature for the Hubbard model (top panel) and
the $t$-$J$ model (bottom panel). For this latter, it has been
also reported the entropy. The Finite Temperature Lanczos (FTL)
data for the Hubbard model ($4 \times 4$) and for the $t$-$J$
model ($N=20$) are taken from Ref.~\protect\onlinecite{Bonca:02}
and Ref.~\protect\onlinecite{Jaklic:00}, respectively. The
agreement is absolutely noteworthy on the whole temperature range.
Again, the discrepancy at low temperatures in the Hubbard model
case is a consequence of the difference regarding the value of the
exchange energy $J$. According to this, we have also reported the
2-site results for $U=24$ and $J=1/6$ ($t$-$J$ model) and obtained
the expected agreement. In the $t$-$J$ model case, we have used a
value of $J$ twice smaller than the one used within the numerical
analysis according to the boundary conditions we applied to the
2-site system. These results show, once more, that the correct and
necessary scales of energies are already present in the 2-site
system.

\begin{figure}[tb!!]
\begin{center}
\includegraphics[width=8cm,keepaspectratio=true]{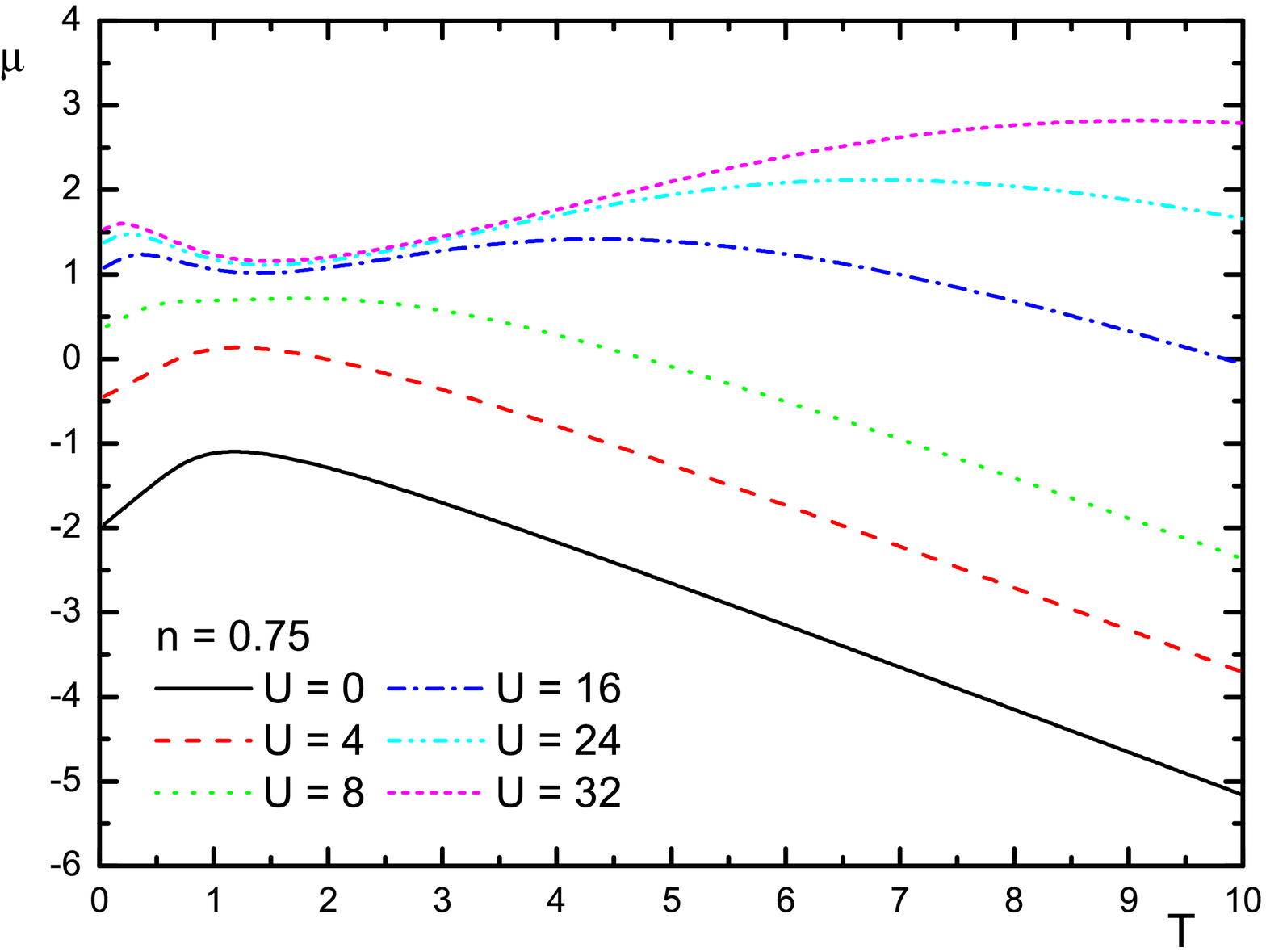}
\includegraphics[width=8cm,keepaspectratio=true]{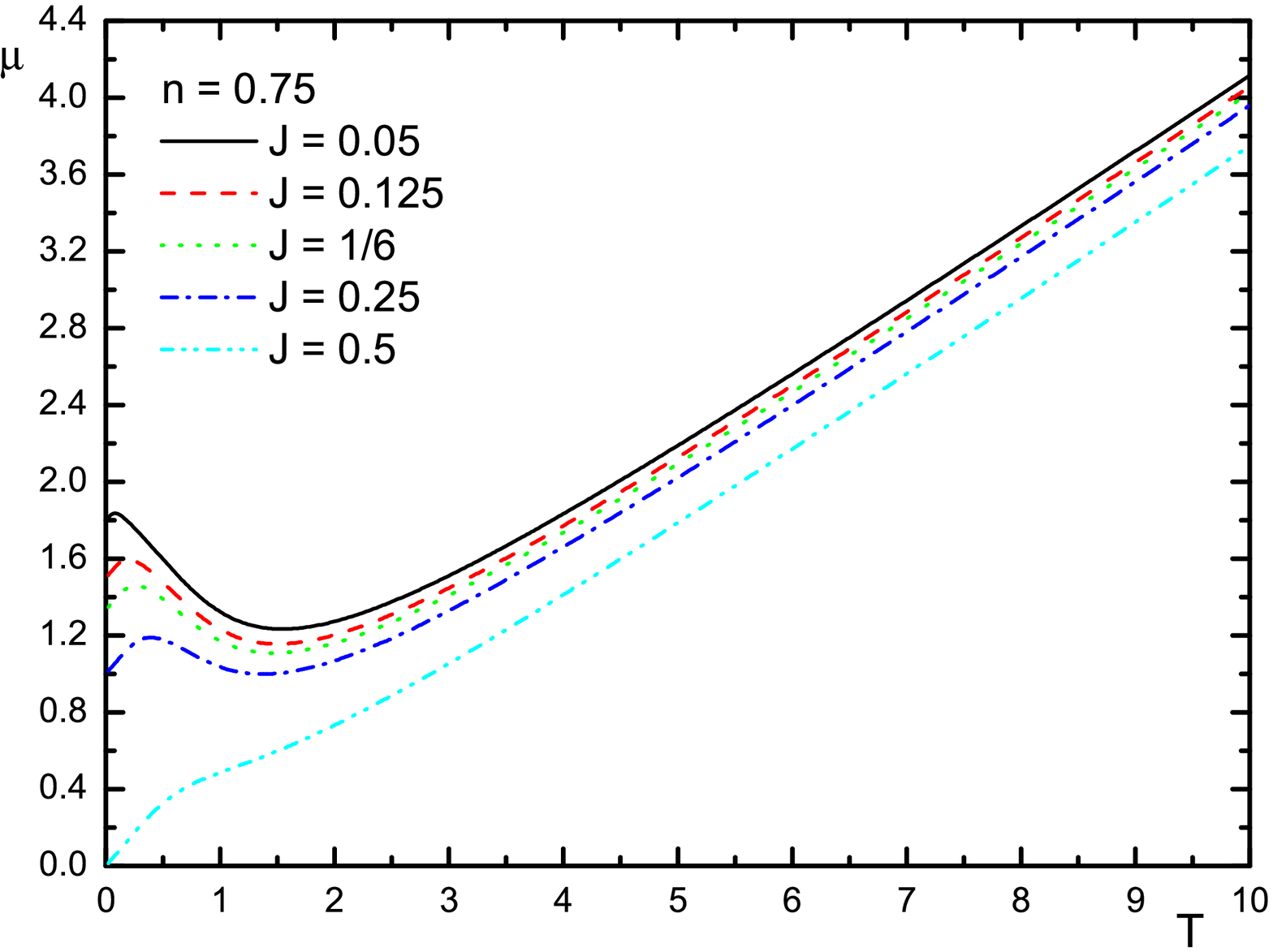}
\includegraphics[width=8cm,keepaspectratio=true]{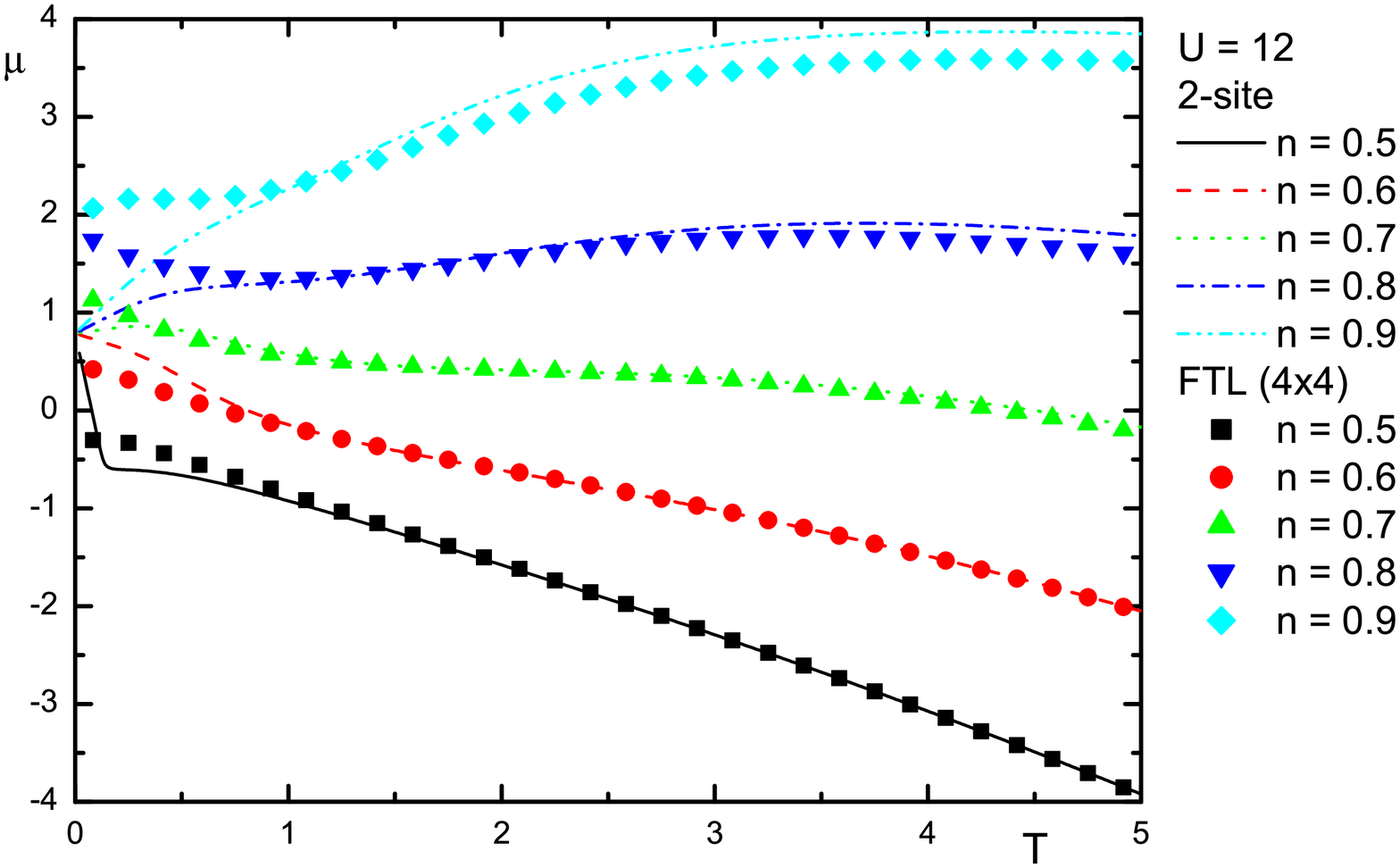}
\end{center}
\caption{The chemical potential $\mu$: (top) as function of the
temperature $T$ for $n=0.75$ and $U=0$, $4$, $8$, $16$, $24$ and
$32$; (middle) as function of the temperature $T$ for $n=0.75$ and
$J=0.05$ , $0.125$, $1/6$, $0.25$ and $0.5$; (bottom) as function
of the temperature $T$ for $U=12$ and $n=0.5$, $0.6$, $0.7$, $0.8$
and $0.9$. The FTL data ($4 \times 4$) are from
Ref.~\protect\onlinecite{Prelovsek}.} \label{Fig2}
\end{figure}

\subsubsection{The temperature dependence of \protect\\
the chemical potential $\mu$ and the double occupancy $D$}

The temperature dependence of the chemical potential $\mu $ is
given in Fig.~\ref{Fig2} (top panel). $\mu $ has only one maximum
for $U \lesssim 8$ and two maxima and a minimum for higher values
of $U$. In the $t$-$J$ model, the chemical potential $\mu $ shows
a behavior at low temperatures in agreement with the one found for
the Hubbard model (i.e., $\lim_{U\gg t}J_{U}=J$). On the contrary,
for high temperatures the {\it divergence} is upward instead of
downward [see Fig.~\ref{Fig2} (bottom panel)].

The temperature dependence of the chemical potential can be
explained as follows. In the Hubbard model and at zero
temperature, there exists a critical value of the Coulomb
interaction, function of the filling, above which the potential
energy becomes larger, in absolute value, than the kinetic one.
This critical value is a decreasing function of the filling: lower
is the doping higher the double occupancy, bigger the potential
energy and smaller, in absolute value, the kinetic energy. Then,
by increasing the filling we increase the energy and the chemical
potential increases. At higher and higher temperatures, the system
behaves like a free system at zero temperature and has a
decreasing chemical potential. In particular, for a diverging
temperature we have a negatively diverging chemical potential
(i.e., $\mu=dE/dn-T dS/dn$) as the entropy is an increasing
function of the filling at very high temperatures (i.e., in the
almost free case) and $n<1$. For intermediate temperatures, the
chemical potential has maxima and minima in coincidence with the
peaks in the specific heat. Any peak in the specific heat marks
the temperature at which the system gets the freedom to occupy a
new state. The {\it spin} (i.e., $T\sim J_{U}$) and the {\it
charge} (i.e., $T\sim U$) peaks reflect in two maxima as they
signal the availability of triplet states and of doubly occupied
states, respectively. Both states has zero kinetic energy. On the
contrary, the {\it kinetic} (i.e., $T\sim t$) peak signals the
availability of further singly occupied states which maximize the
absolute value of the kinetic energy. This reasoning strongly
relies on the region of filling the figure refers to (i.e.,
$n\approx0.75$). For other doping regions the situation can be
quite different owing to the behavior in filling of the double
occupancy, the entropy and the positions and existence of the
peaks in the specific heat. These considerations can also explain
the quite different behavior we have in the $t$-$J$ model. At
$n=0.75$ the $t$-$J$ model is approaching the full-filled system
(i.e., $n=1$), on the contrary the Hubbard model is approaching
the half-filled system. At zero temperature for $n>0.5$,
increasing the filling we lowers, in absolute value, the kinetic
energy as the single electron states, which are the only ones with
a finite kinetic energy, are replaced by the singlet and triplet
states: the chemical potential is positive. Anyway, the exchange
energy can effectively lower its value on increasing $J$. At very
high temperatures (i.e., in the almost {\it free} case), the
entropy is now a decreasing function of the filling and leads to
the positive divergence. At intermediate temperatures, the
reasoning is identical to the one given for the Hubbard model
except for the obvious absence of the {\it charge} peak in the
$t$-$J$ model.

In Fig.~\ref{Fig2} (bottom), it is reported the chemical potential
as function of the temperature $T$ for $U=12$ and $n=0.5$, $0.6$,
$0.7$, $0.8$ and $0.9$. The FTL data ($4 \times 4$) are from
Ref.~\protect\onlinecite{Prelovsek}. The agreement is really
excellent except at low temperatures and for the higher values of
the filling ($n=0.8$ and $n=0.9$). At low temperatures, owing to
the finite level spacing the chemical potential has a step
behavior and, in particular, between $n=0.5$ and $n=0.9$ has
practically no variation (cfr. Fig~\ref{Fig1}). At the higher
values of the filling ($n=0.8$ and $n=0.9$), the number of 2-site
states excited at high temperatures is larger than those needed to
mime the cluster.

The temperature dependence of the double occupancy $D$ at half
filling shows a minimum at around $T \approx 1$ before saturating
at the non-interacting value $n^{2}/4$ for very high
temperatures\cite{Georges:93,Schulte:96}. This minimum coincides
with the {\it kinetic} peak in the specific heat. As already
discussed for the chemical potential, this peak marks the freedom
for the system to occupy further singly occupied states which
have, obviously, zero double occupancy and, therefore, lower its
total value.

\begin{figure}[tb!!]
\includegraphics[width=8cm,keepaspectratio=true]{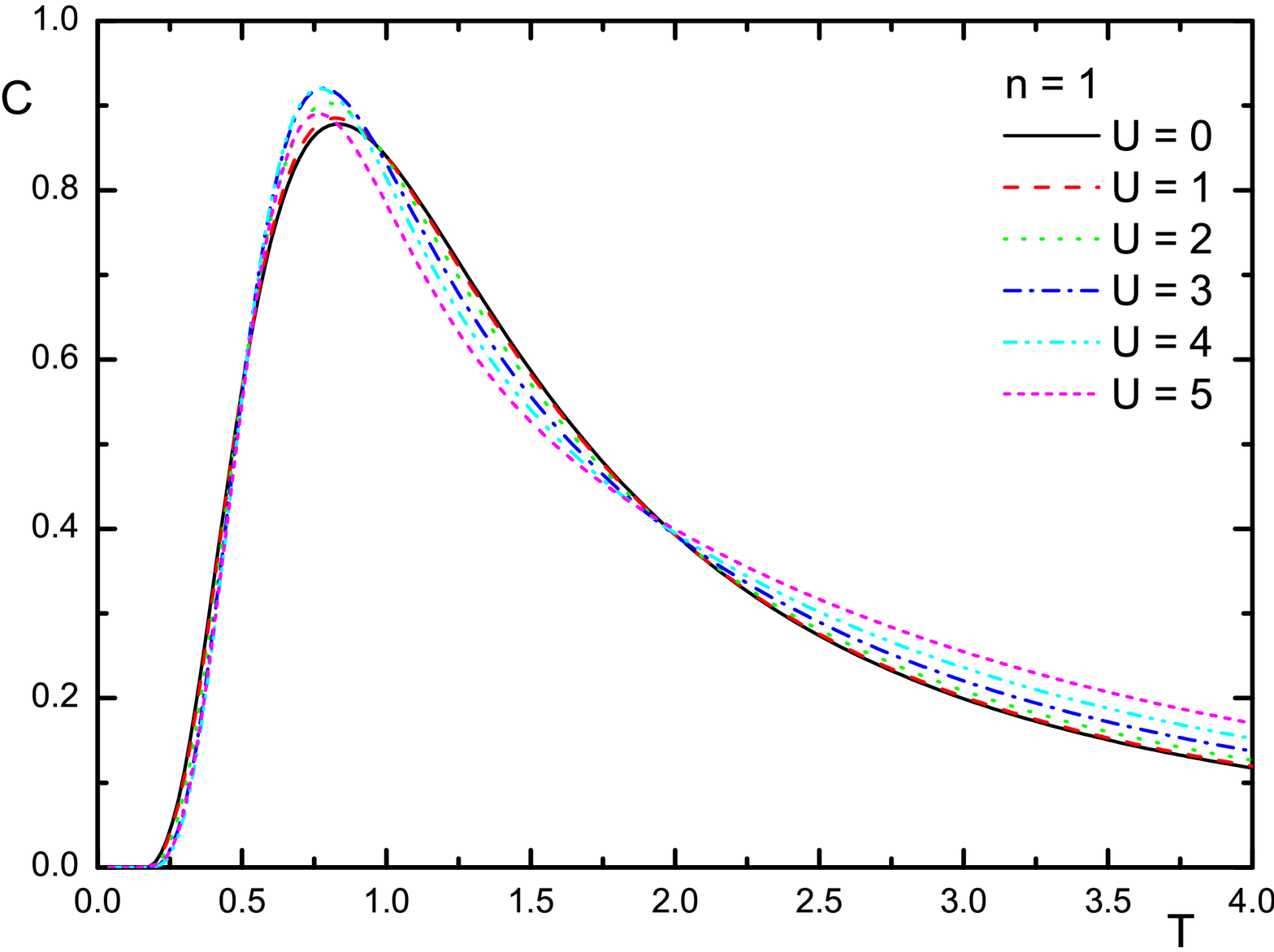}
\includegraphics[width=8cm,keepaspectratio=true]{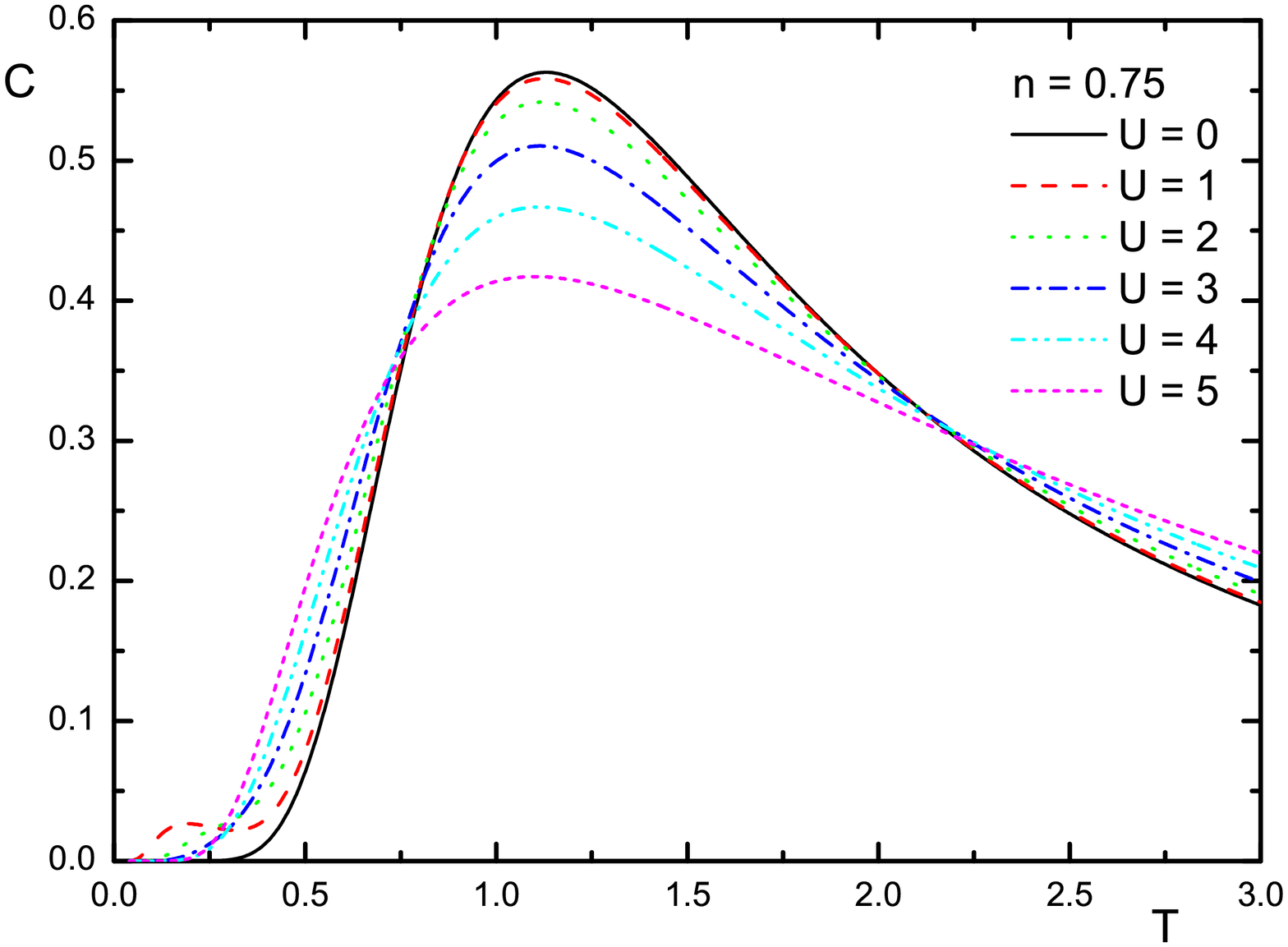}
\caption{The specific heat $C$ as function of the temperature $T$
for $n=1$ (left panel) [$n=0.75$ (right panel)] and $U=0$, $1$,
$2$, $3$, $4$ and $5$.} \label{Fig12}
\end{figure}

\subsubsection{The crossing points}

In Fig.~\ref{Fig12} we report the temperature dependence of the
specific heat for $U \leq 5t$ at half filling (top panel) and
$n=0.75$ (bottom panel). For these low values of the Coulomb
repulsion we identify three crossing points/regions where the
specific heat is almost independent from the value of the Coulomb
repulsion $U$\cite{Vollhardt:97,Chandra:99,Mancini:99a}. To
compute the positions and heights of the crossing points we have
expanded the specific heat $C$ as function of the Coulomb
repulsion $U$. We have got
\begin{subequations}
\begin{align}
& C(T,U,n) = C_{0}(T,n)+C_{2}(T,n) U^2+O\left(U^4\right) \\
& C_{0}(T,n=1) = 2\beta^2 t^2\left[1-\tanh^2\left(\beta t
\right)\right]
\end{align}
\end{subequations}
\begin{widetext}
\begin{equation}
C_{2}(T,n=1) = \frac{\mathrm{e}^{2\beta t}}{8
T^{4}\left(1+\mathrm{e}^{2\beta t}\right)^{6}} \left[48t^2
\mathrm{e}^{2\beta t}\left(1-3\mathrm{e}^{2\beta
t}+\mathrm{e}^{4\beta t}\right) +2t T\left(1+20\mathrm{e}^{2\beta
t}-20\mathrm{e}^{6\beta t}-\mathrm{e}^{8\beta t}\right)
+2T^2\left(1+\mathrm{e}^{2\beta t}\right)^4 \right]
\end{equation}
\end{widetext}

The temperatures $T^*(n)$ at which the crossing point can be
observed are determined by the equation $C_{2}(T,n)=0$. At half
filling and at $n=0.75$, this equation gives the following
results:
\begin{subequations}\label {C0}
\begin{align}
& T^*_1(n=1)\cong 0.499 \quad \Rightarrow \quad C_0 \cong 0.563 \\
& T^*_2(n=1)\cong 0.997 \quad \Rightarrow \quad C_0 \cong 0.841 \\
& T^*_3(n=1)\cong 2.058 \quad \Rightarrow \quad C_0 \cong 0.376
\label {C0H}
\end{align}
\end{subequations}
\begin{subequations}
\begin{align}
& T^*_1(n=0.75)\cong 0.951 \quad \Rightarrow \quad C_0 \cong 0.523 \\
& T^*_2(n=0.75)\cong 2.024 \quad \Rightarrow \quad C_0 \cong 0.342
\end{align}
\end{subequations}
Vollhardt found an approximate formula for $C_0$ and $T^*$, as
function of the dimensionality of the system, for the higher
temperature crossing point at half filling\cite{Chandra:99}. The
values that this formula gives for $d=1$, the
\emph{dimensionality} of the 2-site system within periodic
boundary conditions, are close, as regards $C_0$, to the exact
ones of Eq.~(\ref{C0H}).

It is worth noting that higher is the temperature narrower is the
crossing region; ranging from quite wide regions for the two lower
temperatures to a very sharp point for the higher temperature.
Also the chemical potential (Fig.~\ref{Fig1} (top panel) at
$n\approx0.275, 0.75, 1.25, 1.725$) and the double occupancy show
quite clear crossing points in (no dependence on) temperature.

\begin{figure}[tb!!]
\includegraphics[width=8cm,keepaspectratio=true]{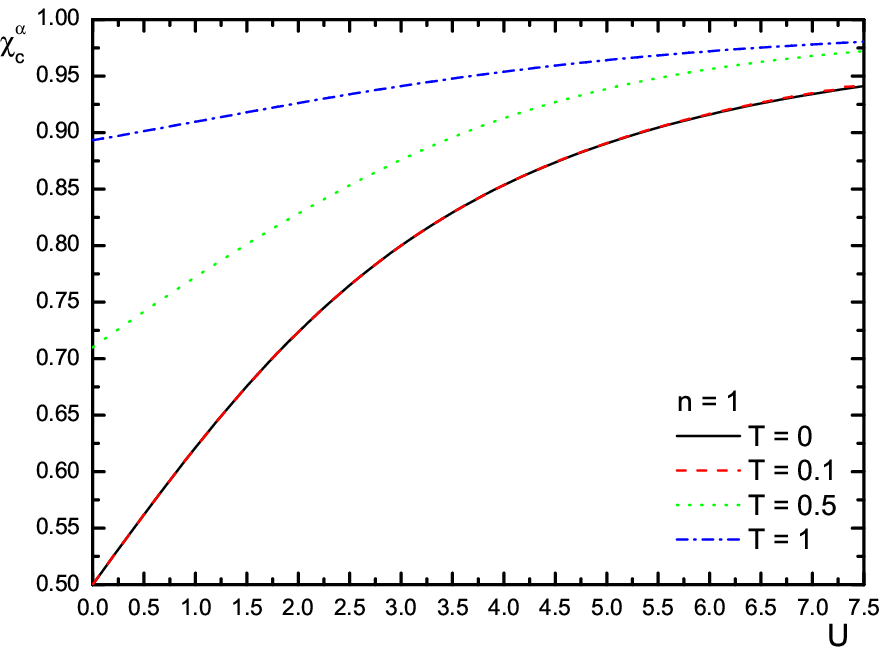}
\includegraphics[width=8cm,keepaspectratio=true]{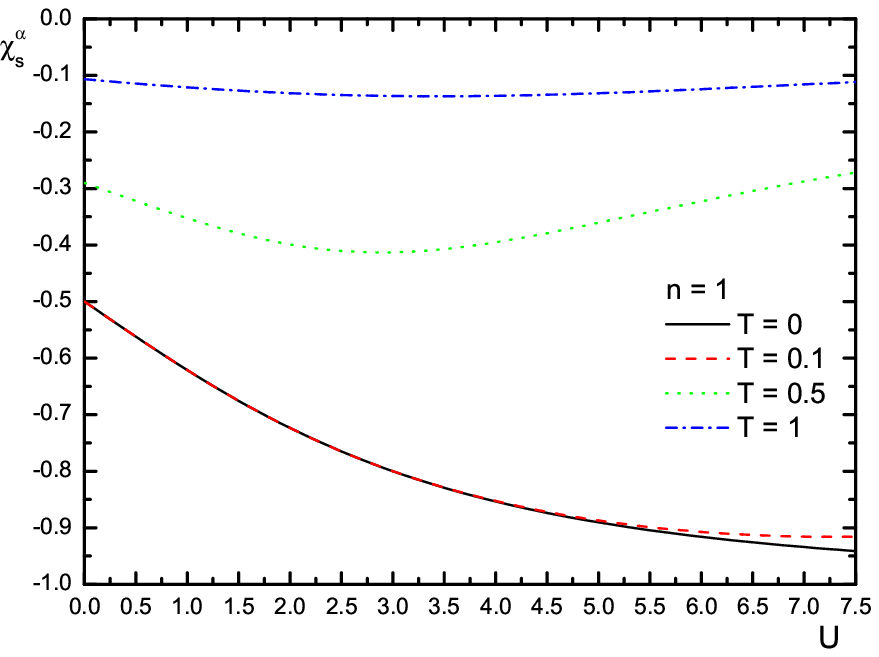}
\caption{The charge $\chi^\alpha_c$ (top) and spin $\chi^\alpha_s$
(bottom) correlation functions as functions of $U$ at $n=1$ and
for $T=0$, $0.1$, $0.5$ and $1$.} \label{Fig15}
\end{figure}

\begin{figure}[tb!!]
\includegraphics[width=8cm,keepaspectratio=true]{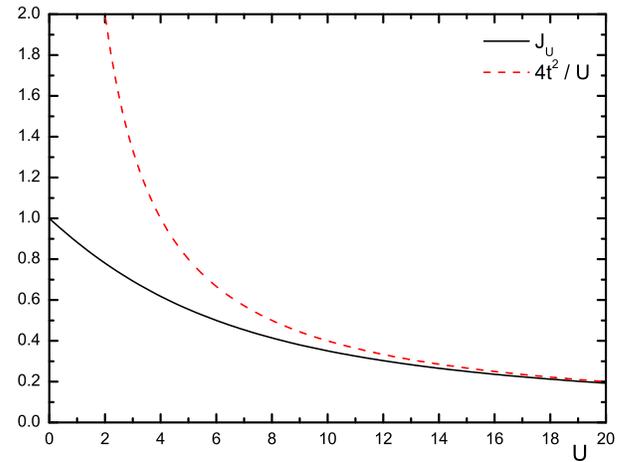}
\caption{Comparison between $J_U$ and its limiting value $4t^2/U$
as functions of $U$.} \label{Fig41}
\end{figure}

\begin{figure}[tb!!]
\includegraphics[width=8cm,keepaspectratio=true]{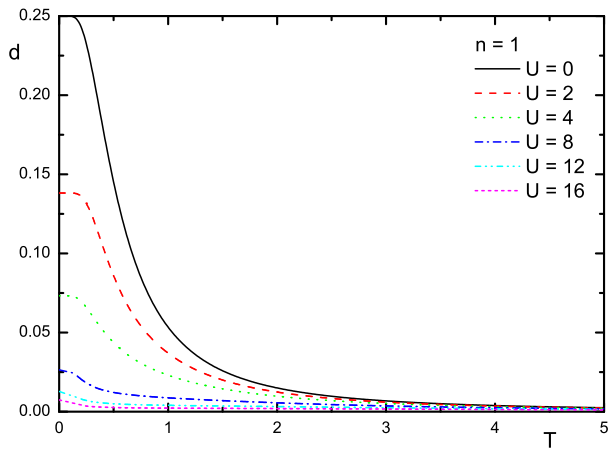}
\caption{The pair correlation function $d$ as function of $T$ at
$n=1$ and for $U=0$, $2$, $4$, $8$, $12$ and $16$.} \label{Fig22}
\end{figure}

\subsection{Charge, spin and pair correlations}

The charge (i.e., $\chi_{c}^{\alpha}$), spin (i.e.,
$\chi_{s}^{\alpha}$) and pair (i.e., $d$) correlation functions
contain information regarding the spatial distributions of the
corresponding quantities. Obviously, we can only consider the
quantum fluctuations in the paramagnetic state, which is the only
admissible equilibrium state on a finite cluster.

The charge correlation function $\chi_{c}^{\alpha}$ is shown in
Fig.~\ref{Fig15} (top) as function of the interaction $U$ for
different values of the temperature $T$ and $n=1$. In the
non-interacting case ($U=0$) and at zero temperature we have
\begin{equation}
\chi_{c}^{\alpha}-n^{2}=\left\{
\begin{tabular}{lll}
$-\frac{1}{2}n^{2}$ & if & $n\leq1$ \\
$-\frac{1}{2}\left( 2-n\right) ^{2}$ & if & $n\geq1$
\end{tabular}
\right.
\end{equation}
as only the singlet states with one electron ($n\leq1$) [one hole
($n\geq1$)] per site contribute. In fact, these are the states
which lower the most the internal energy as they maximize the
absolute value of the kinetic one. In the strongly interacting
limit ($U\rightarrow\infty$) and at zero temperature we have
\begin{equation}
\chi_{c}^{\alpha}-n^{2}=\left\{
\begin{tabular}{lll}
$-n^{2}$ & if & $n\leq0.5$ \\
$-\left( 1-n\right) ^{2}$ & if & $0.5\leq n\leq1.5$ \\
$-\left( 2-n\right) ^{2}$ & if & $n\geq1.5$
\end{tabular}
\right.
\end{equation}
as only the states with a single electron ($n\leq0.5$) [a single
hole ($n\geq1.5$)] or only with one electron per site ($0.5\leq
n\leq1$) [one hole per site ($1 \leq n \leq 1.5$)] contribute. In
fact, no double occupancy is allowed and there is no gain in the
kinetic energy if the singlet states are used as
$J_{U}\rightarrow0$. At intermediate values of the coupling $U$
the double occupied states play a relevant role and we found
results between the limiting cases reported above: on increasing
the coupling $U$ the double occupancy diminishes and consequently
the states with one electron per site increase their contribution
and the value of $\chi_{c}^{\alpha}$ (see Fig.~\ref{Fig15} (top)).
On increasing the temperature $T$, more and more states become
available and the correlation tends to its \textit{reducible}
part, $n^{2}$.

The spin correlation function $\chi_{s}^{\alpha}$ is reported in
Fig.~\ref{Fig15} (bottom) for the same parameters chosen for
$\chi_{c}^{\alpha}$. In the non-interacting case ($U=0$), we have
$\chi_{s}^{\alpha}=\chi_{c}^{\alpha}$: in absence of interaction
the charge and spin behave coherently. On increasing the
interaction potential $U$ the singlet states are favored and the
spin correlations get enhanced ($J_U$ gets closer and closer to
$4t^2/U$ which is the scale of energy of the spin excitations, see
Fig.~\ref{Fig41}). Then, further increasing $U$ the value of $J_U$
tends to zero and the spin correlations get suppressed except at
zero temperature [see Fig.~\ref{Fig15} (bottom)]. A rapid
suppression of the spin fluctuations can be caused also by the
increment of the temperature.

%It is worth noting the energy gap which develops only in the
%interacting case in the filling regions $n<0.5$ and $n>1.5$
%between states with and without spin singlets; only quite high
%(with respect to the actual value of $U$) temperatures can
%overcome the gap and allow non-zero spin fluctuations ($T=0.1$ is
%on the limit for $U=2$).

The pair correlation function $d$ is reported in Fig.~\ref{Fig22}
as function of the temperature $T$ at $n=1$ and for $U=0$, $2$,
$4$, $8$, $12$ and $16$. Obviously, $d$ decreases on increasing
the Coulomb interaction $U$ and is maximum at half-filling where
the singlet state is the favorite one. On increasing the
temperature $T$ (see Fig.~\ref{Fig22}), more and more states
become available and the pair correlation function $d$ tends to
zero.

\begin{figure}[tbp]
\includegraphics[width=8cm,keepaspectratio=true]{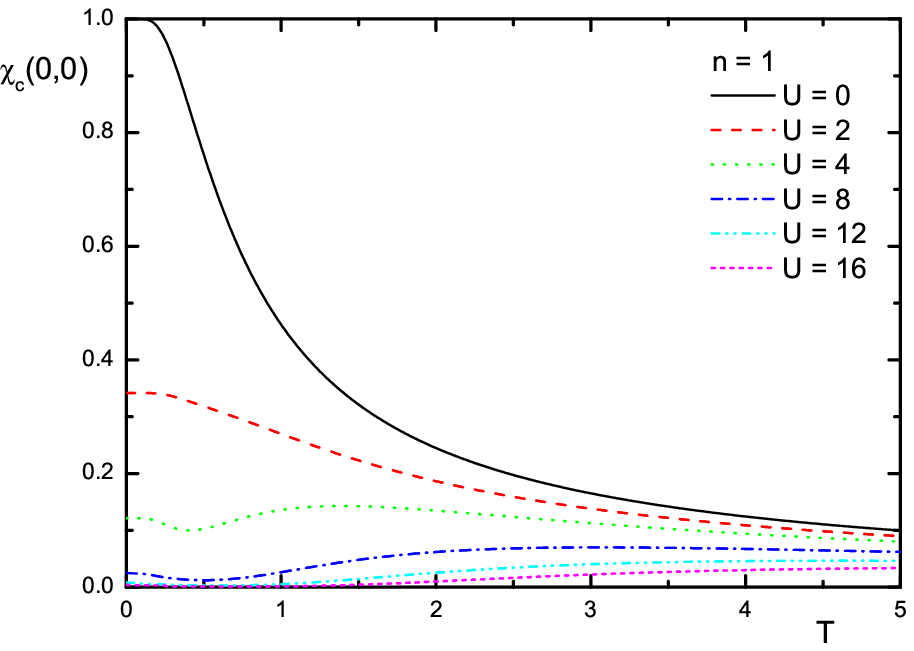}
\includegraphics[width=8cm,keepaspectratio=true]{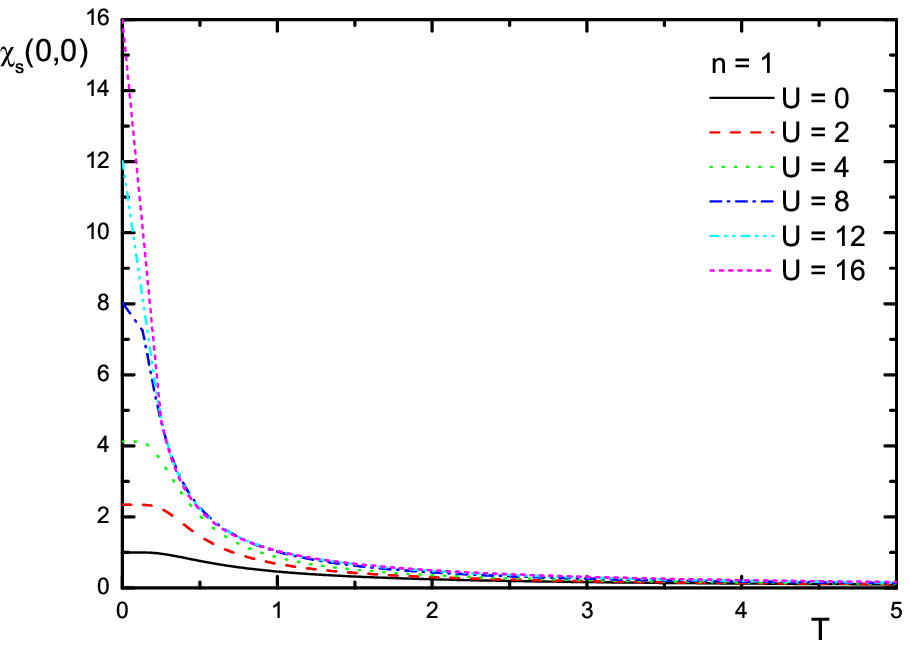}
\caption{The charge $\chi_c(0,0)$ (top) and spin $\chi_s(0,0)$
(bottom) susceptibilities as functions of $T$ at $n=1$ for $U=0$,
$2$, $4$, $8$, $12$ and $16$.} \label{Fig25}
\end{figure}

\subsection{Charge and spin susceptibilities}

The charge and spin susceptibilities $\chi_{\mu}(k,\omega)$ are
given by
\begin{equation}
\chi_{\mu}(k,\omega)=-\mathcal{F}\left\langle \mathcal{R}[n_{\mu
}(i) n_{\mu}(j)]\right\rangle
\end{equation}
where no summation is implied by repeated indices. By means of the
expression of the causal Green's function given in
Sec.~\ref{suscss} is immediate to compute the following static
susceptibilities through the spectral theorem
\begin{align} \label{Suscs}
t \chi_{c}(k,0) & =-C_{cc}^{\alpha}-\frac{U}{2t}d \\
t \chi_{s}(k,0) &
=-C_{cc}^{\alpha}-\frac{1}{12}\frac{U}{t}\chi_{s}^{\alpha}
\end{align}
where $d$ and $\chi_{s}^{\alpha}$ are the first-neighbor pair and
spin correlation functions, respectively.

In Fig.~\ref{Fig25}, the charge $\chi_{c}(k,0)$ and spin
$\chi_{s}(k,0)$ susceptibilities are reported as function of the
temperature $T$ for different values of the Coulomb repulsion $U$
and $n=1$. As it results clear by Eqs.~\ref{Suscs}, the two
susceptibilities result identical in the non-interacting case
$U=0$: once more charge and spin behave coherently in absence of
interaction. On increasing the Coulomb repulsion, the two
susceptibilities behave in opposite manners: the charge
susceptibility is strongly suppressed, in particular at
half-filling where a charge gap develops in the strongly
interacting limit ($U\rightarrow\infty$); the spin susceptibility
is greatly enhanced, in particular at half-filling where the
singlet state is the favorite one. The behavior of the two
susceptibilities as functions of the temperature deserve special
attention as it is directly connected with the formation of a gap
in the corresponding channel. The spin susceptibility shows the
typical paramagnetic behavior: plateau at low temperatures and
Curie tail at high temperatures. The charge susceptibility instead
shows the presence of a gap in the strongly interacting limit
($U\rightarrow\infty$). The small upturn at low temperatures and
medium-high values of the Coulomb repulsion $U$ is due to the
finite level-spacing of the system. In particular, it arises from
the intra-subband gaps characteristic of the scale of energy
$J_U$.

In the $t$-$J$ model, we simply have
\begin{align}
t \chi_{c}\left( k,0\right) & =-C_{11}^{\alpha} \\
t \chi_{s}\left( k,0\right) &
=-C_{11}^{\alpha}-\frac{1}{3}\frac{t}{J} \chi_{s}^{\alpha}
\end{align}

\begin{figure}[tbp]
\includegraphics[width=8cm,keepaspectratio=true]{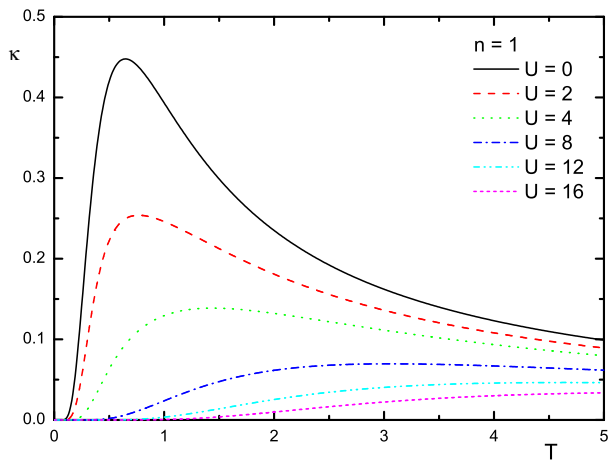}
\caption{The compressibility $\kappa$ as a function of $T$ at
$n=1$ for $U=0$, $2$, $4$, $8$, $12$ and $16$.} \label{Fig31}
\end{figure}

\subsection{Thermal compressibility}

The thermal compressibility is defined as
\begin{equation}
\kappa=\frac{1}{n^{2}}\frac{\partial n}{\partial\mu}
\end{equation}
By using some general quantum statistical relations, which can be
established between the particle density $n$ and the chemical
potential $\mu$ \cite{Feng:00}, we can express the thermal
compressibility in terms of the density-density correlation
function
\begin{equation}
\kappa=\frac{1}{T n^{2}}\left[ \left( n+2D-n^{2}\right) +\left(
\chi _{c}^{\alpha}-n^{2}\right) \right]
\end{equation}

In Fig.~\ref{Fig31}, the thermal compressibility $\kappa$ is
reported as function of the temperature $T$ for different values
of the Coulomb repulsion $U$ and $n=1$. The system is completely
incompressible $\kappa=0$ when no more particles are allowed to
enter the system: the chemical potential, which is a measure of
the energy necessary to insert a new particle in the system,
diverges. Obviously, the system is extremely eager to accept
particles at very low fillings, in order to increase in absolute
value the kinetic energy and lower the total one, and absolutely
incompressible at $n=2$, when all the quantum states are filled.
At zero temperature, the Coulomb repulsion makes incompressible
also the states with commensurate fractional fillings $n=0.5$ and
$n=1.5$. At half-filling, on increasing the Coulomb repulsion, the
compressibility is rapidly suppressed according to the very high
price in energy that should be paid to add one particle to the
singlet state which is the one favored by the Coulomb repulsion.
The most relevant feature for this property is the presence of a
well-defined peak when it is plotted versus temperature. A finite,
but low in comparison with the actual value of the Coulomb
repulsion, temperature permits to overcome the suppression related
to the formation of a charge gap. A further increment of the
temperature makes available more and more states and the system is
driven back to be incompressible.

It is possible to write the compressibility for the Hubbard and
$t$-$J$ models in an identical way
\begin{equation}
\kappa=\frac{2}{T n^{2}}\left(\Gamma_{11c}-n^{2}\right)
\end{equation}
where $\Gamma_{11c}$ is the zero frequency function in the charge
channel and its ergodic value is exactly $n^{2}$. According to
this, the compressibility is a direct measure of the ergodicity of
the charge dynamics.

\begin{figure}[tbp]
\includegraphics[width=8cm,keepaspectratio=true]{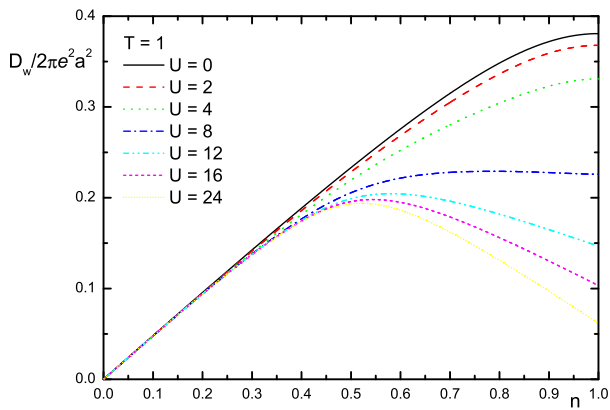}
\caption{The Drude weight $D_w$ as a function of $n$ for $T=1$ and
$U=0$, $2$, $4$, $8$, $12$, $16$ and $24$.} \label{Fig32}
\end{figure}

\subsection{Optical conductivity}\label{Dw}

In the framework of the linear response and by using the
Ward-Takahashi identities\cite{Mancini:98a}, which relate the
current-current propagator to the charge-charge one by exploiting
the charge conservation, the optical conductivity is given by
\begin{equation}
\sigma_{1}(\omega)=D_w \delta(\omega)
\end{equation}
with the Drude weight $D_w$ given by
\begin{equation}
D_w=-4\pi \mathrm{e}^{2} a^{2} t C_{cc}^{\alpha}
\end{equation}
For the 2-site system, the incoherent part of optical conductivity
is zero as no contribution comes from the imaginary part of
retarded current-current propagator.

In Fig.~\ref{Fig32}, the Drude weight $D_w$, normalized by $2\pi
\mathrm{e}^{2} a^{2}$, is reported as function of the filling $n$
for different values of the Coulomb repulsion $U$ and $T=1$. At
half-filling, the Coulomb repulsion tends to suppress the Drude
weight and drives the system to be insulating. However, the Drude
weight vanishes only in the limit $U$ infinite. Higher the
temperature more states with no contribution to the kinetic energy
result available and lower is the Drude weight.

\begin{figure}[tbp]
\includegraphics[width=8cm,keepaspectratio=true]{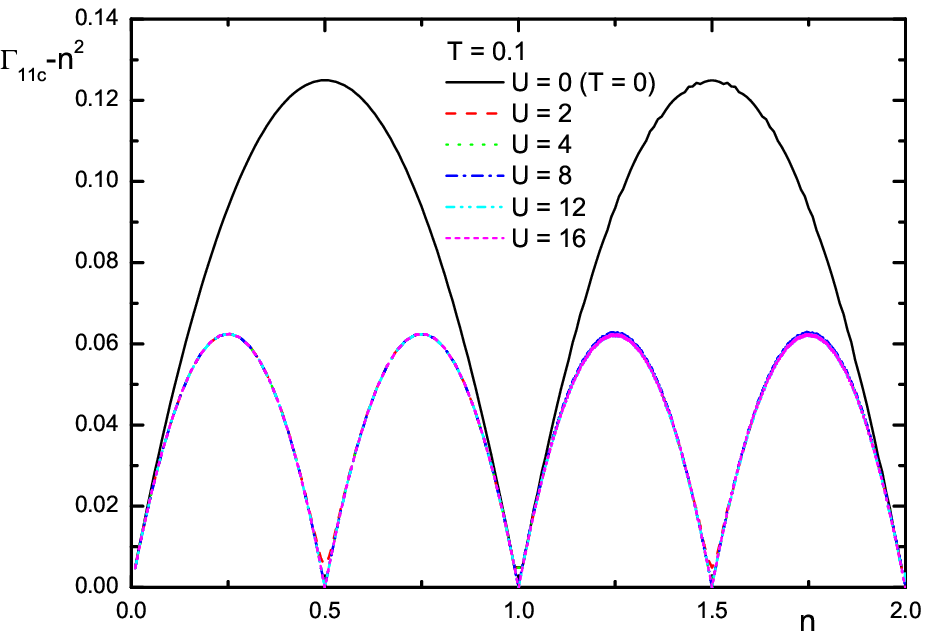}
\includegraphics[width=8cm,keepaspectratio=true]{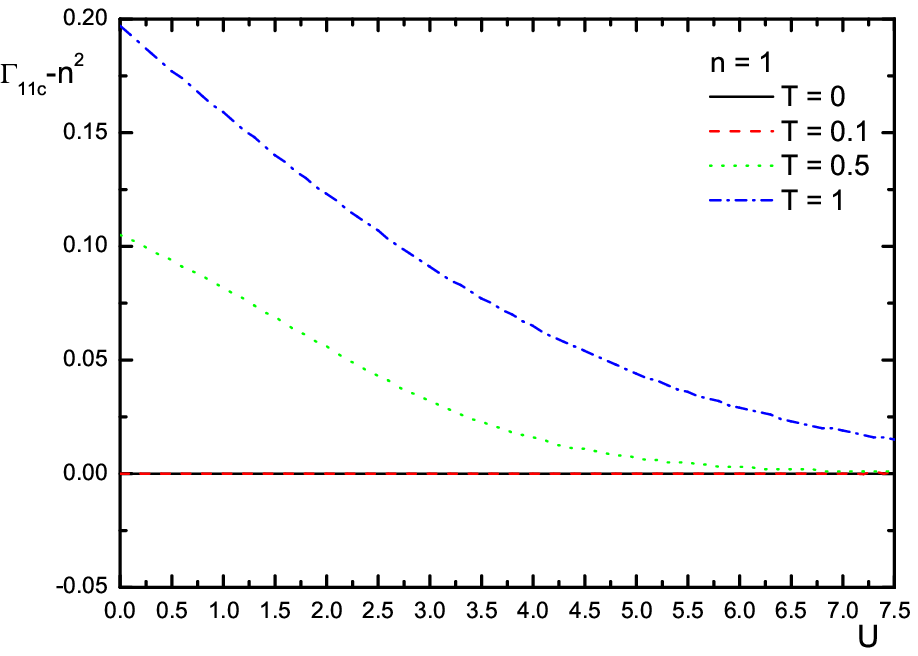}
\caption{The charge zero frequency constant $\Gamma_{11c}$
diminished of $n^2$: (top) as function of $n$ for $T=0.1$ and
$U=0$, $2$, $4$, $8$, $12$ and $16$; (bottom) as function of $U$
at $n=1$ and for $T=0$, $0.1$, $0.5$ and $1$.} \label{Fig35}
\end{figure}

\begin{figure}[tbp]
\includegraphics[width=8cm,keepaspectratio=true]{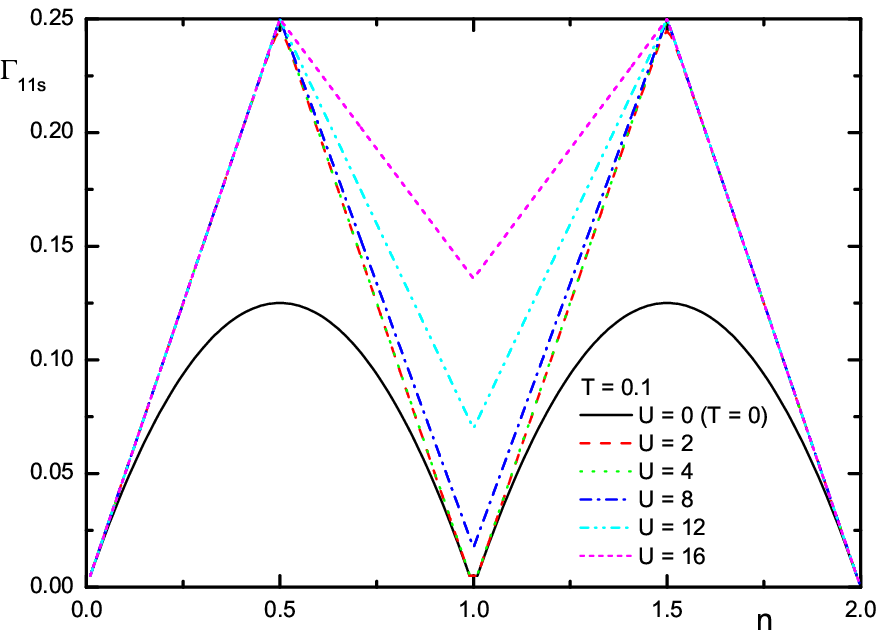}
\includegraphics[width=8cm,keepaspectratio=true]{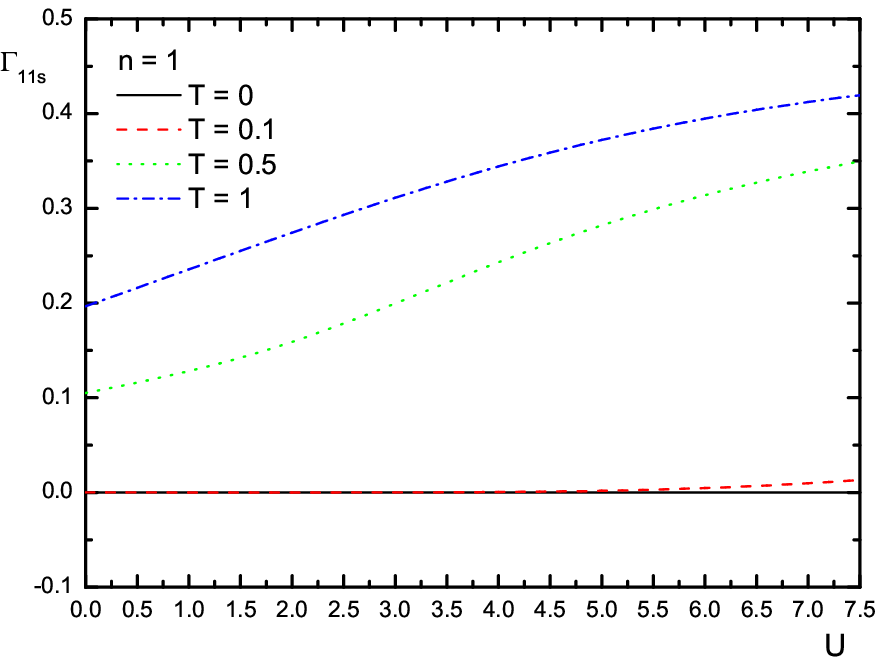}
\caption{The spin zero frequency constant $\Gamma_{11s}$: (top) as
function of $n$ for $T=0.1$ and $U=0$, $2$, $4$, $8$, $12$ and
$16$; (bottom) as function of $U$ at $n=1$ and for $T=0$, $0.1$,
$0.5$ and $1$.} \label{Fig38}
\end{figure}

\section{Ergodicity}

One of the main issues on which this manuscript wishes to draw
attention is the non-ergodicity of the charge and spin dynamics in
the two-site Hubbard model. The non-ergodic dynamics in this
system is due to the finite number of degenerate states available
to the system when dealing with finite temperatures and/or
incommensurate fillings and/or no interaction (cfr.
Table~\ref{HubTh} and Eq.~\ref{Gdef}). The finite level spacing
confines the system to degenerate or non-degenerate states,
according to the filling and interaction strength, making the
dynamic ergodic or non-ergodic according to the degree of
degeneracy of the ground state. The temperature then opens up the
possibility for quite complicate mixtures, only possible according
to our choice to work in the grand-canonical ensemble in order to
get results comparable with larger systems for which the level
spacing is thinner. The coupling to an heat and a particle
reservoirs have given us the possibility to simulate a continuous
tuning of the filling and temperature on a finite system (with a
finite degree of freedom) as it is possible only in a bulk system
with infinite degrees of freedom.

The zero-frequency constants for charge $\Gamma_{11c}$ and spin
$\Gamma_{11s}$ channels have been plotted as functions of the
filling $n$ for different values of the Coulomb repulsion $U$ and
$T=0.1$ and as functions of the interaction $U$ for different
values of temperature $T$ and $n=1$ in Figs.~\ref{Fig35} and
\ref{Fig38}. Their ergodic values are $n^2$ and $0$, respectively.
$\Gamma_{11c}$ assumes its ergodic value only at zero temperature
and commensurate fillings; in particular, only at integer
commensurate fillings for any interaction strength and also at
fractional ones for a finite interaction strength. On the other
hand, $\Gamma_{11s}$ assumes its ergodic value only at zero
temperature and integer commensurate fillings for any interaction
strength. At half-filling, while the charge dynamics is
constrained to be ergodic in the strongly interacting limit
($U\rightarrow\infty$) by the formation of a charge gap, the spin
dynamics gets more and more non-ergodic as larger the interaction
strength is. The states that are left available in this condition
have a quite different behavior once probed with respect to the
charge or spin response.

\section{Conclusions}
We have studied the 2-site Hubbard and $t$-$J$ models by means of
the Green's function and equations of motion formalism. The main
results can be so summarized:
\begin{itemize}
  \item We have got a complete basis of eigenoperators for the
  fermionic and bosonic sectors which could be used to get a controlled
  approximation in the study of the lattice case by means of any
  approximation that strongly relies on the choice of the basic
  field.
  \item We have identified the eigenoperator responsible for the
  appearance of the exchange scale of energy $J$.
  \item We have illustrated, once more\cite{Mancini:00}, as the
  local algebra constrains can properly fix the representation
  and easily give the values of the bosonic correlations that
  appear as internal parameters in the fermionic sector and
  of the zero frequency functions that appear in the bosonic
  sector. This also permits, while studying the fermionic sector,
  to avoid the opening of the bosonic one (i.e., the charge, the spin
  and pair channels) and all the heavy calculations required
  to solve it.
  \item We have explored the possibility of the 2-site systems
  to mime the behavior of larger clusters, as regards some of their physical properties, by
  using an higher temperature and by exploiting the qualitative
  properties of the level spacing. We have to report that the exact
  results of the 2-site system managed to reproduce the
  Mott-Hubbard \emph{MIT} of the bulk, the
  behavior of some local quantities (i.e., the double occupancy,
  the local magnetic moment and the kinetic energy) and of some thermodynamic
  quantities (i.e., the energy, the specific heat and the entropy)
  of larger clusters. This shows that the necessary energy scales
  are already present in the 2-site system. According to this, as already said
  in the first point above, we strongly believe that the operatorial basis that exactly
  solves this system can also give excellent results if used for the bulk.
  \item The study of the specific heat has given many valuable
  information regarding: the scales of energy present in the
  system, their origin, interaction and possibility of resolution,
  the range of parameters for which the $t$-$J$ model faithfully
  reproduces the low energy/temperature behavior of the Hubbard
  model, the explanation for the temperature dependence of the
  local properties, the existence of crossing points.
  \item We have shown how relevant is the determination of the
  zero-frequency constants in order to correctly compute the
  bosonic Green's functions. In particular, we have shown that,
  for the 2-site system, they assume values very far from the
  ergodic ones.
\end{itemize}

\begin{acknowledgments}
We wish to thank A. Moreo\cite{Duffy:97b} and P.
Prelovsek\cite{Prelovsek,Bonca:02,Jaklic:00} for providing us with
the numerical data.
\end{acknowledgments}

\appendix

\section{The eigenproblem}
\label{App:Eigen}

The eigenproblem of a given \emph{grand canonical} Hamiltonian $H$
is solved once the latter is diagonalized on the Fock space of the
system under study, that is
\begin{subequations}
\begin{align} \label{eigenstuff}
H\left| n\right\rangle &=E_{n}\left| n\right\rangle \\
N\left| n\right\rangle &=N_{n}\left| n\right\rangle
\end{align}
\end{subequations}
where $N=\sum_{i}n(i)$ is the total number operator and $\left|
n\right\rangle$ is a complete orthonormal basis.

Once the eigenproblem is solved, we can compute the thermal
average of an operator $\Phi$ by means of the following expression
\begin{equation} \label{tave}
\left\langle \Phi \right\rangle =\frac{1}{Z}\sum_{n}\left\langle
n\right| \Phi \left| n\right\rangle \mathrm{e}^{-\beta  E_{n}}
\end{equation}
where $Z=\sum_{n}\mathrm{e}^{-\beta  E_{n}}$ is the grand
canonical partition function and $\beta $ is the inverse
temperature.

If a finite minimal energy $E_{{\rm min}}$ exists then
\begin{equation}
\lim_{T\mapsto 0}\left\langle \Phi \right\rangle =\frac{1}{M}
\sum_{n | E_{n}=E_{{\rm min}}}\left\langle n\right| \Phi \left|
n\right\rangle
\end{equation}
where $M$ is the number of eigenstates $\left| n\right\rangle $
with $ E_{n}=E_{{\rm min}}$.

On the basis of the knowledge of the set of eigenstates and
eigenvalues of $H$ (\ref{eigenstuff}) it is possible to derive the
expressions for Green's functions and correlation functions. Let
$\psi(i)$ be a field operator in the Heisenberg scheme
$\psi(i)=\psi(\mathbf{i},t)=\mathrm{e}^{\mathrm{i}Ht}\psi(\mathbf{i})\mathrm{e}^{-\mathrm{i}Ht}$;
we do not specify the nature, fermionic or bosonic, of $\psi(i)$
($\psi(i)$ can be, for instance, either $c(i)$ or $n(i)$). By
considering the two-time thermodynamic Green's functions
\cite{Bogolubov:59,Zubarev:60,Zubarev:74}, let us define the
causal function
\begin{multline} \label{Gt}
G^{C(\eta)}_{\psi\psi^\dagger}(i,j)=\theta\left(t_i-t_j\right)\left\langle\psi(i)\psi^\dagger(j)\right\rangle
\\ -\eta\theta\left(t_j-t_i\right)\left\langle\psi^\dagger(j)\psi(i)\right\rangle
\end{multline}
the retarded and advanced functions
\begin{equation} \label{Gra}
G^{R,A(\eta)}_{\psi\psi^\dagger}(i,j)=\pm\theta\left[\pm\left(t_i-t_j\right)\right]
\left\langle\left[\psi(i),\psi^\dagger(j)\right]_\eta\right\rangle
\end{equation}
and the correlation function
\begin{equation} \label{Gc}
C_{\psi\psi^\dagger}(i,j)=\left\langle\psi(i)\psi^\dagger(j)\right\rangle
\end{equation}
Here $\eta=\pm1$; usually, it is convenient to take $\eta=1$
($\eta=-1$) for a fermionic (bosonic) field $\psi$ (i.e., for a
composite field constituted of an odd (even) number of original
fields) in order to exploit the canonical (anti)commutation
relations of the constituting original fields; but, in principle,
both choices are possible. Accordingly, we define
\begin{equation}
\left[A,B\right]_\eta=
\begin{cases}
\left\{A,B\right\}=AB+BA & \text{for } \eta=1 \\
\left[A,B\right]=AB-BA & \text{for } \eta=-1
\end{cases}
\end{equation}
By using the definition of thermal average (\ref{tave}) it is
possible to derive the following expressions for the Green's
functions (\ref{Gt}, \ref{Gra}) and correlation functions
(\ref{Gc}) in terms of eigenstates and eigenvalues
\begin{multline}\label{Gnt}
G^{C(\eta)}_{\psi\psi^\dagger}(\mathbf{i},\mathbf{j},\omega)
=\Gamma_{\psi\psi^\dagger}(\mathbf{i},\mathbf{j})
\left[(1+\eta)\mathcal{P}\frac{1}{\omega}-\mathrm{i}\pi(1-\eta)\delta(\omega)\right] \\
+\frac{1}{Z}\sum_{\substack{n,m \\ E_n \neq E_m}}\left(\frac{
A^{n,m}_{\psi\psi^\dagger}(\mathbf{i},\mathbf{j})\mathrm{e}^{-\beta
E_n}}{\omega+E_n-E_m+\mathrm{i}\delta}+\frac{\eta
A^{n,m}_{\psi\psi^\dagger}(\mathbf{i},\mathbf{j})\mathrm{e}^{-\beta
E_m}}{\omega+E_n-E_m-\mathrm{i}\delta}\right)
\end{multline}
\begin{multline}\label{Gnra}
G^{R,A(\eta)}_{\psi\psi^\dagger}(\mathbf{i},\mathbf{j},\omega)
=\Gamma_{\psi\psi^\dagger}(\mathbf{i},\mathbf{j})
\frac{1+\eta}{\omega\pm\mathrm{i}\delta} \\
+\frac{1}{Z}\sum_{\substack{n,m \\ E_n \neq E_m}}\frac{
A^{n,m}_{\psi\psi^\dagger}(\mathbf{i},\mathbf{j})\left(\mathrm{e}^{-\beta
E_n}+\eta \mathrm{e}^{-\beta
E_m}\right)}{\omega+E_n-E_m\pm\mathrm{i}\delta}
\end{multline}
\begin{multline}\label{Gnc}
C_{\psi\psi^\dagger}(\mathbf{i},\mathbf{j},\omega)
=2\pi\Gamma_{\psi\psi^\dagger}(\mathbf{i},\mathbf{j})
\delta(\omega)\\
+\frac{2\pi}{Z}\sum_{\substack{n,m \\ E_n \neq E_m}}
\mathrm{e}^{-\beta E_n}
A^{n,m}_{\psi\psi^\dagger}(\mathbf{i},\mathbf{j})\delta(\omega+E_n-E_m)
\end{multline}
where
\begin{equation}
A^{n,m}_{\psi\psi^\dagger}(\mathbf{i},\mathbf{j})=\langle n |
\psi(\mathbf{i}) | m \rangle \langle m | \psi^\dagger(\mathbf{j})
| n \rangle
\end{equation}
and the zero frequency constant
$\Gamma_{\psi\psi^\dagger}(\mathbf{i},\mathbf{j})$ has the
following representation
\begin{equation}\label{Gdef}
\Gamma_{\psi\psi^\dagger}(\mathbf{i},\mathbf{j})=\frac{1}{Z}\sum_{\substack{n,m
\\ E_n = E_m}} \mathrm{e}^{-\beta E_n}
A^{n,m}_{\psi\psi^\dagger}(\mathbf{i},\mathbf{j})
\end{equation}

It is worth noticing that, in this presentation (i.e., in terms of
eigenstates and eigenvalues), the Green's functions and
correlation functions are fully determined up to the value of the
chemical potential, which is present in the expressions of the
eigenvalues. As usual, the computation of the chemical potential
requires the inversion of the expression, in terms of eigenstates
and eigenvalues, for the number of particle per site
(\ref{eqmun}). In the main text instead, we presented expressions
for the same quantities (Green's and correlation functions) in
terms of eigenenergies of eigenoperators and of correlation
functions of these latter. In this case, more parameters appeared
($\Delta$ and $p$ in the fermionic sector and $\chi_{s}^{\alpha}$,
$d$, $\gamma$ and the zero frequency constant $\Gamma_{11\mu}$ in
the bosonic sector) that have been computed self-consistently as
we usually do for the chemical potential. The two presentations,
although equivalent (i.e., they obviously lead to the same
results), require two different self-consistent procedures to come
to the computation of the physical quantities. The reason of this
occurrence resides in the level of knowledge of the representation
we have in the two cases. In the first case, we have the full
knowledge of the states of the system and we should only fix their
occupancy with respect to the average number of particle per site
we wish to fix; in the second case, as we have no direct knowledge
of the states before fixing the counting we have to reduce the
Hilbert space to the correct one (i.e., the one of the system
under analysis), that is, we have to impose constraints in order
to select only those states that enjoy the correct \emph{symmetry}
properties (e.g., we have to discard states with site double
occupied by electrons with parallel spins). At any rate, we want
to emphasize once more that all the results obtained in this paper
by means of the Green's function formalism exactly coincide, as it
should be, with those computable by means of the thermal averages.

\begin{table*}[tb!!]
\begin{tabular}{||l||l|l|l|l||}
 \hline \hline $n$ & $|n\rangle$ & $E_n$ & $N_n$ & $S_{zn}$ \\
 \hline \hline $1$ & $\left| 0,0\right\rangle$ & $0$ & $0$ & $0$ \\
 \hline $2$ & $\frac{1}{\sqrt{2}}\left[ \left| \uparrow,0\right\rangle -\left| 0,\uparrow
 \right\rangle \right]$ & $-\mu +2t$ & $1$ & $\frac12$ \\
 \hline $3$ & $\frac{1}{\sqrt{2}}\left[ \left| \downarrow,0\right\rangle -\left|
 0,\downarrow \right\rangle \right]$ & $-\mu+2t$ & $1$ & $-\frac12$ \\
 \hline $4$ & $\frac{1}{\sqrt{2}}\left[ \left| \uparrow,0\right\rangle +\left| 0,\uparrow
 \right\rangle \right]$ & $-\mu -2t$ & $1$ & $\frac12$ \\
 \hline $5$ & $\frac{1}{\sqrt{2}}\left[ \left| \downarrow,0\right\rangle +\left| 0,\downarrow
 \right\rangle \right]$ & $-\mu -2t$ & $1$ & $-\frac12$ \\
 \hline $6$ & $\left| \uparrow ,\uparrow \right\rangle$ & $-2\mu$ & $2$ & $1$ \\
 \hline $7$ & $\left| \downarrow ,\downarrow \right\rangle$ & $-2\mu$ & $2$ & $-1$ \\
 \hline $8$ & $\frac{1}{\sqrt{2}}\left[ \left| \uparrow ,\downarrow\right\rangle
 +\left| \downarrow ,\uparrow\right\rangle \right]$ & $-2\mu$ & $2$ & $0$ \\
 \hline $9$ & $\alpha _{1}\left[ \left|\uparrow ,\downarrow \right\rangle -\left|
 \downarrow ,\uparrow \right\rangle\right] -\alpha _{2} \left[ \left| \uparrow \downarrow
 ,0\right\rangle +\left|0,\uparrow\downarrow \right\rangle \right]$ & $-2\mu -4J_{U}$ & $2$ & $0$ \\
 \hline $10$ & $\frac{1}{\sqrt{2}}\left[ \left| \uparrow \downarrow,0\right\rangle
 -\left| 0,\uparrow \downarrow \right\rangle \right]$ & $-2\mu +U$ & $2$ & $0$ \\
 \hline $11$ & $\alpha _{2}\left[ \left| \uparrow ,\downarrow\right\rangle -\left|
 \downarrow ,\uparrow \right\rangle \right] +\alpha _{1}\left[ \left| \uparrow
 \downarrow ,0\right\rangle +\left| 0,\uparrow \downarrow\right\rangle \right]$ & $-2\mu +U+4J_{U}$ & $2$ & $0$ \\
 \hline $12$ & $\frac{1}{\sqrt{2}}\left[ \left| \uparrow \downarrow,\uparrow
 \right\rangle +\left| \uparrow ,\uparrow \downarrow\right\rangle \right]$ & $-3\mu +2t+U$ & $3$ & $\frac12$ \\
 \hline $13$ & $\frac{1}{\sqrt{2}}\left[ \left| \uparrow \downarrow,\downarrow
 \right\rangle +\left| \downarrow ,\uparrow \downarrow\right\rangle \right]$ & $-3\mu +2t+U$ & $3$ & $-\frac12$ \\
 \hline $14$ & $\frac{1}{\sqrt{2}}\left[ \left| \uparrow \downarrow,\uparrow
 \right\rangle -\left| \uparrow ,\uparrow \downarrow\right\rangle \right]$ & $-3\mu -2t+U$ & $3$ & $\frac12$ \\
 \hline $15$ & $\frac{1}{\sqrt{2}}\left[ \left| \uparrow \downarrow,\downarrow
 \right\rangle -\left| \downarrow ,\uparrow \downarrow\right\rangle \right]$ & $-3\mu -2t+U$ & $3$ & $-\frac12$ \\
 \hline $16$ & $\left| \uparrow \downarrow ,\uparrow \downarrow\right\rangle$ & $-4\mu +2U$ & $4$ & $0$ \\
 \hline \hline
\end{tabular}
\caption{The eigenstates and eigenenergies of the 2-site Hubbard
and $t$-$J$ (first nine with $\alpha_1=1/\sqrt{2}$, $\alpha_2=0$
and $J_U=J$) models. In the last column is reported the
corresponding eigenvalue of $S_z$ that, together with $E_n$ the
eigenvalue of the Hamiltonian, completely characterized the states
(i.e., the presence of an external magnetic field will completely
lift the degeneracy).} \label{HubTh}
\end{table*}

The eigenstates of the systems under study are given by linear
combinations of the vectors spanning their Fock spaces. These
vectors can be displayed as $\left| a, b\right\rangle$ where $a$
and $b$ denote the occupancy of each site, respectively. In
particular, we have $0$ for an empty site, $\uparrow $ ($
\downarrow $) for a single occupied site by a spin-up (spin-down)
electron and $ \uparrow \downarrow $ for a double occupied site.
This latter is not allowed in the $t$-$J$ model (see
Tab.~\ref{HubTh}).

\subsection{The Hubbard model}

The eigenstates and eigenenergies of the 2-site Hubbard model are
reported in Tab.~\ref{HubTh}. The coefficients $\alpha _{1}$ and
$\alpha _{2}$ are determined by the orthonormality of the
eigenstates and have the following expressions:
\begin{align}
\alpha _{1} &=\frac{\left(U+4J_{U}\right)\sqrt{2}}{2\sqrt{\left(U+4J_{U}\right)^{2}+16t^{2}}} \\
\alpha _{2}
&=-\frac{2t\sqrt{2}}{\sqrt{\left(U+4J_{U}\right)^{2}+16t^{2}}}
\end{align}
We also have $\alpha _{1}^{2}+\alpha _{2}^{2}=1/2$.

Expressions (\ref{Gnt}), (\ref{Gnra}) and (\ref{Gnc}) show that
the Green's functions and the correlation functions are completely
determined (up to the value of the chemical potential) once the
matrices $A^{n,m}_{\psi\psi^\dagger}(\mathbf{i},\mathbf{j})$ are
known. We here give the results for some relevant operators
($\mathbf{i}$, $\mathbf{j}=\mathbf{1}$, $\mathbf{2}$;
$\Delta_{a,b}=\delta_{a,n}\delta_{b,m}$):

\begin{description}
 \item[\textbf{Operator $\xi(i)$}]
 \begin{align}
  &A^{n,m}_{\xi_\uparrow\xi^\dagger_\uparrow}(\mathbf{1},\mathbf{1})=B^{n,m}_{\xi_\uparrow\xi^\dagger_\uparrow}+C^{n,m}_{\xi_\uparrow\xi^\dagger_\uparrow}\\
  &A^{n,m}_{\xi_\uparrow\xi^\dagger_\uparrow}(\mathbf{1},\mathbf{2})=-B^{n,m}_{\xi_\uparrow\xi^\dagger_\uparrow}+C^{n,m}_{\xi_\uparrow\xi^\dagger_\uparrow}\\
  &A^{n,m}_{\xi_\downarrow\xi^\dagger_\downarrow}(\mathbf{1},\mathbf{1})=B^{n,m}_{\xi_\downarrow\xi^\dagger_\downarrow}+C^{n,m}_{\xi_\downarrow\xi^\dagger_\downarrow}\\
  &A^{n,m}_{\xi_\downarrow\xi^\dagger_\downarrow}(\mathbf{1},\mathbf{2})=-B^{n,m}_{\xi_\downarrow\xi^\dagger_\downarrow}+C^{n,m}_{\xi_\downarrow\xi^\dagger_\downarrow}
 \end{align}
\noindent where
 \begin{multline}
  B^{n,m}_{\xi_\uparrow\xi^\dagger_\uparrow}= \frac12\left(\Delta_{1,2}+\Delta_{4,6}+\frac12\Delta_{5,8}+\alpha_1^2\Delta_{3,9}\right) \\
  +\frac12\left(\alpha_2^2\Delta_{3,11}+\alpha_2^2\Delta_{9,14}+\frac12\Delta_{10,12}+\alpha_1^2\Delta_{11,14}\right)
 \end{multline}
 \begin{multline}
  C^{n,m}_{\xi_\uparrow\xi^\dagger_\uparrow}=
  \frac12\left(\Delta_{1,4}+\Delta_{2,6}+\frac12\Delta_{3,8}+\alpha_1^2\Delta_{5,9}\right)\\
  +\frac12\left(\alpha_2^2\Delta_{5,11}+\alpha_2^2\Delta_{9,12}+\frac12\Delta_{10,14}+\alpha_1^2\Delta_{11,12}\right)
 \end{multline}
 \begin{multline}
  B^{n,m}_{\xi_\downarrow\xi^\dagger_\downarrow}= \frac12\left(\Delta_{1,3}+\Delta_{5,7}+\frac12\Delta_{4,8}+\alpha_1^2\Delta_{2,9}\right) \\
  +\frac12\left(\alpha_2^2\Delta_{2,11}+\alpha_2^2\Delta_{9,15}+\frac12\Delta_{10,13}+\alpha_1^2\Delta_{11,15}\right)
 \end{multline}
 \begin{multline}
  C^{n,m}_{\xi_\downarrow\xi^\dagger_\downarrow}= \frac12\left(\Delta_{1,5}+\Delta_{3,7}+\frac12\Delta_{2,8}+\alpha_1^2\Delta_{4,9}\right) \\
  +\frac12\left(\alpha_2^2\Delta_{4,11}+\alpha_2^2\Delta_{9,13}+\frac12\Delta_{10,15}+\alpha_1^2\Delta_{11,13}\right)
 \end{multline}
 \item[\textbf{Operator $\eta(i)$}]
 \begin{align}
  &A^{n,m}_{\eta_\uparrow\eta^\dagger_\uparrow}(\mathbf{1},\mathbf{1})=B^{n,m}_{\eta_\uparrow\eta^\dagger_\uparrow}+C^{n,m}_{\eta_\uparrow\eta^\dagger_\uparrow}\\
  &A^{n,m}_{\eta_\uparrow\eta^\dagger_\uparrow}(\mathbf{1},\mathbf{2})=-B^{n,m}_{\eta_\uparrow\eta^\dagger_\uparrow}+C^{n,m}_{\eta_\uparrow\eta^\dagger_\uparrow}\\
  &A^{n,m}_{\eta_\downarrow\eta^\dagger_\downarrow}(\mathbf{1},\mathbf{1})=B^{n,m}_{\eta_\downarrow\eta^\dagger_\downarrow}+C^{n,m}_{\eta_\downarrow\eta^\dagger_\downarrow}\\
  &A^{n,m}_{\eta_\downarrow\eta^\dagger_\downarrow}(\mathbf{1},\mathbf{2})=-B^{n,m}_{\eta_\downarrow\eta^\dagger_\downarrow}+C^{n,m}_{\eta_\downarrow\eta^\dagger_\downarrow}
 \end{align}
\noindent where
 \begin{multline}
  B^{n,m}_{\eta_\uparrow\eta^\dagger_\uparrow}= \frac12\left(\Delta_{7,15}+\Delta_{15,16}+\frac12\Delta_{5,10}+\alpha_1^2\Delta_{3,11}\right) \\
  +\frac12\left(\alpha_2^2\Delta_{3,9}+\alpha_2^2\Delta_{11,12}+\frac12\Delta_{8,14}+\alpha_1^2\Delta_{9,12}\right)
 \end{multline}
 \begin{multline}
  C^{n,m}_{\eta_\uparrow\eta^\dagger_\uparrow}= \frac12\left(\Delta_{7,13}+\Delta_{13,16}+\frac12\Delta_{3,10}+\alpha_1^2\Delta_{5,11}\right) \\
  +\frac12\left(\alpha_2^2\Delta_{5,9}+\alpha_2^2\Delta_{11,14}+\frac12\Delta_{8,12}+\alpha_1^2\Delta_{9,14}\right)
 \end{multline}
 \begin{multline}
  B^{n,m}_{\eta_\downarrow\eta^\dagger_\downarrow}= \frac12\left(\Delta_{6,14}+\Delta_{14,16}+\frac12\Delta_{4,10}+\alpha_1^2\Delta_{2,11}\right) \\
  +\frac12\left(\alpha_2^2\Delta_{2,9}+\alpha_2^2\Delta_{11,13}+\frac12\Delta_{8,15}+\alpha_1^2\Delta_{9,13}\right)
 \end{multline}
 \begin{multline}
  C^{n,m}_{\eta_\downarrow\eta^\dagger_\downarrow}= \frac12\left(\Delta_{6,12}+\Delta_{12,16}+\frac12\Delta_{2,10}+\alpha_1^2\Delta_{4,11}\right) \\
  +\frac12\left(\alpha_2^2\Delta_{4,9}+\alpha_2^2\Delta_{11,15}+\frac12\Delta_{8,13}+\alpha_1^2\Delta_{9,15}\right)
 \end{multline}
 \item[\textbf{Operator $\mathrm{n}(i)=c^\dagger(i)c(i)$}]
 \begin{align}
  &A^{n,m}_{\mathrm{n}\mathrm{n}}(\mathbf{1},\mathbf{1})=B^{n,m}_{\mathrm{n}\mathrm{n}}+C^{n,m}_{\mathrm{n}\mathrm{n}}+D^{n,m}_{\mathrm{n}\mathrm{n}}\\
  &A^{n,m}_{\mathrm{n}\mathrm{n}}(\mathbf{1},\mathbf{2})=B^{n,m}_{\mathrm{n}\mathrm{n}}-C^{n,m}_{\mathrm{n}\mathrm{n}}-D^{n,m}_{\mathrm{n}\mathrm{n}}
 \end{align}
\noindent where
 \begin{multline}
  B^{n,m}_{\mathrm{n}\mathrm{n}}= \frac14\left(\Delta_{2,2}+\Delta_{3,3}+\Delta_{4,4}+\Delta_{5,5}+4\Delta_{6,6}\right) \\
  + \left(\Delta_{7,7}+\Delta_{8,8}+\Delta_{9,9}+\Delta_{10,10}+\Delta_{11,11}\right) \\
  + \frac14\left(9\Delta_{12,12}+9\Delta_{13,13}+9\Delta_{14,14}+9\Delta_{15,15}+16\Delta_{16,16}\right)
 \end{multline}
 \begin{multline}
  C^{n,m}_{\mathrm{n}\mathrm{n}}= \frac14\left(\Delta_{2,4}+\Delta_{3,5}+\Delta_{4,2}+\Delta_{5,3}\right) \\
  + \frac14\left(\Delta_{12,14}+\Delta_{13,15}+\Delta_{14,12}+\Delta_{15,13}\right)
 \end{multline}
 \begin{multline}
  D^{n,m}_{\mathrm{n}\mathrm{n}}= 2\left(\alpha_2^2\Delta_{9,10}+\alpha_2^2\Delta_{10,9}+\alpha_1^2\Delta_{10,11}+\alpha_1^2\Delta_{11,10}\right)
 \end{multline}
 \item[\textbf{Operator $n_3(i)$}]
 \begin{align}
  &A^{n,m}_{n_3n_3}(\mathbf{1},\mathbf{1})=B^{n,m}_{n_3n_3}+C^{n,m}_{n_3n_3}+D^{n,m}_{n_3n_3}\\
  &A^{n,m}_{n_3n_3}(\mathbf{1},\mathbf{2})=B^{n,m}_{n_3n_3}-C^{n,m}_{n_3n_3}-D^{n,m}_{n_3n_3}
 \end{align}
\noindent where
 \begin{multline}
  B^{n,m}_{n_3n_3}= \frac14\left(\Delta_{2,2}+\Delta_{3,3}+\Delta_{4,4}+\Delta_{5,5}+4\Delta_{6,6}\right) \\
  + \frac14\left(4\Delta_{7,7}+\Delta_{12,12}+\Delta_{13,13}+\Delta_{14,14}+\Delta_{15,15}\right)
 \end{multline}
 \begin{equation}
  C^{n,m}_{n_3n_3}=C^{n,m}_{n n}
 \end{equation}
 \begin{multline}
  D^{n,m}_{n_3n_3}= 2\left(\alpha_2^2\Delta_{8,11}+\alpha_2^2\Delta_{11,8}+\alpha_1^2\Delta_{8,9}+\alpha_1^2\Delta_{9,8}\right)
 \end{multline}
\end{description}

\subsection{The $t$-$J$ model}

The eigenstates and eigenenergies of the 2-site $t$-$J$ model
coincides with the first nine of the Hubbard model with
$\alpha_1=1/\sqrt{2}$, $\alpha_2=0$ and $J_U=J$. Actually, in the
strong coupling regime (i.e., $U \gg t$) we have
$\alpha_1\rightarrow 1/\sqrt{2}$, $\alpha_2\rightarrow 0$ and
$J_U\rightarrow 4t^{2}/U$. In particular, this latter value
($4t^{2}/U$) is the one we get in the strong coupling regime when
deriving the $t$-$J$ model from the Hubbard one. All other states
of the Hubbard model can not be realized in the $t$-$J$ model
owing to the exclusion of the double occupancies.

Following the same prescription (only first nine states,
$\alpha_1=1/\sqrt{2}$, $\alpha_2=0$ and $J_U=J$), it is possible
to get $A^{n,m}_{\xi\xi^\dagger}(\mathbf{i},\mathbf{j})$,
$A^{n,m}_{\nu\nu}(\mathbf{i},\mathbf{j})$ and
$A^{n,m}_{\nu_3\nu_3}(\mathbf{i},\mathbf{j})$ form the
corresponding expressions given for the Hubbard model
($A^{n,m}_{\xi\xi^\dagger}(\mathbf{i},\mathbf{j})$,
$A^{n,m}_{nn}(\mathbf{i},\mathbf{j})$ and
$A^{n,m}_{n_3n_3}(\mathbf{i},\mathbf{j})$, respectively).

\subsection{The properties}

Given the expressions of
$A^{n,m}_{\psi\psi^\dagger}(\mathbf{i},\mathbf{j})$, we can now
compute the relevant physical quantities.

\subsubsection{The Hubbard model}

The partition function is given by
\begin{align}
Z & =1+2\mathrm{e}^{-\beta E_{2}}+2\mathrm{e}^{-\beta
E_{4}}+3\mathrm{e}^{-\beta E_{6}}+\mathrm{e}^{-\beta
E_{9}}+\mathrm{e}^{-\beta E_{10}} \nonumber \\ &
+\mathrm{e}^{-\beta E_{11}}+2\mathrm{e}^{-\beta
E_{12}}+2\mathrm{e}^{-\beta E_{14}}+\mathrm{e}^{-\beta E_{16}}
\end{align}

The chemical potential $\mu$ can be computed inverting the
following expression for the particle number per site
\begin{align} \label{eqmun}
n &=\frac{1}{Z}\left(\mathrm{e}^{-\beta E_{2}}+\mathrm{e}^{-\beta
E_{4}}+3\mathrm{e}^{-\beta
E_{6}}+\mathrm{e}^{-\beta E_{9}}+\mathrm{e}^{-\beta E_{10}}\right. \nonumber \\
&\left. +\mathrm{e}^{-\beta E_{11}}+3\mathrm{e}^{-\beta
E_{12}}+3\mathrm{e}^{-\beta E_{14}}+2\mathrm{e}^{-\beta
E_{16}}\right)
\end{align}
This expression determines the chemical potential as a function of
$n$, $T$ and $U$.

The self-consistent parameter $\Delta $ and $p$ are given by
\begin{equation}
\Delta =\frac{1}{2Z}\left(\mathrm{e}^{-\beta
E_{2}}-\mathrm{e}^{-\beta E_{4}}-\mathrm{e}^{-\beta
E_{13}}+\mathrm{e}^{-\beta E_{15}}\right)
\end{equation}
\begin{align} \label{pthe}
p &=\frac{1}{2Z}\left(3\mathrm{e}^{-\beta
E_{6}}-\mathrm{e}^{-\beta E_{9}}+\mathrm{e}^{-\beta
E_{10}}-\mathrm{e}^{-\beta E_{11}}+2\mathrm{e}^{-\beta E_{12}} \right. \nonumber \\
& \left. +2\mathrm{e}^{-\beta E_{14}}+2\mathrm{e}^{-\beta
E_{16}}\right)
\end{align}

The correlation functions $C^\alpha_{12}$ and $C^\alpha_{cc}$ are
given by
\begin{equation}
C^\alpha_{12}=\frac{\alpha_1\alpha_2}{Z}\left(\mathrm{e}^{-\beta
E_{9}}-\mathrm{e}^{-\beta E_{11}}\right)
\end{equation}
\begin{align}
C^\alpha_{cc} &=\frac{1}{2Z}\left[\mathrm{e}^{-\beta
E_{2}}-\mathrm{e}^{-\beta
E_{4}}+4\alpha_1\alpha_2\left(\mathrm{e}^{-\beta
E_{9}}-\mathrm{e}^{-\beta E_{11}}\right)\right. \nonumber \\
&\left. +\mathrm{e}^{-\beta E_{12}}-\mathrm{e}^{-\beta
E_{14}}\right]
\end{align}

The double occupancy per site is given by
\begin{align}
\emph{D}&=\frac{1}{2Z}\left(2\alpha _{2}^{2}\mathrm{e}^{-\beta
E_{9}}+\mathrm{e}^{-\beta E_{10}}+2\alpha
_{1}^{2}\mathrm{e}^{-\beta E_{11}}+2\mathrm{e}^{-\beta E_{12}} \right. \nonumber \\
&\left. +2\mathrm{e}^{-\beta E_{14}}+2\mathrm{e}^{-\beta
E_{16}}\right)
\end{align}

Spin ($\chi^\alpha_s$), charge ($\chi^\alpha_c$) and pair ($d$)
correlation functions are given by
\begin{equation}
\chi^\alpha_s=\frac{3}{Z}\left(\mathrm{e}^{-\beta
E_{6}}-2\alpha_{1}^{2}\mathrm{e}^{-\beta E_{9}}-2\alpha
_{2}^{2}\mathrm{e}^{-\beta E_{11}}\right)
\end{equation}
\begin{align}
\chi^\alpha_c&=\frac{1}{Z}\left(3\mathrm{e}^{-\beta
E_{6}}+2\alpha_{1}^{2}\mathrm{e}^{-\beta
E_{9}}+2\alpha_{2}^{2}\mathrm{e}^{-\beta E_{11}} \right. \nonumber \\
&\left. +4\mathrm{e}^{-\beta E_{12}}+4\mathrm{e}^{-\beta
E_{14}}+4\mathrm{e}^{-\beta E_{16}}\right)
\end{align}
\begin{equation}
d=\frac{1}{2Z}\left(2\alpha _{2}^{2}\mathrm{e}^{-\beta
E_{9}}-\mathrm{e}^{-\beta
E_{10}}+2\alpha_{1}^{2}\mathrm{e}^{-\beta E_{11}}\right)
\end{equation}

The zero-frequency constant
$\Gamma_{110}=\Gamma_{\mathrm{n}\mathrm{n}}(\mathbf{1},\mathbf{2})$
and $\Gamma_{113}=\Gamma_{n_3n_3}(\mathbf{1},\mathbf{2})$ are
given by
\begin{align}
\Gamma_{110}&=\frac{1}{2Z}\left(\mathrm{e}^{-\beta
E_{2}}+\mathrm{e}^{-\beta
E_{4}}+6\mathrm{e}^{-\beta E_{6}}+2\mathrm{e}^{-\beta E_{9}}+2\mathrm{e}^{-\beta E_{10}} \right. \nonumber \\
&\left. +2\mathrm{e}^{-\beta E_{11}}+9\mathrm{e}^{-\beta
E_{12}}+9\mathrm{e}^{-\beta E_{14}}+8\mathrm{e}^{-\beta
E_{16}}\right)
\end{align}
\begin{equation}
\Gamma_{113}=\frac{1}{2Z}\left(\mathrm{e}^{-\beta
E_{2}}+\mathrm{e}^{-\beta E_{4}}+4\mathrm{e}^{-\beta
E_{6}}+\mathrm{e}^{-\beta E_{12}}+\mathrm{e}^{-\beta
E_{14}}\right)
\end{equation}

\subsection{The $t$-$J$ model}

The partition function is given by
\begin{equation}
Z =1+2\mathrm{e}^{-\beta E_{2}}+2\mathrm{e}^{-\beta
E_{4}}+3\mathrm{e}^{-\beta E_{6}}+\mathrm{e}^{-\beta E_{9}}
\end{equation}

The chemical potential $\mu $ can be computed through the
following expression for the particle number per site
\begin{equation}
n =\frac{1}{Z}(\mathrm{e}^{-\beta E_{2}}+\mathrm{e}^{-\beta
E_{4}}+3\mathrm{e}^{-\beta E_{6}}+\mathrm{e}^{-\beta E_{9}})
\end{equation}
which can be inverted and give
\begin{widetext}
\begin{equation}
\mu =T\ln \frac{(2n-1)\cosh (2\beta t)+\sqrt{(2n-1)^{2}\cosh
^{2}(2\beta t)+n(1-n)(3+\mathrm{e}^{4\beta
J})}}{(1-n)(3+\mathrm{e}^{4\beta J})}
\end{equation}
\end{widetext}

The self-consistent parameter $C^\alpha_{11}$ and $\chi^\alpha$
are given by
\begin{equation}
C^\alpha_{11} =\frac{1}{2Z}\left(\mathrm{e}^{-\beta
E_{2}}-\mathrm{e}^{-\beta E_{4}}\right)
\end{equation}
\begin{equation}
\chi^\alpha_\mu =\frac{2}{Z}\left(3\mathrm{e}^{-\beta
E_{6}}-\mathrm{e}^{-\beta E_{9}}\right)
\end{equation}

Spin ($\chi^\alpha_s$) and charge ($\chi^\alpha_c$) correlation
functions are given by
\begin{equation}
\chi^\alpha_s =\frac{1}{Z}\left(\mathrm{e}^{-\beta
E_{6}}-\mathrm{e}^{-\beta E_{9}}\right)
\end{equation}
\begin{equation}
\chi^\alpha_c =\frac{1}{Z}\left(3\mathrm{e}^{-\beta
E_{6}}+\mathrm{e}^{-\beta E_{9}}\right)
\end{equation}

The zero-frequency constant
$\Gamma_{110}=\Gamma_{\mathrm{n}\mathrm{n}}(\mathbf{1},\mathbf{2})$
and $\Gamma_{113}=\Gamma_{n_3n_3}(\mathbf{1},\mathbf{2})$ are
given by
\begin{align}
\Gamma_{110}&=\frac{1}{2Z}\left(\mathrm{e}^{-\beta
E_{2}}+\mathrm{e}^{-\beta E_{4}}+6\mathrm{e}^{-\beta
E_{6}}+2\mathrm{e}^{-\beta E_{9}}\right)
\end{align}
\begin{equation}
\Gamma_{113}=\frac{1}{2Z}\left(\mathrm{e}^{-\beta
E_{2}}+\mathrm{e}^{-\beta E_{4}}+4\mathrm{e}^{-\beta E_{6}}\right)
\end{equation}

\bibliographystyle{apsrev}
\bibliography{biblio}

\begin{thebibliography}{42}
\bibitem[{ZFC()}]{ZFC}
\bibinfo{note}{The zero-frequency constants naturally appear in the
  determination of the two-particle and, more generally, bosonic Green's
  functions. An exhaustive discussion about their origin and the way to
  determine them is given in Ref.~\protect\onlinecite{Mancini:00}.}
\end{thebibliography}

\end{document}